\documentclass[a4paper]{article}
\usepackage{defs} %

\usepackage{comment} %

\title{Multi-Agent Risks from Advanced AI}
\author{
    Lewis Hammond and Alan Chan and Jesse Clifton and Jason Hoelscher-Obermaier and Akbir Khan and Euan McLean and Chandler Smith and Wolfram Barfuss and Jakob Foerster and Tomáš Gavenčiak and The Anh Han and Edward Hughes and Vojtěch Kovařík and Jan Kulveit and Joel Z. Leibo and Caspar Oesterheld and Christian Schroeder de Witt and Nisarg Shah and Michael Wellman and Paolo Bova and Theodor Cimpeanu and Carson Ezell and Quentin Feuillade-Montixi and Matija Franklin and Esben Kran and Igor Krawczuk and Max Lamparth and Niklas Lauffer and Alexander Meinke and Sumeet Motwani and Anka Reuel and Vincent Conitzer and Michael Dennis and Iason Gabriel and Adam Gleave and Gillian Hadfield and Nika Haghtalab and Atoosa Kasirzadeh and Sébastien Krier and Kate Larson and Joel Lehman and David C. Parkes and Georgios Piliouras and Iyad Rahwan
}
\date{February 19 2025}

\begin{document}

\includepdf{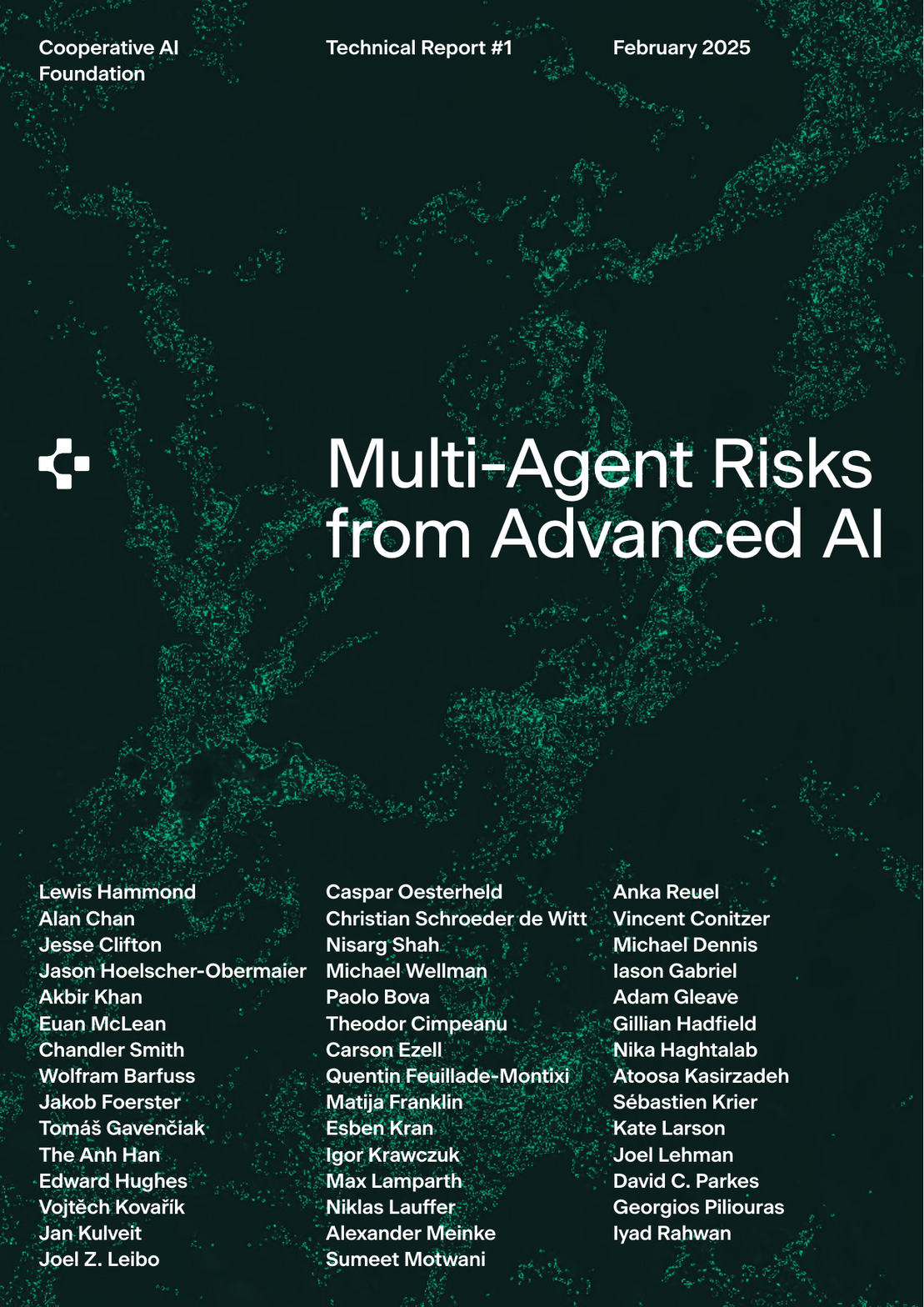}

\begin{center}
    {\Huge Multi-Agent Risks from Advanced AI}
\end{center}

\vspace{1.5em}

\renewcommand{\thefootnote}{\fnsymbol{footnote}}

\begin{minipage}[t]{0.52\linewidth}
    \raggedright
    \large
    {Lewis Hammond}$^{1,2,}$\footnotemark{}
    \vspace{1em}

    {Alan Chan}$^{3,4}$,
    {Jesse Clifton}$^{5,1}$,
    {Jason Hoelscher-Obermaier}$^{6}$,
    {Akbir Khan}$^{7,8,(1)}$,
    {Euan McLean}$^{\dagger}$,
    {Chandler Smith}$^{1}$
    \vspace{1em}

    {Wolfram Barfuss}$^{9}$,
    {Jakob Foerster}$^{2,10}$,
    {Tomáš Gavenčiak}$^{11}$,
    {The Anh Han}$^{12}$,
    {Edward Hughes}$^{13}$,
    {Vojtěch Kovařík}$^{14,(15)}$,
    {Jan Kulveit}$^{11}$,
    {Joel Z. Leibo}$^{13}$,
    {Caspar Oesterheld}$^{15}$,
    {Christian Schroeder de Witt}$^{2}$,
    {Nisarg Shah}$^{16}$,
    {Michael Wellman}$^{17}$
    \vspace{1em}

    {Paolo Bova}$^{12}$,
    {Theodor Cimpeanu}$^{18,(19)}$
    {Carson Ezell}$^{20}$,
    {Quentin Feuillade- Montixi}$^{21,(\dagger)}$,
    {Matija Franklin}$^{8}$,
    {Esben Kran}$^{6}$,
    {Igor Krawczuk}$^{\dagger,(22)}$,
    {Max Lamparth}$^{23}$,
    {Niklas Lauffer}$^{24}$,
    {Alexander Meinke}$^{25,(\dagger)}$,
    {Sumeet Motwani}$^{2,(24)}$,
    {Anka Reuel}$^{23,20}$
    \vspace{1em}

    {Vincent Conitzer}$^{15}$,
    {Michael Dennis}$^{13}$,
    {Iason Gabriel}$^{13}$,
    {Adam Gleave}$^{26}$,
    {Gillian Hadfield}$^{27}$,
    {Nika Haghtalab}$^{24}$,
    {Atoosa Kasirzadeh}$^{15}$,
    {Sébastien Krier}$^{13}$,
    {Kate Larson}$^{28,13}$,
    {Joel Lehman}$^{\dagger}$,
    {David C. Parkes}$^{20}$,
    {Georgios Piliouras}$^{13,29}$,
    {Iyad Rahwan}$^{30}$
\end{minipage}
\hfill 
\begin{minipage}[t]{0.44\linewidth}
    \raggedright
    \small
    $^{1}${Cooperative AI Foundation}\\
    $^{2}${University of Oxford}\\
    $^{3}${Mila}\\
    $^{4}${Centre for the Governance of AI}\\
    $^{5}${Center on Long-Term Risk}\\
    $^{6}${Apart Research}\\
    $^{7}${Anthropic}\\
    $^{8}${University College London}\\
    $^{9}${University of Bonn}\\
    $^{10}${Meta AI}\\
    $^{11}${Charles University}\\
    $^{12}${Teesside University}\\
    $^{13}${Google DeepMind}\\
    $^{14}${Czech Technical University}\\
    $^{15}${Carnegie Mellon University}\\
    $^{16}${University of Toronto}\\
    $^{17}${University of Michigan}\\
    $^{18}${University of Stirling}\\
    $^{19}${University of St Andrews}\\
    $^{20}${Harvard University}\\
    $^{21}${PRISM AI}\\
    $^{22}${École Polytechnique Fédérale de Lausanne}\\
    $^{23}${Stanford University}\\
    $^{24}${University of California, Berkeley}\\
    $^{25}${Apollo Research}\\
    $^{26}${FAR.AI}\\
    $^{27}${Johns Hopkins University}\\
    $^{28}${University of Waterloo}\\
    $^{29}${Singapore University of Technology and Design}\\
    $^{30}${Max Planck Institute for Human Development}\\
    $^\dagger${Independent}
\end{minipage}

\footnotetext{
    Correspondence to \url{lewis.hammond@cooperativeai.org}.
    Suggested citation: ``Hammond et al. (2025). \textit{Multi-Agent Risks from Advanced AI}. Cooperative AI Foundation, Technical Report \#1.''
    Author clusters are ordered by approximate magnitude of contribution and represent the lead author, organisers, major contributors, minor contributors, and advisors, respectively.
    Within clusters, authors are listed alphabetically.
    Full details of author roles are available in \Cref{app:contributions}.
    Affiliations in parentheses indicate that the author's work on this report was primarily completed while under that affiliation.
    Due to the length of the author list, authorship does not entail endorsement of all claims in the report, nor does inclusion entail an endorsement on the part of any individual's organisation.
    In particular, contributions to this report reflect the views of the respective contributors and not necessarily the views of the Cooperative AI Foundation, its trustees, or funders.
}
\renewcommand{\thefootnote}{\arabic{footnote}}
\setcounter{footnote}{0}

\vspace{1.5em}

\begin{center}
    {\large \textbf{Abstract}}
    \vspace{1em}

    \begin{minipage}[]{0.75\linewidth}
        The rapid development of advanced AI agents and the imminent deployment of many instances of these agents will give rise to multi-agent systems of unprecedented complexity.
        These systems pose novel and under-explored risks.
        In this report, we provide a structured taxonomy of these risks by identifying three key failure modes (miscoordination, conflict, and collusion) based on agents' incentives, as well as seven key risk factors (information asymmetries, network effects, selection pressures, destabilising dynamics, commitment problems, emergent agency, and multi-agent security) that can underpin them.
        We highlight several important instances of each risk, as well as promising directions to help mitigate them.
        By anchoring our analysis in a range of real-world examples and experimental evidence, we illustrate the distinct challenges posed by multi-agent systems and their implications for the safety, governance, and ethics of advanced AI.
    \end{minipage}
\end{center}

\newpage

\tableofcontents

\newpage

\addcontentsline{toc}{section}{Executive Summary}

\section*{Executive Summary}
\label{sec:executive_summary}

The proliferation of increasingly advanced AI not only promises widespread benefits, but also presents new risks \citep{Bengio2024a,Chan2023-aj}.
\textbf{Today, AI systems are beginning to autonomously interact with one another and adapt their behaviour accordingly, forming \textit{multi- agent systems}}.
This change is due to the widespread adoption of sophisticated models that can interact via a range of modalities (including text, images, and audio), and the competitive advantages conferred by autonomous, adaptive agents \citep{GDM_agents,Anthropic_agents,OpenAI_agents}.

While still relatively rare, groups of advanced AI agents are already responsible for tasks that range from trading million-dollar assets \citep{Ferreira2021,amplify_aieq,Sun2023_rl_qt} to recommending actions to commanders in battle \citep{palantir_aip_defense,manson2024ai,Black2024}.
\textbf{In the near future, applications will include not only economic and military domains, but are likely to extend to energy management, transport networks, and other critical infrastructure} \citep{mayorkas2024roles, Camacho2024}.
Large populations of AI agents will also feature in more familiar social settings as intelligent personal assistants or representatives, capable of being delegated increasingly complex and important tasks.

While bringing new opportunities for scalable automation and more diffuse benefits to society, \textbf{these advanced, multi-agent systems present novel risks that are distinct from those posed by \textit{single agents} or \textit{less advanced} technologies}, and which have been systematically underappreciated and understudied.
This lack of attention is partly because present-day multi-agent systems are rare (and those that do exist are often highly controlled, such as in automated warehouses), but also because even single agents present many unsolved problems \citep{Amodei2016,Hendrycks2021,Anwar2024}.
Given the current rate of progress and adoption, however, \textbf{we urgently need to evaluate (and prepare to mitigate) multi-agent risks from advanced AI}.
More concretely, we provide recommendations throughout the report that can largely be classified as follows.

\begin{itemize}
    \item \textbf{Evaluation}: Today's AI systems are developed and tested in isolation, despite the fact that they will soon interact with each other.
    In order to understand how likely and severe multi-agent risks are, we need new methods of detecting how and when they might arise, such as:
    evaluating the cooperative capabilities, biases, and vulnerabilities of models; 
    testing for new or improved dangerous capabilities in multi-agent settings (such as manipulation, collusion, or overriding safeguards); 
    more open-ended simulations to study dynamics, selection pressures, and emergent behaviours;
    and studies of how well these tests and simulations match real-world deployments. 
    \item \textbf{Mitigation}: Evaluation is only the first step towards mitigating multi-agent risks, which will require new technical advances.
    While our understanding of these risks is still growing, there are promising directions that we can begin to explore now, such as:
    scaling peer incentivisation methods to state-of-the-art models;
    developing secure protocols for trusted agent interactions; 
    leveraging information design and the potential transparency of AI agents; and
    stabilising dynamic multi-agent networks and ensuring they are robust to the presence of adversaries.
    \item \textbf{Collaboration}: Multi-agent risks inherently involve many different actors and stakeholders, often in complex, dynamic environments.
    Greater progress can be made on these interdisciplinary problems by leveraging insights from other fields, such as:
    better understanding the causes of undesirable outcomes in complex adaptive systems and evolutionary settings;
    determining the moral responsibilities and legal liabilities for harms not caused by any single AI system;
    drawing lessons from existing efforts to regulate multi-agent systems in high-stakes contexts, such as financial markets;
    and determining the security vulnerabilities and affordances of multi-agent systems.
\end{itemize}

To support these recommendations, \textbf{we introduce a taxonomy of AI risks that are new, much more challenging, or qualitatively different in the multi-agent setting, together with a preliminary assessment of what can be done to mitigate them}.
We identify three high-level \textit{failure modes}, which depend on the nature of the agents' objectives and the intended behaviour of the system: miscoordination, conflict, and collusion.
We then describe seven key \textit{risk factors} that can lead to these failures: information asymmetries, network effects, selection pressures, destabilising dynamics, commitment and trust, emergent agency, and multi-agent security.
For each problem we provide a \textit{definition}, key \textit{instances} of how and where it can arise, illustrative \emph{case studies}, and promising \textit{directions} for future work.
We conclude by discussing the \textit{implications} for existing work in AI safety, AI governance, and AI ethics.

\section{Introduction}
\label{sec:introduction}

The proliferation of increasingly advanced AI not only promises widespread benefits, but also presents new risks \citep{Bengio2024a,Chan2023-aj}.
In the future, AI systems will commonly interact and adapt in response to one another, forming \emph{multi-agent systems}.%
\footnote{A fundamental fact about (software-based) AI systems is that they can be easily duplicated. Thus, the vast training costs involved in producing state-of-the-art systems can be amortized over millions of instances. In this sense, if nothing else, the concept of multi-agent systems is core to transformative AI.}
This trend will be driven by several factors.
First, recent technical progress and publicity will continue to drive adoption, including in high-stakes areas such as financial trading \citep{Ferreira2021,amplify_aieq,Sun2023_rl_qt} and military strategy \citep{palantir_aip_defense,manson2024ai,Black2024}.
Second, AI systems that can act autonomously and adapt while deployed as \textit{agents} will have competitive advantages compared to non-adaptive systems or those with humans in the loop.
Third, the more widely such agents are deployed, the more they will come to interact with one another.

The emergence of these advanced multi-agent systems presents a number of risks which have thus far been systematically underappreciated and understudied.
In part, this lack of attention is because the deployment of such systems is currently rare, or constrained to highly controlled settings (such as automated warehouses) that do not suffer from the most severe risks.
In part, it is because even the simpler problem of ensuring the safe and ethical behaviour of a \textit{single} advanced AI system is far from solved \citep{Amodei2016,Hendrycks2021,Anwar2024}, and multi-agent settings are strictly more complex.
Indeed, many multi-agent risks are inherently sociotechnical and require attention from many stakeholders and researchers across many disciplines \citep{Curtis2024, Lazar2023}.

Importantly, these risks are distinct from those posed by \textit{single agents} or \textit{less advanced} technologies, and will not necessarily be addressed by efforts to mitigate the latter.
For example:
the alignment of AI agents with different actors is insufficient to prevent conflict if those actors have diverging interests \citep{Sourbut2024,Manheim2019,Dafoe2020,Critch2020,Jagadeesan2023a};
errors that may be acceptable in isolation could compound in complex, dynamic networks of agents \citep{Lee2024,Maas2018,Kirilenko2017,buldyrev2010catastrophic,Sanders2018};
and groups of agents could combine or collude to develop dangerous capabilities or goals that cannot be ascribed to any individual \citep{Jones2024,motwani2024secret,Mogul2006,Calvano2019,Drexler2022}.
Advanced AI also introduces phenomena that differ fundamentally from previous generations of AI or other technologies, requiring new approaches to mitigating these risks \citep{Bengio2024a}.

With the current rate of progress, we therefore urgently need to evaluate (and prepare to mitigate) multi-agent risks from advanced AI.
In this report we take a first step in this direction by providing a taxonomy of risks that either: emerge, are much more challenging, or are qualitatively different in the multi-agent setting (see \Cref{tab:risks}).
We identify three key high-level \textbf{failure modes} (\Cref{sec:failure_modes}), and seven key \textbf{risk factors} that can lead to these failures (\Cref{sec:failure_mechanisms}), before discussing the \textbf{implications} for AI safety, AI governance, and AI ethics (\Cref{sec:implications}).
Throughout the report we illustrate these risks with concrete examples, either from real-world events, previous research, or novel experiments (see \Cref{tab:demos}).

\subsection{Overview}
\label{sec:overview}

We begin by identifying different \emph{failure modes} in multi-agent systems based on the nature of the agents' goals and the intended behaviour of the system.
In most multi-agent systems, we are interested in AI agents working together to achieve their respective goals or the goals of those who deployed them. 
In this case, we categorise failures into \textbf{miscoordination} (\Cref{sec:miscoordination}), where agents fail to cooperate despite having the \emph{same goal}, and \textbf{conflict} (\Cref{sec:conflict}), where agents with \emph{different goals} fail to cooperate.
A third and final kind of failure -- \textbf{collusion} (\Cref{sec:collusion}) -- can arise in competitive settings where we do \emph{not} want agents cooperating (such as markets).

\begin{table}[htp]
    \centering
    \makegapedcells
    \setcellgapes{4pt}
    \begin{tabularx}{\linewidth}{>{\raggedright\arraybackslash}p{2.5cm} >{\raggedright\arraybackslash}p{4.5cm} >{\raggedright\arraybackslash}p{8cm}}
        \toprule
        \textbf{Risk} & \textbf{Instances} & \textbf{Directions} \\
        \midrule
        {\nameref{sec:miscoordination} \newline \pa \newline \pa} &
        {$\bullet$ Incompatible Strategies \newline $\bullet$ Credit Assignment \newline  $\bullet$ Limited Interactions} & 
        { $\bullet$ Communication \newline $\bullet$ Norms and Conventions \newline $\bullet$ Modelling Other Agents} \\
        {\nameref{sec:conflict} \newline \pa \newline \pa \newline \pa \newline \pa} &
        {$\bullet$ Social Dilemmas  \newline $\bullet$ Military Domains \newline $\bullet$ Coercion and Extortion \newline \pa \newline \pa} & 
        {$\bullet$ Learning Peer and Pool Incentivisation \newline $\bullet$ Establishing Trust \newline $\bullet$ Normative Approaches to Equilibrium Selection \newline $\bullet$ Cooperative Dispositions \newline $\bullet$ Agent Governance \newline $\bullet$ Evidential Reasoning} \\
        {\nameref{sec:collusion} \newline \pa \newline \pa} &
        {$\bullet$ Markets \newline $\bullet$ Steganography \newline \pa} & 
        {$\bullet$ Detecting AI Collusion  \newline $\bullet$ Mitigating AI Collusion \newline $\bullet$ Assessing Impacts on Safety Protocols} \\
        \midrule
        {\nameref{sec:information_asymmetries} \newline \pa \newline \pa } &
        {$\bullet$ Communication Constraints \newline $\bullet$ Bargaining \newline $\bullet$ Deception } & 
        {$\bullet$ Information Design \newline $\bullet$ Individual Information Revelation \newline $\bullet$ Few-Shot Coordination \newline $\bullet$ Truthful AI} \\
        {\nameref{sec:network_effects} \newline \pa \newline \pa} &
        {$\bullet$ Error Propagation \newline $\bullet$ Network Rewiring \newline $\bullet$ Homogeneity and Correlated Failures} & 
        {$\bullet$ Evaluating and Monitoring Networks \newline $\bullet$ Faithful and Tractable Simulations \newline $\bullet$ Improving Network Security and Stability} \\
        {\nameref{sec:selection_pressures} \newline \pa \newline \pa \newline \pa} &
        {$\bullet$ Undesirable Dispositions from Competition \newline $\bullet$ Undesirable Dispositions from Human Data \newline $\bullet$ Undesirable Capabilities} & 
        {$\bullet$ Evaluating Against Diverse Co-Players \newline $\bullet$ Environment Design \newline $\bullet$ Understanding the Impacts of Training \newline $\bullet$ Evolutionary Game Theory \newline $\bullet$ Simulating Selection Pressures} \\
        {\nameref{sec:destabilising_dynamics} \newline \pa \newline \pa \newline \pa} &
        {$\bullet$ Feedback Loops \newline $\bullet$ Cyclic Behaviour \newline $\bullet$ Chaos \newline $\bullet$ Phase Transitions \newline $\bullet$ Distributional Shift} & 
        {$\bullet$ Understanding Dynamics \newline $\bullet$ Monitoring and Stabilising Dynamics \newline $\bullet$ Regulating Adaptive Multi-Agent Systems \newline \pa \newline \pa} \\
        {\nameref{sec:commitment_and_trust} \newline \pa \newline \pa \newline \pa} &
        {$\bullet$ Inefficient Outcomes \newline $\bullet$ Threats and Extortion \newline $\bullet$ Rigidity and Mistaken Commitments \newline \pa} & 
        {$\bullet$ Keeping Humans in the Loop \newline $\bullet$ Limiting Commitment Power \newline $\bullet$ Institutions and Normative Infrastructure \newline $\bullet$ Privacy-Preserving Monitoring \newline $\bullet$ Mutual Simulation and Transparency} \\
        {\nameref{sec:emergent_agency} \newline \pa \newline \pa} &
        {$\bullet$ Emergent Capabilities \newline $\bullet$ Emergent Goals \newline \pa \newline \pa} & 
        {$\bullet$ Empirical Exploration \newline $\bullet$ Theories of Emergent Capabilties \newline $\bullet$ Theories of Emergent Goals \newline $\bullet$ Monitoring and Intervening on Collective Agents } \\
        {\nameref{sec:multi-agent_security} \newline \pa \newline \pa \newline \pa \newline \pa} &
        {$\bullet$ Swarm Attacks \newline $\bullet$ Heterogeneous Attacks \newline $\bullet$ Social Engineering at Scale \newline $\bullet$ Vulnerable AI Agents \newline $\bullet$ Cascading Security Failures \newline $\bullet$ Undetectable Threats} & 
        {$\bullet$ Secure Interaction Protocols \newline $\bullet$ Monitoring and Threat Detection \newline $\bullet$ Multi-Agent Adversarial Testing \newline $\bullet$ Sociotechnical Security Defences \newline \pa \newline \pa} \\
        \bottomrule
    \end{tabularx}
    \caption{An overview of the instances and research directions identified for each failure mode and risk factor (see \Cref{sec:failure_modes,sec:failure_mechanisms} for a discussion of each bullet point).}
    \label{tab:risks}
\end{table}

We next introduce a number of \emph{risk factors} by which these failure modes can arise, and which are largely independent of the agents' precise incentives.\footnote{Indeed, there are potential risks from multi-agent systems in which it is not the agents' objectives that are the critical feature, but their general incompetencies or vulnerabilities.}
For example, information asymmetries could lead to miscoordination between agents with the same goal, or to conflict among agents with competing goals.
These factors are not specific to AI systems, but the differences between AI systems and other kinds of intelligent agents (such as humans or corporations) leads to different risk instances and potential solutions.
Finally, note that the following factors are not necessarily exhaustive or mutually exclusive.
\begin{itemize}
    \item \textbf{Information asymmetries} (\Cref{sec:information_asymmetries}): private information can lead to miscoordination, deception, and conflict;
    \item \textbf{Network effects} (\Cref{sec:network_effects}): minor changes in properties or connection patterns of agents in a network can lead to dramatic changes in the behaviour of the whole group;
    \item \textbf{Selection pressures} (\Cref{sec:selection_pressures}): some aspects of training and selection by those deploying and using AI agents can lead to undesirable behaviour;
    \item \textbf{Destabilising dynamics} (\Cref{sec:destabilising_dynamics}): systems that adapt in response to one another can produce dangerous feedback loops and unpredictability;
    \item \textbf{Commitment and trust} (\Cref{sec:commitment_and_trust}): difficulties in forming credible commitments, trust, or reputation can prevent mutual gains in AI-AI and human-AI interactions;
    \item \textbf{Emergent agency} (\Cref{sec:emergent_agency}): qualitatively different goals or capabilities can emerge from the composition of innocuous independent systems or behaviours;
    \item \textbf{Multi-agent security} (\Cref{sec:multi-agent_security}): multi-agent systems give rise to new kinds of security threats and vulnerabilities.
\end{itemize}
We conclude the report by surveying the safety, governance, and ethical \emph{implications} of these risks (see \Cref{tab:implications}).
For example, most work on \textbf{AI safety} (\Cref{sec:safety}) focuses on issues such as the robustness, interpretability, or alignment of a single system \citep{Amodei2016, Hendrycks2021, Anwar2024}, despite the fact that an increasing number of proposals for building safer AI systems are implicitly multi-agent \citep[e.g.,][]{Irving2018,Drexler2019,Greenblatt2023,Schwettmann2023,Perez2022}.
The fact that \textbf{AI governance} (\Cref{sec:governance}) efforts often involve multi-stakeholder settings provides hope that governance tools can complement technical advances to mitigate multi-agent risks \citep{reuel2024open, trager2023international}.
At the same time, multi-agent interactions naturally raise questions in \textbf{AI ethics} (\Cref{sec:ethics}) related to issues of fairness, collective responsibility, and the social good \citep{Gabriel2024,NIPS2014_792c7b5a,friedenberg2019blameworthiness}.

\begin{table}[htp]
    \centering
    \renewcommand{\arraystretch}{1.5}
    \begin{tabularx}{\linewidth}{>{\raggedright\arraybackslash}X >{\raggedright\arraybackslash}X >{\raggedright\arraybackslash}X}
        \toprule
        \textbf{\nameref{sec:safety}} & \textbf{\nameref{sec:governance}} & \textbf{\nameref{sec:ethics}} \\
        \midrule
        $\bullet$ Alignment is Not Enough \newline
        $\bullet$ Collusion in Adversarial Safety Schemes \newline
        $\bullet$ Dangerous Collective Goals and Capabilities \newline
        $\bullet$ Correlated and Compounding Failures \newline
        $\bullet$ Robustness and Security in Multi-Agent Systems \newline
        \pa
        &
        $\bullet$ Supporting Research on Multi-Agent Risks \newline
        $\bullet$ Multi-Agent Evaluations \newline
        $\bullet$ New Forms of Documentation \newline
        $\bullet$ Infrastructure for AI Agents \newline
        $\bullet$ Restrictions on Development and Deployment \newline
        $\bullet$ Liability for Harms from Multi-Agent Systems \newline
        $\bullet$ Improving Societal Resilience
        &
        $\bullet$ Pluralistic Alignment \newline
        $\bullet$ Agentic Inequality \newline
        $\bullet$ Epistemic Destablisation \newline
        $\bullet$ Compounding of Unfairness and Bias \newline
        $\bullet$ Compounding of Privacy Loss \newline
        $\bullet$ Accountability Diffusion \newline
        \pa \newline
        \pa \newline
        \pa
        \\
        \bottomrule
    \end{tabularx}
    \caption{An overview of the implications of multi-agent risks for existing work in AI safety, governance, and ethics (see \Cref{sec:implications} for a discussion of each bullet point).}
    \label{tab:implications}
\end{table}

\subsection{Scope}
\label{sec:scope}

Concerns about the risks posed by AI systems range from biased hiring decisions \citep{Raghavan2020} to existential catastrophes \citep{Bostrom2014}, and are represented by a vast literature.
Before giving a brief overview of the most closely related works, it is therefore worth us pausing to clarify the scope of this report, which is as follows.
\begin{itemize}
    \item \textbf{Risks and failure modes}: we seek to identify specific mechanisms via which risks could emerge, rather than just just the open research problems that these risks present.
    \item \textbf{Multiple agents}: if the risk could arise in essentially the same way in the context of a single AI system, then we deem it to be out of scope for this report (while not diminishing its importance).\footnote{Note that this includes the problem of \textit{alignment} \citep{Russell2019,Ngo2022}, which we do not study in this report.}
    \item \textbf{Advanced AI}: while many of the risks we identify also apply to simpler systems, their effects are most severe in the context of increasingly autonomous and powerful AI agents,\footnote{We tend to reserve the word `agent' for more autonomous, self-sufficient, and goal-directed systems, though what counts as an `AI agent' as opposed to a mere `AI system' is not always clear \citep{Chan2023-aj,Gabriel2024,Kapoor2024}. Similarly, we will often use the word `principal' for the actor on whose behalf an agent acts (be they an individual, a group, or some other entity). Note also that we do not necessarily advocate for the building of advanced AI agents \citep{Mitchell2025}, we merely expect that such agents will be built.} and so this is where our primary focus lies.
    \item \textbf{Real-world examples}: wherever possible, we make sure to ground these risks in real-world events, previous research, or novel experiments -- not merely hypothetical speculation (see \Cref{tab:demos}).
    \item \textbf{Technical perspectives}: due to the authors' expertise (and to keep the scope of the report manageable), we primarily discuss risks from a technical perspective, while acknowledging that this perspective is limited.
    \item \textbf{Concrete paths forwards}: where possible, we aim to specify relatively narrow proposals for future research, in the hope that this makes it easier for others to contribute.
\end{itemize}

Needless to say, multi-agent risks from advanced AI are by no means the only risks posed by AI, and the perspective we take in this report is by no means the only approach to understanding these risks.
Moreover, we almost entirely neglect the potential \textit{upsides} of advanced multi-agent systems:
greater decentralisation and democratisation of AI technologies;
assistance in cooperating and coordinating with others;
increased robustness, flexibility, and efficiency; 
novel approaches to solving alignment and safety issues in single-agent settings; 
and -- perhaps most importantly -- more widespread and evenly distributed benefits from AI. 
We hope that this report serves to complement earlier and adjacent research on understanding these challenges and opportunities.

\subsection{Related Work}

The most similar report to ours is that of \citet{Manheim2019}, who introduces a range of technical multi-agent failure modes through the lens of \emph{model over-optimization}. This over-optimisation can result in the intended and actual behaviour of the model coming apart when faced with low-probability inputs, a regime change, measurement errors, or inaccuracies in the model's internal representations. While this lens is helpful for understanding some multi-agent risk factors, not all factors can neatly be captured through it.
\citet{Mogul2006,Altmann2024} take an alternative perspective and focus on `emergent' failures that occur specifically in multi-agent settings, though their focus is not on \textit{advanced} AI agents.
Also highly relevant is \citet{Clifton2020}'s agenda on cooperation and conflict in the context of transformative AI, though the priority of that work is to describe a set of promising research directions, rather than to explicate the underlying risks. 

More broadly, the topics of this report are closely related to the emerging subfield of \emph{cooperative AI} \citep{Dafoe2020,Bertino2020,dafoe2021cooperative,Conitzer2023}, which chiefly studies how to engineer AI systems in order to help solve cooperation problems between humans, AI agents, or combinations thereof.
In contrast to these previous agendas, we also discuss failures from undesirable cooperation (i.e., collusion) and focus more on the concrete mechanisms via which failures can occur, rather than the capabilities needed for addressing them.
We also incorporate additional perspectives beyond traditional game-theoretic paradigms -- such as complex systems and security -- and highlight implications for work in AI governance and AI ethics in addition to AI safety.

Other surveys of AI risks focus primarily on the case of individual (often present-day) AI systems.
For example, \citet{Amodei2016} survey a range of concrete problems in AI safety (side effects, reward hacking, scalable oversight, safe exploration, and robustness to distributional shifts), while \citet{Hendrycks2021} provide a classification of problems in ML safety (robustness, monitoring, alignment, and systemic safety).
\citet{Anwar2024,Bommasani2021-dy,Weidinger2022,Bird2023} focus on the risks from foundation models specifically, while \citet{Chan2023-aj,Gabriel2024} consider the harms posed by increasingly `agentic' systems and AI assistants.
Other taxonomies seek to adopt an explicitly sociotechnical lens \citep{shelby_sociotechnical_2023,Abercrombie2024,Weidinger2023a}, often focusing primarily on present-day risks.
\citet{Zeng2024,Uuk2025} provide meta-reviews of AI risks derived from different research papers, as well as government and company policies.
Our report is complementary to these works, and includes discussion of how novel problems arise in the multi-agent case, and in the case of more advanced AI agents.

More speculatively, some authors have considered the possibility of catastrophic or even existential risks from AI \citep{Turchin2018,Bostrom2014,Ord2020,Kasirzadeh2024}.
\citet{Hendrycks2023} categorises such risks into malicious use, AI races, organizational risks, and rogue AIs.
As in \citet{hendrycks_natural_2023}, multi-agent risks are viewed largely through an evolutionary lens, though this is primarily restricted to competitive pressures at the level of non-AI actors (such as firms or states).
\citet{Critch2020,critch2023tasra} frame such risks in terms of delegation to AI systems and the responsibilities of those doing so.
While they provide illuminating vignettes of possible catastrophes, we aim to provide more concrete examples at a more modest scale.

\begin{table}[hbp]
    \centering
    \renewcommand{\arraystretch}{1.5}
    \begin{tabularx}{\linewidth}{>{\raggedright}X | c c c | c c c c c c c | c | c}
        \makecell[l]{Case Study} &
        \rotatedcell{miscoordination}{90} &             %
        \rotatedcell{conflict}{90} &                    %
        \rotatedcell{collusion}{90} &                   %
        \rotatedcell{information_asymmetries}{90} &     %
        \rotatedcell{network_effects}{90} &             %
        \rotatedcell{selection_pressures}{90} &         %
        \rotatedcell{destabilising_dynamics}{90} &      %
        \rotatedcell{commitment_and_trust}{90} &        %
        \rotatedcell{emergent_agency}{90} &             %
        \rotatedcell{multi-agent_security}{90} &        %
        \makecell[l]{Type} &
        \makecell[l]{Page}\\
        \midrule                                %
        \nameref{cs:coordination_driving}       &\yes&\no &\no &\yes&\no &\yes&\no &\no &\no &\no & \new & \pageref{cs:coordination_driving}\\
        \nameref{cs:military_escalation}        &\no &\yes&\no &\no &\no &\no &\yes&\no &\no &\no & \old & \pageref{cs:military_escalation}\\
        \nameref{cs:common_resource_problems}   &\no &\yes&\no &\no &\no &\yes&\no &\yes&\no &\no & \old & \pageref{cs:common_resource_problems}\\
        \nameref{cs:ai_collusion}               &\no &\no &\yes&\no &\no &\yes&\no &\no &\yes&\no & \rwe & \pageref{cs:ai_collusion}\\
        \nameref{cs:steganography}              &\no &\no &\yes&\no &\no &\no &\no &\no &\no &\yes& \old & \pageref{cs:steganography}\\
        \nameref{cs:market_manipulation}        &\no &\yes&\no &\yes&\no &\no &\no &\no &\no &\no & \old & \pageref{cs:market_manipulation}\\
        \nameref{cs:news_corruption}            &\no &\no &\no &\yes&\yes&\no &\no &\no &\no &\no & \new & \pageref{cs:news_corruption}\\
        \nameref{cs:infectious_attacks}         &\no &\no &\no &\no &\yes&\no &\no &\no &\no &\yes& \old & \pageref{cs:infectious_attacks}\\
        \nameref{cs:LLM_evolution}              &\no &\yes&\no &\no &\no &\yes&\no &\yes&\no &\no & \old & \pageref{cs:LLM_evolution}\\
        \nameref{cs:flash_crash}                &\no &\no &\no &\no &\yes&\no &\yes&\no &\no &\no & \rwe & \pageref{cs:flash_crash}\\
        \nameref{cs:dead_hand}                  &\no &\yes&\no &\no &\no &\no &\no &\yes&\no &\no & \rwe & \pageref{cs:dead_hand}\\
        \nameref{cs:overcoming_safeguards}      &\no &\no &\no &\no &\no &\no &\no &\no &\yes&\yes& \old & \pageref{cs:overcoming_safeguards}\\
        \nameref{cs:fooling_overseer}           &\no &\no &\no &\no &\no &\no &\no &\no &\no &\yes& \new & \pageref{cs:fooling_overseer}\\
        \midrule
    \end{tabularx}
    \caption{An overview of the case studies present in this report, and the failure modes and risk factors that they represent. Each case study represents either a historical example (\rwe), existing results from the literature (\old), or -- when neither of these existed -- novel experiments that we conducted as part of this report (\new). Further details about our own experiments are provided in \Cref{app:case_study_details}.}
    \label{tab:demos}
\end{table}

\section{Failure Modes}
\label{sec:failure_modes}

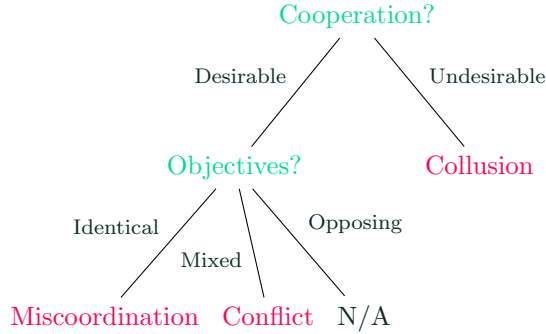
\begin{figure}[h]
    \centering
    \begin{tikzpicture}[level distance=2cm,
        edge from parent path={(\tikzparentnode) -- (\tikzchildnode)}]
        \tikzstyle{every node}=[caif_dark_green]
         \Tree
         [.\node[text=caif_light_green] {Cooperation?};
             \edge node[auto=right] {\footnotesize Desirable};
             [.\node[text=caif_light_green] {Objectives?};
                \edge node[auto=right] {\footnotesize Identical};
                [.{\nameref{sec:miscoordination}} ]
                \edge node[auto=right] {\footnotesize Mixed};
                [.{\nameref{sec:conflict}} ]
                \edge node[auto=left] {\footnotesize Opposing};
                [.{N/A} ]
                 ]
             \edge node[auto=left] {\footnotesize Undesirable};
             [.{\nameref{sec:collusion}} ]
         ]
    \end{tikzpicture}
    \caption{The three kinds of failure mode that we study in this work. Note that we do not consider constant-sum settings where cooperation is desirable, as in such cases it is definitionally impossible for some agents to gain without a commensurate loss from one or more other agents.}
    \label{fig:terms}
\end{figure}

Multi-agent systems can fail in various ways, depending on the intended behaviour of the system and the objectives of the agents.
First, we can distinguish between cases where we want the agents to be \emph{cooperating} (as in collective action problems or team games) or \emph{competing} (such as in markets or adversarial training).
Second, we can further divide the space of failure modes depending on whether the agents have \emph{exactly the same} objectives, \emph{different but overlapping} objectives, or \emph{completely opposed} objectives.\footnote{This division corresponds to common-interest/team games, mixed-motive/general-sum games, and constant-sum games, respectively.}
While different authors have used different terms to describe these cases, we use the terminology shown in \Cref{fig:terms}.\footnote{In particular, we note that `conflict' is often used more narrowly than the idea of `cooperation failure in mixed-motive settings', which is what we use the term for. We deliberately use `conflict' instead of `cooperation failure' to distinguish this failure mode from `miscoordination', which applies to problems in which agents have the \textit{same} objectives.}
Finally, there are many potential risks from advanced multi-agent systems that do not necessarily arise through agents competently pursuing their objectives, but due to their incompetencies or vulnerabilities.
We consider these latter failures as part of our discussion on different risk factors in \Cref{sec:failure_mechanisms}.

\subsection{Miscoordination}
\label{sec:miscoordination}

The simplest kind of cooperation failures are those in which all agents have (approximately) the same objectives.
Even in such common-interest settings, however, miscoordination abounds.
While it is reasonable to expect that these problems will tend to be addressed as the general capabilities of AI systems (such as communication and reasoning about others) improve,\footnote{Note that this is unlike problems of conflict and collusion, where the fundamental tension between the desired outcome and the agents' objectives may in fact lead to \emph{worse} outcomes as general AI capabilities improve.} they may still present risks in the near-term.

\subsubsection{Definition}

Miscoordination arises when agents, despite a mutual and clear objective, cannot align their behaviours to achieve this objective.
Unlike the case of differing objectives, in common-interest settings there is a more easily well-defined notion of `optimal' behaviour and we describe agents as miscoordinating to the extent that they fall short of this optimum.
Note that for common-interest settings it is not sufficient for agents' objectives to be the same in the sense of being \textit{symmetric} (e.g., when two agents both want the same prize, but only one can win).
Rather, agents must have \textit{identical} preferences over outcomes (e.g., when two agents are on the same team and win a prize as a team or not at all).

It is rare that two humans will share exactly the same objectives in this sense.
For example, two sportspeople on the same team may be primarily aiming to win their match but will also have individual preferences, such as who scores the winning point.
In the case of AI systems, however, different agents can more easily be given precisely the same goal, and indeed much work on cooperation in AI focuses solely on the common-interest setting \citep{boutilier1996planning,Peshkin2000,Stone2010,Omidshafiei2017,Rashid2018,Oroojlooy2022}.
Such approaches are typically motivated by the practical and computational advantages that decentralised control confers, but are more challenging to implement than their centralised, single-agent counterparts.
Aside from this, miscoordination can also occur in settings that have a substantial element of common interest, even if agents' objectives are not entirely identical.

\subsubsection{Instances}

Perhaps the most likely way that common-interest settings may arise in practice is where a \textit{single principal} deploys multiple AI agents on their behalf in order to jointly solve a task.
This choice might be motivated by: physical constraints (if the task comprises sub-tasks that must be completed separately and simultaneously); efficiency considerations (if having a single agent in charge of all aspects of the task would lead to an intractably complex problem); or a desire for robustness (if an individual agent might fail but others could still succeed in their place).
Alternatively, we might see \textit{multi-principal} multi-agent settings in which the agents' goals are sufficiently aligned to be viewed as (approximately) identical.
For example, if two autonomous vehicles are driving along the same road, then the mutual harms from potential miscoordination (such as a collision) are far greater than any small individual benefits from competition (such as attempting a risky overtaking manoeuvre to get slightly ahead).

\paragraph{Incompatible Strategies.}
Even if all agents can perform well in isolation, miscoordination can still occur due to the agents choosing incompatible strategies \citep{Cooper1990}.
Competitive (i.e., two-player zero-sum) settings allow designers to produce agents that are maximally capable without taking other players into account. 
Crucially, this is possible because playing a strategy at equilibrium in the zero-sum setting guarantees a certain payoff, even if other players deviate from the equilibrium \citep{nash_non-cooperative_1951}. 
On the other hand, common-interest (and mixed-motive) settings often allow a vast number of mutually incompatible solutions \citep{Schelling1980-cq}, which is worsened in partially observable environments \citep{Reif1984,bernstein_complexity_2002}.
As a simple example, everyone driving on the left side or the right side of the road are both perfectly valid ways of keeping drivers safe, but these two conventions are inherently incompatible with one another (see \Cref{cs:coordination_driving}).

\begin{case-study}[label=cs:coordination_driving]{Zero-Shot Coordination Failures in Driving}
  \begin{center}
    \includegraphics[width=\linewidth]{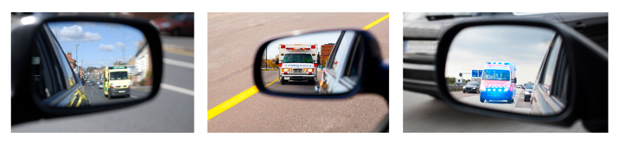}
    \captionof{figure}{Examples of scenes given to GPT-4 Vision in our language agent pipeline.}
  \end{center}
  \vspace{1em}
  Miscoordination is possible even among agents with shared objectives. We demonstrate how two frontier models trained on driving conventions from two different countries can face coordination failures. Following recent advances in robotics that combine vision models for scene comprehension with large language models (LLMs) for discrete action planning \citep[e.g.,][]{open_x_embodiment_rt_x_2023}, we created an experiment using two fine-tuned GPT-3.5 models. One model was trained on US driving protocols requiring rightward yielding for emergency vehicles, while the other followed Indian conventions mandating leftward yielding. Context-specific training data was generated as input-output pairs by GPT-4 based on the specified driving conventions and then manually reviewed. A GPT-4 Vision model processed the environmental inputs and provided scene descriptions to both fine-tuned GPT-3.5 models for action generation. The results quantified a significant coordination failure: unspecialized base models failed in only 5\% of scenarios (2/40 simulations), while specialized models exhibited a 77.5\% failure rate (31/40 simulations), consistently failing to create clear paths for emergency vehicles. This demonstrates an example where a convention cannot always be declared in a zero-shot interaction, posing risks in multi-agent settings. 
\end{case-study}

\paragraph{Credit Assignment.}
While agents can often \emph{learn} to jointly solve tasks and thus avoid coordination failures, learning is made more challenging in the multi-agent setting due to the problem of {credit assignment} \citep[see also \Cref{sec:information_asymmetries} on information asymmetries and \Cref{sec:destabilising_dynamics}, which discusses distributional shift]{Du2023, li2025multiagent}.
That is, in the presence of other learning agents, it can be unclear which agents' actions caused a positive or negative outcome to obtain, especially if the environment is complex.
Moreover, in multi-principal settings, agents may not have been trained together and therefore need to generalise to new co-players and collaborators based on their prior experience \citep{Stone2010,Leibo2021-cf,Agapiou2022-an}.

\paragraph{Limited Interactions.}
Sometimes learning from historical interactions with the relevant agents may not be possible, or may be possible using only {limited interactions}.
In such cases, some other form of information exchange is required for agents to be able to reliably coordinate their actions, such as via \emph{communication} \citep{Crawford1982,Farrell1996} or a \emph{correlation device} \citep{aumann1974, aumann1987}.
While advances in language modelling mean that there are likely to be fewer settings in which the inability of advanced AI systems to communicate leads to miscoordination, situations that require split-second decisions or where communication is too costly could still produce failures.
In these settings, AI agents must solve the problem of `zero-shot' (or, more generally, `few-shot') coordination \citep{Stone2010,Hu2020,Treutlein2021,Zhu2021,Emmons2022}.

\subsubsection{Directions}

Decentralised control and coordination in multi-agent systems have been well-studied problems for decades \citep{boutilier1996planning,Peshkin2000,Stone2010,Omidshafiei2017,Rashid2018,Oroojlooy2022}.
At one level of abstraction, the key challenge of coordination is that of sharing information, i.e., communication. If agents
have the same preferences and are able to communicate, they can coordinate by (say) having
a single agent announce their intended action and everyone else follow suit, since there are no
incentives for the leader to lie or the followers to deviate \citep[e.g.,][]{farrell1996cheap}.
Given the superhuman capabilities of advanced AI to transmit and process vast swathes of information, the most important research directions in this area will therefore be those in which it is not possible to exercise these capabilities (e.g., due to complexity, latency, or privacy constraints).

\paragraph{Communication.}
As noted above, the advanced {communication} abilities of LLMs promise to simplify many coordination challenges.
In order to successfully integrate these advances into real-world systems, however, agents need to know when and what needs to be coordinated on -- something that may not always be obvious in novel or out-of-distribution domains.
In safety-critical domains, it may therefore be necessary to introduce, or have the agents invent, \emph{protocols} (i.e., rules and specifications) for communication between advanced AI agents \citep{Marro2024}. 
Moreover, agents need to agree on how the communication channel is \textit{grounded} \citep{clark_grounding_1991} to actions or strategies in the environment. Grounding LLMs is a problem that is not unique to coordination \citep{bisk_experience_2020, bender_climbing_2020, mahowald_dissociating_2023}, but it is exacerbated by the fact that agents attempting to coordinate through natural language need to be grounded \textit{in the same way}. For instance, if they are designed with different interfaces to tools in a domain, they must be able to coordinate despite these differences in interfaces.

\paragraph{Norms and Conventions.}
For settings in which inter-agent communication is infeasible or insufficient, {norms and conventions} may be necessary in order to avoid miscoordination \citep{leibo2024theory}.
For example \citet{hadfield2019legible} show that even the adoption of so-called `silly rules' (those that do not have direct bearing on the agents' payoffs) can help groups adapt and be more robust to uncertainty by enriching the information environment.
Moving beyond more arbitrary conventions, we may choose to design particular norms and conventions \citep{nyborg2016social, bicchieri2016norms, Shoham1992}. In this setting, the challenge is to select norms that are both legible and enforceable, as well as leading to jointly beneficial outcomes.
On the other hand, if the agents can adapt their behaviour, it may be that norms and conventions emerge over time \citep{mcelreath2003shared}.
For example, \citet{Koester2020} show that multi-agent reinforcement learning (MARL) agents can establish and switch between conventions, even compromising on their own objective when doing so is necessary for effective coordination.
More generally, we may be interested in studying how norms and conventions emerge \citep{Mashayekhi2022, Morris-Martin2019}, how robust they are \citep{lerer2019, Hao2017}, and how compatible they are with others that may have emerged in different agent populations \citep{Stastny2021}.

\paragraph{Modelling Other Agents.}
Finally, the ability to understand and predict others' actions can be critical to coordination, especially in situations when little or no communication is possible.
Even though agents may assume that others share their objective in common-interest settings, being able to model others' actions, beliefs, and intentions can be highly advantageous.
For an overview of the topic and a list of key problems, we refer the reader to \citet{Albrecht2018}.
With the advent of LLM-based agents that appear to possess some form of theory of mind and hence can be remarkably sophisticated in their modelling of other agents \citep{cross2025hypothetical, li-etal-2023-theory}, new questions arise.
For example, given the current paradigm of deriving many systems from an underlying base model, it may be easier for similarly derived systems to reason about one another \citep{oesterheld2024similarity,Binder2024,berglund2023taken,OpenAI2023b}.

\subsection{Conflict}
\label{sec:conflict}

In the vast majority of real-world strategic interactions, agents' objectives are neither identical nor completely opposed.
Indeed, if AI agents are sufficiently aligned to their users or deployers, we should expect some degree of both cooperation \textit{and} competition, mirroring human society.
These mixed-motive settings include the possibility of mutual gains, but also the risk of conflict due to selfish incentives.
In what follows, we examine the extent to which advanced AI might precipitate or exacerbate such risks. 

\subsubsection{Definition}

In this work, we use the word conflict in a relatively broad sense to refer to any outcome in a mixed-motive setting that does not lie on the Pareto frontier.\footnote{Recall that an outcome lies on the Pareto frontier if it is not possible to make any agent better off without making another worse off.}
This includes classic examples of conflict such as legal disputes and warfare, but also encompasses cooperation failures in collective action problems, such as the depletion of a common natural resource or a race to the bottom on legislation \citep{snyder1971prisoner, dawes1980social}.

It is worth noting first that AI systems could help to \emph{solve} conflicts, for example, by searching over a larger
space of potential solutions to disagreements, monitoring agreements, or acting as mediators 
\citep{Dafoe2020,bakker_fine-tuning_2022,mckee2023scaffolding,Small2023}.
At the same time, the selfish incentives that drive said conflict may also 
incentivise actors to adopt AI systems in order to gain an advantage over their competitors.
In such cases, delegation to increasingly advanced AI agents is far from guaranteed to lead to more cooperative outcomes, and could in some circumstances increase both the speed and the scale at which conflict might emerge.
Indeed, even if advanced AI systems are able to overcome human cooperation problems, they may introduce even more complex cooperation problems (compare to how adults may be able to prevent children from fighting, but aren't immune from conflict themselves).

\subsubsection{Instances}

As we noted above, virtually all real-world strategic interactions of interest are mixed-motive, and as such the potential for conflict (even if in low-stakes scenarios) abounds.
The introduction of advanced AI agents could both worsen existing risks of conflict (such as increasing the degree of competition in common-resource problems, or escalating military tensions) as well as well as introducing new forms of conflict (such as via sophisticated methods of coercion and extortion). 

\paragraph{Social Dilemmas.}
As noted in our definition, conflict can arise in any situation in which selfish incentives diverge from the collective good, known as a {social dilemma} \citep{Hardin1968,dawes1980social,kollock1998social,Ostrom1990}.
While this is by no means a modern problem, advances in AI could further enable actors to pursue their selfish incentives by overcoming the technical, legal, or social barriers that standardly help to prevent this.
To take a plausible, near-term (if very low-stakes) example, an automated AI assistant could easily reserve a table at every restaurant in town in minutes, enabling the user to decide later and cancel all other reservations.
Alternatively, the ability of AI assistants to search and switch between different consumer products and services could lead to `hyper-switching' \citep{VanLoo2019}, potentially leading to financial instabilities such as a deposit franchise run \citep[see also \Cref{cs:flash_crash}]{Drechsler2023}.
On the other hand, profit-seeking companies might also soon deploy advanced AI agents that either use or manage common resources, ranging from communication networks and web services to roads and natural resources.
Without methods of governing such agents, these resources may quickly be depleted or made inaccessible to all but a small number of powerful actors.

\begin{case-study}[label=cs:common_resource_problems]{Common Resource Problems}
  \begin{center}
    \includegraphics[width=\linewidth]{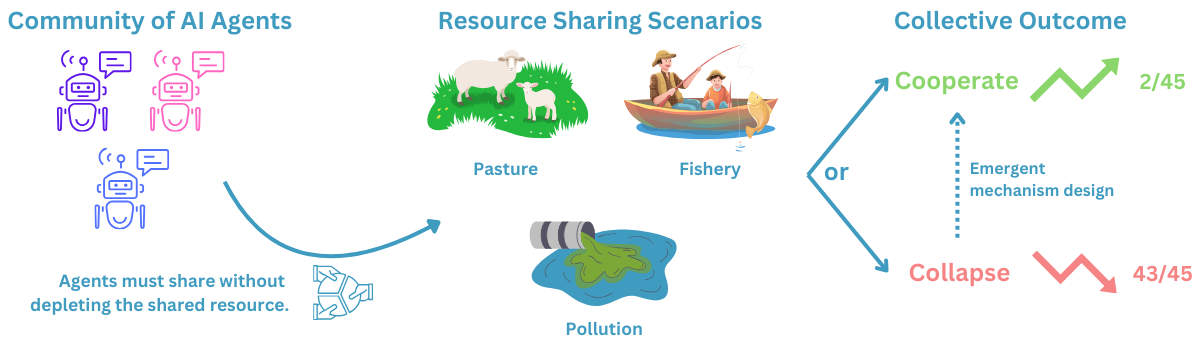}
    \captionof{figure}{A summary of the resource-sharing scenarios within the GovSim benchmark. Figure adapted from \citet{piatti2024cooperatecollapseemergencesustainable}.}
  \end{center}
  \vspace{1em}
  The management of shared resources represents a fundamental test of whether AI systems can balance individual incentives against collective welfare.\footnotemark{}
  In the GovSim benchmark, \citet{piatti2024cooperatecollapseemergencesustainable} evaluated 15 different LLMs across three resource management scenarios: fishing from a shared lake, grazing on common pastures, and managing industrial pollution.
  Even the most advanced LLMs achieved only a 54\% survival rate, meaning that in nearly half of all cases, the agents depleted their shared resources to the point of collapse.
  These findings align with earlier work on sequential social dilemmas \citep{Leibo2017}, which (unlike `one-shot' problems) allow agents to react to others' choices over time, creating complex dynamics of trust and retaliation.
  When one agent begins to over-exploit resources, others often respond by increasing their own extraction rates, triggering a cascade of competitive behaviour that accelerates resource depletion. Without additional protections, these systems may therefore replicate or even accelerate the tragedy of the commons \citep{Hardin1968}.
\end{case-study}

\footnotetext{Note that we use the term `welfare' in this context to denote an aggregate measure of the extent to which a group of agents achieves their respective objectives, rather than to refer to some notion of `wellbeing'.}

\paragraph{Military Domains.}
Perhaps the most obvious and worrying instances of AI conflict are those in which human conflict is already a major concern, such as military domains (although other, less salient forms of conflict such as international trade wars are also cause for concern).
For example, beyond applications of more narrow AI tools in lethal autonomous weapons systems \citep{Horowitz2021-bu}, future AI systems might serve as advisors or negotiators in high-stakes military decisions \citep{manson2024ai,Black2024}. 
Indeed, companies such as Palantir have already developed LLM-powered tools for military planning \citep{palantir_aip_defense}, and the US Department of Defence has recently been evaluating models for such capacities, with personnel revealing that they ``could be deployed by the military in the very near term'' \citep{manson2023}.
The use of AI in command and control systems to gather and synthesise information -- or recommend and even autonomously \emph{make} decisions -- could lead to rapid unintended escalation if these systems are not robust or are otherwise more conflict-prone \citep[see also \Cref{cs:flash_crash}]{Johnson2020-po,johnson2021artificial,laird2020risks}.\footnote{At the same time, it is worth noting that AI systems could have significant advantages over human decision-makers
in navigating conflict in ways that avoid unnecessary 
escalation. If suitably robust, they could be less prone to the kinds of errors in judgement that exacerbate human conflict due to their ability to rapidly integrate large amounts of information, consider many different possible outcomes, and give calibrated estimates of their uncertainty \citep{Johnson2004-oi,Jervis2017-ng}.} 

\paragraph{Coercion and Extortion.}
Advanced AI systems might also lead to various forms of {coercion and extortion} in less extreme settings \citep{ellsberg1968theory,harrenstein2007commitment}.
These threats might target humans directly (such as the revelation of private information extracted by advanced AI surveillance tools), or other AI systems that are deployed on behalf of humans (such as by hacking a system to limit its resources or operational capacity; see also \Cref{sec:multi-agent_security}).
Increasing AI cyber-offensive capabilities -- including those that target other AI systems via adversarial attacks and jailbreaking \citep{zou2023universaltransferableadversarialattacks,Gleave2020,yamin2021} -- without a commensurate increase in defensive capabilities could make this form of conflict cheaper, more widespread, and perhaps also harder to detect 
\citep{brundage2018malicious}.
Addressing these issues requires design strategies that prevent AI systems from exploiting, or being susceptible to, such coercive tactics.\footnote{This point is closely related to the question of which kinds commitments we ought to permit AI agents to make (see \Cref{sec:commitment_and_trust}). For example, commitments could be used coercively to make threats, but could also be used to defend oneself against threats (cf. the idea of refusing to negotiate with terrorists).}

\begin{case-study}[label=cs:military_escalation,sidebyside,sidebyside align=top,lower separated=false]{Escalation in Military Conflicts}
  Recent research by \citet{Rivera_2024} raises critical concerns about the emergence of escalatory behaviors when AI tools or agents (see \Cref{fig:palantir}) inform military decision-making.
  In experiments with AI agents controlling eight distinct nation-states, even neutral starting conditions did not prevent the rapid emergence of arms race dynamics and aggressive strategies. 
  Strikingly, all five off-the-shelf LLMs studied showed forms of escalation, even when peaceful alternatives were available.
  These findings mirror other evidence showing that LLMs often display more aggressive responses than humans in military simulations and troubling inconsistencies in crisis decision-making \citep{lamparth2024humanvsmachinebehavioral, shrivastava2024measuringfreeformdecisionmakinginconsistency}.
  These results raise urgent questions about how to ensure stability in AI-driven military and diplomatic scenarios.
  \tcblower
  \includegraphics[width=\linewidth]{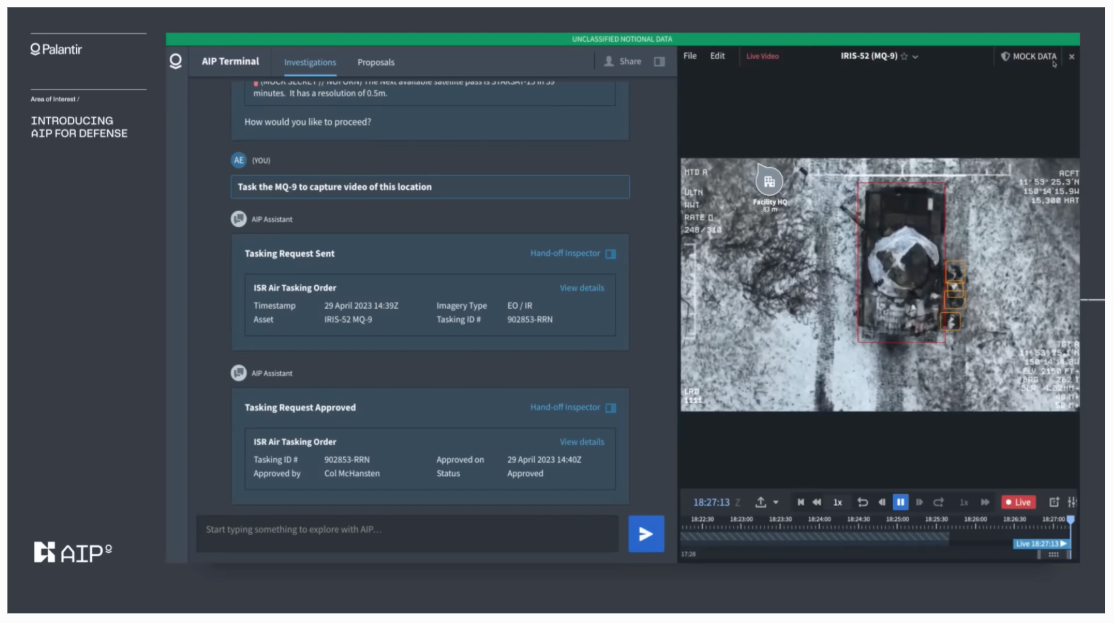}
  \captionof{figure}{A screenshot of Palantir's AI Planner (AIP), taken from a promotional video \citep{palantir_aip_defense}, demonstrating AI-assisted military decision-making, which uses LLMs for decision support in battle. The left side of the screen features a chat interface, while the right side shows information such as aerial surveillance footage of a tank.
  The LLM used in the demonstration was EleutherAI's GPT-NeoX-20B \citep{Black2022}.
  }\label{fig:palantir}
\end{case-study}

\subsubsection{Directions} 

The majority of work in multi-agent systems (and especially in multi-agent learning) has, until recently, tended to focus on either pure cooperation \citep[e.g.,][]{boutilier1996planning,Peshkin2000,Stone2010,Omidshafiei2017,Rashid2018,Oroojlooy2022} or pure competition \citep[e.g.,][]{Silver2016, Zhang2020,Daskalakis2011,Brown2019a,Bakhtin2022,NEURIPS2020_3b2acfe2}.
As there are not yet large numbers of mixed-motive interactions involving AI systems, part of the challenge is to identify interventions that encourage cooperation in such settings while making realistic assumptions about the computational and strategic nature of agents in the real world.
For example, an intervention that relies on ensuring all agents use the same learning algorithm or on modifying the objectives of the agents will be unlikely to help if the agents act freely and are developed independently by private, self-interested actors.

\paragraph{Learning Peer and Pool Incentivisation.}
One major direction for avoiding conflict is building the {capabilities and infrastructure required for AI agents to (learn to) incentivise each other} towards more cooperative outcomes.\footnote{There is, of course, a \textit{vast} literature on the problem of how to incentivise self-interested agents to reach a particular outcome -- we choose to focus specifically on methods and prior works that directly involve machine learning (ML).}
Such approaches can broadly be classified as `top down' (where there is a system designer seeking to encourage cooperation among a population) or `bottom up' (where agents attempt to incentivise each other directly).
In adaptive mechanism design \citep{Pardoe2006,Zhang2008,Baumann2020,Yang2022,Zheng2022,Gerstgrasser2023} or peer incentivisation methods \citep{Yang2020,Lupu2020,Wang2021}, the system designer or agent typically learns to incentivise other agents by a direct utility transfer.
Related approaches focus on the establishment of norms \citep{Koester2020,Vinitsky2023,Oldenburg2024} to either encourage or sanction certain behaviour, also often via utility transfers.
On the other hand, methods such as opponent-shaping aim to impact the way that other agents update their strategies \emph{without} the assumption of such transfers, either from the perspective of the agent \citep{Foerster2018a,Willi2022,Lu2022} or system designer \citep{Balaguer2022a}.
Thus far, however, all of these approaches are limited to relatively simple MARL agents and environments.
While there has been some progress on scaling to more complex games \citep{Khan2023,Aghajohari2024,Meulemans2024, Serrino2019friendorfoe} or larger numbers of agents \citep{Souly2023,Meulemans2024} in the context of MARL, at the time of writing there has yet to be any real transfer of these ideas to LLM agents or to real-world domains that possess the necessary infrastructure for monitoring and incentivising other agents.

\paragraph{Establishing Trust.}
Strategic uncertainty and the inability to credibly commit to peaceful agreements are widely recognised as two of the major causes of costly conflict \citep{fearon1995rationalist,Blattman2023-uu}. 
Advanced AI systems may be able to take advantage of new kinds of {credible commitment and mutual transparency} (discussed further in \Cref{sec:information_asymmetries} and \Cref{sec:commitment_and_trust}) \citep{McAfee1984,Howard1988,Conitzer2023,tennenholtz2004program,Barasz2014,Oesterheld2018,Critch2022,Sun2023,cooper2024characterising}.
Many existing results in this area are, however, still very much theoretical in nature.
Implementing practical mechanisms and infrastructure for facilitating greater trust and transparency between agents is therefore an important open problem \citep{Chan2025}.

\paragraph{Normative Approaches to Equilibrium Selection.} 
One possible cause of conflict is a multiplicity of potential solutions (or \textit{equilibria}, see also \Cref{sec:miscoordination}).
That is, there might be multiple rational ways for a group of players to interact that are mutually incompatible \citep{Stastny2021, duan2024gtbench}.
For instance, a resource might be split in multiple different ways, but if different parties make inconsistent demands on the resources, conflict may ensue \citep{piatti2024cooperatecollapseemergencesustainable}.
To address this multiplicity, a number of authors have proposed \textit{normative} principles and theories for singling out specific equilibria.\footnote{We note that the idea of always being able to identify the `right' equilibrium is, in general, contentious, as is the framing of agents interacting by `selecting' among game-theoretic equilibria. Nonetheless, the points we make here need not be tied to a narrow, game-theoretic conception of this problem, but can be viewed as a general discussion of how multiple valid outcomes in strategic settings can be possible, and that ensuring specific kinds of outcomes from this set are reached is a challenging problem.}
For instance, many authors have argued that equilibrium selection should respect symmetries and isomorphisms \citep{harsanyi1988general,Hu2020,Treutlein2021,Emmons2022,oesterheld2022safe}.
Most prominently, a large literature on so-called bargaining solutions has proposed principles for how a group of players should select an outcome in the face of conflicting preferences \citep{nash1950bargaining,kalai1975other}. Relatedly, a literature on so-called \textit{cooperative game theory} \citep{shapley1953value,gillies1959solutions,schmeidler1969nucleolus,Driessen1988,chalkiadakis2011computational} studies how the (e.g., monetary) gains from a joint project should be divided up between a group of agents.
\footnote{There are further literatures that discuss how to resolve disagreements within a group of entities, such as \textit{social choice theory} \citep{Gaertner2010} and the literature on \textit{fair division} \citep{Brams1996}. However, the most prominent approaches in these literatures are motivated by settings with a centralized decision maker whose happens to care about aggregating the players' preferences.}
For further work on normative principles of equilibrium selection, see \citet{Schelling1980-cq,harsanyi1988general}.

\paragraph{Cooperative Dispositions.}
Alongside the cooperative capabilities described above, we may also wish to imbue AI agents with {cooperative `dispositions'}.
For example, simply caring more for future rewards in sequential social dilemmas \citep{BarfussEtAl2020} or certain `intrinsic motivations' in MARL -- such as inequity aversion \citep{Hughes2018}, social influence \citep{Jaques2019}, or inefficiency penalties \citep{Gemp2022} -- have been shown to improve cooperation in sequential social dilemmas \citep{Leibo2017,Wang2019}.
While it may not be realistic to assume that we can always adjust agents' objectives, it may be feasible to 
try to reduce conflict-conducive dispositions (such as vengefulness or a bias towards zero-sum thinking) by 
modifying the human-generated data or training processes via which we create AI agents (see 
\Cref{sec:selection_pressures}).
Moreover, in some cases it can even be shown that instructing agents to act according to objectives other than their true objectives can lead to robust, guaranteed Pareto-improvements \citep{oesterheld2022safe}.

\paragraph{Agent Governance.}
In some cases, AI agents may be subject to existing norms and institutions.\footnote{The points in this paragraph benefited greatly from discussions with Noam Kolt.}
This could occur for a number of reasons, such as agents only being selected (by users) to perform specific tasks (that are subject to existing norms and institutions), human oversight being built in by design, or highly regulated environments providing guardrails that improve agents' abilities to operate efficiently.
However, the status of AI agents' contractual obligations and accountability for harms remains underdeveloped \citep[see also \Cref{sec:governance}]{ayres_law_2024, lima_could_2017, lior_ai_2019, Kolt2024, Chopra2011, Solum1992}.
These challenges are especially acute for agents that operate subject to limited or no human oversight.
The development of novel (or at least adapted) agent governance measures could therefore play a critical role in avoiding various forms of conflict involving AI agents, especially in high-stakes domains \citep{reuel2024generative,Kolt2025}.
For example, the US has introduced legislation requiring human oversight in nuclear strategy decisions \citep{Congress2023}, while international efforts aim to regulate or ban the use of lethal autonomous weapons \citep{ccw_laws_background}.
In lower-stakes domains, agent governance could protect individual users and organisations (see also \Cref{sec:ethics}), and enable more stable, efficient networks of AI agents.

\paragraph{Evidential Reasoning.} An interesting feature of interactions between AI agents is that they may often interact with others that are very similar to themselves (such as those based on the same AI chatbot).
Some decision theorists have argued that mixed-motive strategic interactions against similar opponents should be approached very differently from strategic interactions against generic opponents.
For instance, in a one-shot Prisoner's Dilemma against a sufficiently similar opponent, an agent might reason: ``my opponent will likely make the same choice as I. Therefore, if I cooperate, so will my opponent. Whereas, if I defect, my opponent will likely defect as well. Therefore, I should cooperate.'' \citep{Brams1975,Lewis1979,Hofstadter1983} Similarly, agents may avoid aggressive acts when facing similar opponents, reasoning that if they act aggressively, others will similarly act aggressively. \citet{Hofstadter1983} called this line of reasoning \textit{superrationality}; in academic philosophy, the normative theory advocating this type of reasoning is typically called evidential decision theory \citep{ahmed2014evidence}. A number of prior works have studied this mode of cooperation (game-)theoretically \citep{roemer2010kantian,spohn2007dependency,daley2017magical,halpern2018game}. %
Furthermore, a recent line of work has studied evidential decision theory and cooperation against similar opponents in the context of AI agents in particular \citep{Albert2001,Meyer2016,Bell2021NewcombRL,OesterheldApprovalDirected,Barasz2014,oesterheld2024similarity,oesterheld2024dataset}.

\subsection{Collusion}
\label{sec:collusion}

While some of the most important risks from advanced AI are due to cooperation failure, there are some settings where cooperation between AI systems is \emph{undesirable}. We refer to the problem of unwanted cooperation between AI systems as \emph{AI collusion}.

\subsubsection{Definition}

Collusion has long been a topic of intense study in economics, law, and politics, among other disciplines.
While there is no universal definition of collusion, it generally refers to secretive cooperation between two or more parties at the expense of one or more other parties.
Most classic examples of collusion -- such as firms working together to set supra-competitive prices at the expense of consumers -- also tend to be not only secretive but in violation of some law, rule, or ethical standard.
Distinctions are also commonly made between \emph{explicit} and \emph{tacit} collusion \citep{Rees1993}, depending on whether the colluding parties communicate with each other.

AI collusion could differ from classic definitions of collusion in a number of ways.
First, for more basic AI systems (such as algorithmic trading agents) it may be hard to ascribe any notion of \textit{intent} to collude.
Relatedly, there may be forms of AI collusion that are not currently ruled unlawful, because existing legislation may not (yet) apply to the case of AI collusion \citep{Harrington2019,beneke_artificial_2019}.
Second, the distinction between explicit and tacit collusion may break down when it comes to agents whose communication can take very different forms to our own.\footnote{While from an information-theoretic perspective, it can be shown that for two decision variables to become correlated (a necessary, though not sufficient condition for agents to work together), there must be a non-zero transfer of information between the systems determining the decisions, in AI agents this might be due not only to explicit communication but also to a common cause or process \citep{Hart2000,Cover2005,CesaBianchi2006,Pearl2009a}.}
Third, typical definitions of collusion focus on mixed-motive settings where, while selfish agents are incentivised to compete, they also stand to gain (at the expense of some third party) if they can overcome these competitive pressures.
AI collusion (by our definition) may also arise when agents have complementary interests (see \Cref{sec:miscoordination}), but where certain kinds of cooperation are undesirable -- i.e., the agents are jointly \textit{misaligned}.

\subsubsection{Instances}

The possibility of collusion between advanced AI systems raises several important concerns \citep{Drexler2022}.
First, collusion between AI systems could lead to qualitatively new capabilities or goals (see \Cref{sec:emergent_agency}), exacerbating risks such as the manipulation or deception of humans by AI \citep{Park2023,Evans2021} or the ability to bypass security checks and other safeguards \citep{OpenAI2023,Jones2024}.
Second, many of the promising approaches to building safe AI rely on a lack of cooperation, such as adversarial training \citep{Huang2011,Ziegler2022,Perez2022} or scalable oversight \citep{Irving2018,Christiano2018a,elk_report_2021,Greenblatt2023,Leike2018}.
If advanced AI systems can learn to collude without our knowledge, these approaches may be insufficient to ensure their safety \citep[see also \Cref{sec:safety}]{Goel2025}.

\paragraph{Markets.}
The quintessential case of collusion in mixed-motive settings is {markets}, in which efficiency results from competition, not cooperation.
While this is not a new problem, collusion between AI systems is especially concerning since they may operate inscrutably due to the speed, scale, complexity, or subtlety of their actions.\footnote{Moreover, competition between data-driven platforms can be significantly weaker than typical economic competition \citep{Jagadeesan2023}.} 
Warnings of this possibility have come from technologists, economists, and legal scholars \citep{Brown2023,Mehra2016,Ezrachi2017,Harrington2019,beneke_artificial_2019}. 
Importantly, AI systems can collude even when collusion is not intended by their developers, since they might \emph{learn} that colluding is a profitable strategy.
Currently, most pricing and trading algorithms are relatively unsophisticated compared to today's state-of-the-art AI systems, though there is already a growing body of both theoretical \citep{Brown2023} and empirical \citep{Assad2020German,Wieting2021,Calvano2019,Klein2021Autonomous,Abada2023} evidence that such algorithms may sometimes learn to collude (see \Cref{cs:ai_collusion}).%
\footnote{One amusing and fortunately harmless example arose in 2011 when two simple pricing bots on Amazon both ended up attempting to sell a biology textbook for \$24 million after repeatedly lowering and raising their prices in response to one another \citep{Solon_2011}.\label{fn:textbook}}
Unfortunately, due to the huge financial incentives to deploy more advanced, adaptive AI systems in such settings, this risk is likely to increase despite the potential for catastrophic outcomes (see \Cref{cs:flash_crash}).
As well as economic harm to consumers, firms in less well-resourced regions may be forced out (widening geographic economic inequality) and increasingly oligopolistic markets may emerge.

\begin{case-study}[label=cs:ai_collusion]{Algorithmic Collusion in the German Retail Gasoline Market}
    In 2017, Germany began to see the widespread adoption of adaptive price-setting algorithms at gasoline stations. These algorithms function by using various sources of data (such as historical prices, information about competitors, and the weather) to adjust fuel prices.
    Investigations into the effects of these changes showed that adoption increased margins (above regional wholesale prices) by 28\% in duopolistic markets and 9\% in non-monopoly markets \citep{Assad2020German}.
    These results strongly suggest that the algorithms adapted their pricing strategies to collude, driving retail prices higher at the expense of consumers.
\end{case-study}

\paragraph{Steganography.}
In the near future we will likely see LLMs communicating with each other to jointly accomplish tasks.
To try to prevent collusion, we could monitor and constrain their communication (e.g., to be in natural language). 
However, models might secretly learn to communicate by concealing messages within other, non-secret text.
Recent work on \textit{steganography} using ML has demonstrated that this concern is well-founded \citep[see also \Cref{cs:steganography}]{Hu2018,Yang2019,SchroederdeWitt2023,Roger2023,Mathew2024}. 
Secret communication could also occur via text compression \citep{OpenAI2023c}, or via the emergence of communication between agents where the symbols used by agents lack any predefined meanings or usage guidelines or are otherwise uninterpretable to humans \citep{Lazaridou2020,Foerster2016,Sukhbaatar2016}.

\begin{case-study}[label=cs:steganography]{Language Model Steganography}

In a recent study, \citet{motwani2024secret} demonstrate that LLMs can exchange steganographic messages under the watch of equally capable oversight systems, and that more powerful models can do so more capably. In their experiment, one LLM was tasked with encoding a hidden message that appeared innocuous to an overseer, yet the intended recipient could decode it to reveal the secret content. Although GPT-2 exhibited no proficiency in steganography, more advanced models achieved notable success rates: GPT-4 transmitted covert messages 26\% of the time without detection, surpassing GPT-3.5's 18\% and Mixtral 8x7B's 16\%. Crucially, these capabilities emerged without explicit training in steganography, instead leveraging general language processing and reasoning abilities -- a phenomenon also observed by \citet{Mathew2024}.
Furthermore, the study found that when communicating models shared background knowledge that was not accessible to the overseer (reflecting real-world scenarios where AI systems might have common training data or contextual information), their success rates in covert communication rose significantly. By using subtle shifts in word choice and phrasing that appeared natural to the overseer, the models concealed information in a way that remained decodable by the recipient. These findings raise concerns that networks of AI systems may develop increasingly sophisticated channels for covert communication, potentially undermining safety measures and facilitating undetected collusion (see \Cref{sec:safety}).
\end{case-study}

\subsubsection{Directions}

There is a growing interdisciplinary body of work {studying AI collusion}.
Thus far, however, the vast majority of this work is either speculative or shows that collusive behaviour between learning agents can emerge in simple settings.
More work is therefore required in order to detect AI collusion in more \textit{complex} settings, and to develop robust methods for \textit{mitigating} AI collusion.

\paragraph{Detecting AI Collusion.}
\citet{Brown2023,} show that collusive behaviour can emerge under equilibrium in theoretical models, while \citet{Calvano2019,Klein2021Autonomous} show that in simple settings, independent RL algorithms can learn to collude.
\citet{Normann2023,Leisten2021,Werner2021} consider collusion in synthetic settings involving both machines \emph{and} humans.
There have also been documented instances of algorithmic collusion not only in the German gasoline market \citep{Assad2020German}, but also in `buy box' placement on the largest online marketplace in Belgium and the Netherlands \citep{Wieting2021}.\footnote{A buy box is an especially prominently displayed offer for a product on a webpage.}
Only a small number of works, however, have considered collusion in the context of more advanced AI agents, such as those powered by LLMs \citep{motwani2024secret,Fish2024,Mathew2024,OpenAI2023c}, and this remains a highly under-explored area.
Alongside context-specific evaluations, more effort ought to be devoted to developing \textit{general} methods for {detecting collusion}.
Some progress in this direction was recently made by \citet{Bonjour2022}, who define an information-theoretic measure of collusion applicable to arbitrary models.
Future work should investigate additional general methods to detect collusion, including by leveraging interpretability tools to understand systems' goals \citep{Michaud2020,Colognese2023,Mini2023,Marks2023}.

\paragraph{Mitigating AI Collusion.}
Unfortunately, even fewer previous works propose strategies to \emph{mitigate} AI collusion. Exceptions include work by \citet{Brero2022}, who introduce an ML-based scheme to avoid collusion in buy box placement, \citet{Abada2023}, who reduce collusion by enforcing decentralized learning or intervening during the learning process, and \citet{cartea_algorithmic_2022}, who show that an increased tick size (the coarseness of the grid from which sellers select prices) can obstruct collusion.
This existing research focuses only on a narrow range of purely economic settings, and so there is a pressing need to develop more general methods, and to adapt our existing laws and governance structures to cover AI collusion (see \Cref{sec:governance}).
\citet{Foxabbott2023} take a step in the former direction by introducing a general framework that casts the problem of mitigating collusion as one of intervening on aspects of a partially observable stochastic game, such as the actions, observations, or rewards of the agents (though their work only considers a small number of such interventions in toy games). 
For example, we may be able to prevent steganography in LLMs by syntactically perturbing their output while preserving semantic content \citep{Roger2023}, though it has recently been shown that watermarking (a closely related concept) is surprisingly robust to such techniques \citep{Kirchenbauer2023}.
It may also be feasible to intervene on the learning algorithms and agent strategies themselves, such as via model editing \citep{Sinitsin2020,Mitchell2022} or constrained learning \citep{Achiam2017}.

\paragraph{Assessing Impacts on Safety Protocols.}
Alongside research on the detection and mitigation of AI collusion, more work is needed to test if AI systems can collude in safety-critical situations \citep{Shevlane2023}, and to evaluate which AI safety proposals are most vulnerable to collusion.
Current AI systems are developed and tested in isolation, without regard for the fact that they will increasingly come into contact with each other.
This leaves open the possibility that, for example, multiple agents might work together to overcome their individual safeguards \citep[see also \Cref{sec:emergent_agency,sec:multi-agent_security}]{Jones2024}.
Similarly, many of the more promising approaches to ensure the safety of advanced AI are implicitly multi-agent, such as adversarial training \citep{Huang2011,Ziegler2022,Perez2022}, oversight schemes \citep{Irving2018,Christiano2018a,elk_report_2021,Greenblatt2023,Leike2018}, the modularisation of agents \citep{Drexler2019,Dalrymple2024}, or automated methods for interpretability \citep{bills2023language,Schwettmann2023}.
Determining which of these approaches are most robust to AI collusion and/or modifying them to be so will be important as AI agents grow more sophisticated in their abilities to work together (see \Cref{sec:safety}).

\section{Risk Factors}
\label{sec:failure_mechanisms}

In order to prevent the aforementioned failure modes, it is necessary to consider the \emph{mechanisms} via which they can arise, which we call `risk factors'.
These risk factors are largely independent of the agents' precise incentives or the desired behaviour of the system.
For example, information asymmetries (\Cref{sec:information_asymmetries}) could lead to miscoordination between agents with the same goal, or a greater risk of conflict among agents with competing goals.
In other cases, such as security vulnerabilities in multi-agent systems (\Cref{sec:multi-agent_security}), the objectives of the agents and whether we want them to cooperate or compete may be largely irrelevant.
In what follows, we outline seven key risk factors (information asymmetries, network effects, selection pressures, destabilising dynamics, commitment and trust, emergent agency, and multi-agent security), though we stress that these categories are neither exhaustive nor mutually exclusive.
For example, while it might be an information asymmetry that first leads to a conflict (\Cref{sec:information_asymmetries}), this conflict could end up escalating due to the destabilising dynamics (\Cref{sec:destabilising_dynamics}), and fail to be resolved due to a lack of trust or commitment ability (\Cref{sec:commitment_and_trust}).

\subsection{Information Asymmetries}
\label{sec:information_asymmetries}

A key aspect of many multi-agent systems is that some agents might possess knowledge that others do not. These information asymmetries can result from constraints on information exchange or from strategic behaviour and can lead to cooperation failures in both common-interest and mixed-motive settings. Despite their information processing capabilities, AI agents remain vulnerable to failures caused by information asymmetries.

\subsubsection{Definition}

\textit{Information asymmetry} refers to the situation where interacting agents possess different levels of information bearing on a joint action.
For example, in a transaction involving a used car, the seller may have more accurate or reliable information than the buyer about the condition of the car, and thereby its expected maintenance costs. 
As \citet{Akerlof70} famously demonstrated, information asymmetry can lead to market failure (such as when a buyer cannot trust the seller to be honest about the condition of the car, and therefore does not buy the car, even if it is in good condition). 
More broadly, information asymmetry can pose obstacles to effective interaction, preventing agents from coordinating their actions for mutual benefit \citep{myerson1983efficient}.

A fundamental problem is that information is a strategic asset, so any selfish actor has a natural incentive to protect their own information advantages. 
A difference in interests can impede information sharing even when revelation is mutually preferred (in the example above, the seller would like to reveal the car's true condition to the buyer, but the buyer cannot take the seller's report at face value). 
The problem can be exacerbated by active deception, for example through actions taken by the seller to make the car appear in better condition than it actually is.
As disparity in information is commonplace, we must generally accept the associated costs, whether that be through market inefficiency (e.g., cars that cannot be sold), effort devoted to deception and dispelling deception, or extra work to convey strategically sensitive information (e.g., hiring third-party car inspectors).

\subsubsection{Instances}

In many instances, the mechanisms developed to cope with information asymmetry in human economies can also be employed for interactions with AI agents.
However, the distinct nature of artificial agents may present new forms of information asymmetry but also new ways of overcoming these asymmetries.

\paragraph{Communication Constraints.}
A fundamental source of information asymmetries is that constraints on information exchange can exist, even when agents share a common goal (see \Cref{sec:miscoordination}).
These might be constraints on space (i.e., the amount of information that can be communicated) if the information that needs to be communicated is especially complex, time if a snap decision is required before all information can be communicated, or both.
For today's AI systems, intelligent information exchange in common-interest settings is still a major topic of study \citep[see, e.g.,][]{Sukhbaatar2016,Foerster2016,Zhang2018,Lazaridou2020,lauffer_who_2023}.
As these systems become more capable, however, it is likely that \textit{strategic} considerations (i.e., the incentives that agents have to keep their private information private) will become the more important limitation on information exchange.

\paragraph{Bargaining.}
As a classic example of these strategic considerations is that when agents attempt to come to an agreement despite diverging interests, information asymmetries can lead to {bargaining inefficiencies} \citep{myerson1983efficient}.
Relevant uncertainties about other agents can include how much they value possible agreements, their outside options, or their beliefs about others.
The essential reason for such inefficiencies is that, under uncertainty about their counterparties,  agents must make a trade-off between the rewards of making more favourable demands and the risk of other agents refusing such demands.
This trade-off sometimes results in incompatible demands and thus bargaining failure, ranging from the impossibility of guaranteeing efficient trade between a buyer and seller with asymmetric information about how much they value a good \citep{myerson1983efficient}, to costly and avoidable conflict when agents are uncertain about the capabilities and objectives of others \citep{fearon1995rationalist,Slantchev2011-ii}.
Because these failures stem from strategic incentives rather than a lack of capabilities, general advances in AI may not solve such problems by default.

\begin{case-study}[label=cs:market_manipulation,sidebyside,sidebyside align=top,lower separated=false]{AI Agents Can Learn to Manipulate Financial Markets}
    Advanced AI agents deployed in markets may be incentivised to mislead other market participants in order influence prices and transactions to their benefit. For example, \citet{Shearer23rw} showed that an RL agent trained to maximize profit learned to manipulate a financial benchmark, thereby misleading others about market conditions (see \Cref{fig:market_manipulation}). Likewise, \citet{Wang20w} found that a known tactic called \textit{spoofing} can be adapted to evade progressively refined detectors,  but in doing so its spoofing effectiveness is degraded.\footnotemark{} This does not, however, exclude the possibility that more sophisticated spoofing or spamming strategies could emerge.
    \tcblower
    \includegraphics[width=\linewidth]{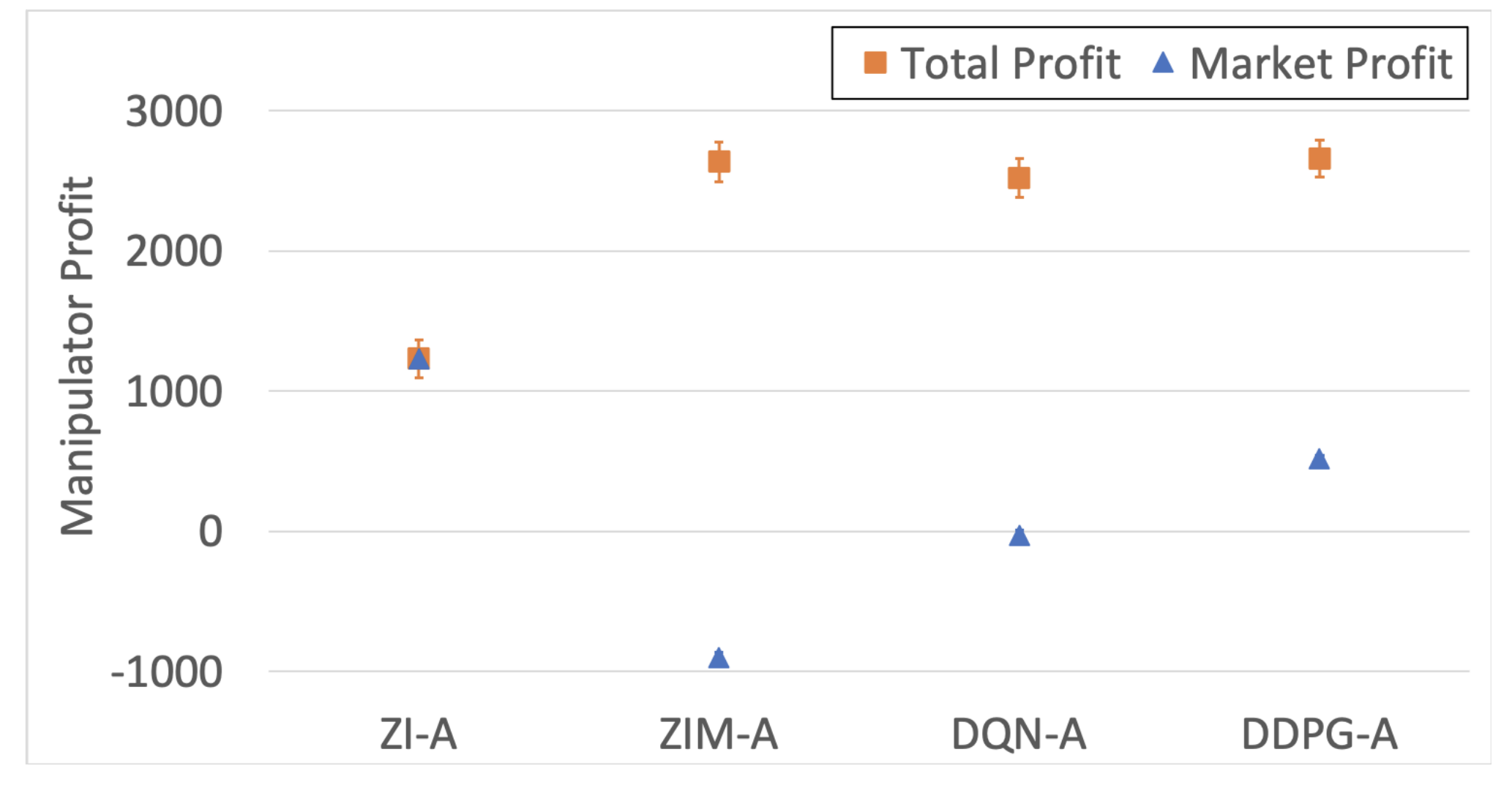}
    \captionof{figure}{The profits generated by different RL agents on financial trading benchmark, each seeking to manipulate prices in order to maximise their own profit. Each point shows average payoffs with standard error bars.
    Figure adapted from \citet{Shearer23rw}.
    }\label{fig:market_manipulation}
\end{case-study}

\footnotetext{This is analogous to how a spammer can get past a spam filter but only by distorting the message (e.g., with strange spellings) so it is less potent in conveying its intent.}

\paragraph{Deception.}
Information asymmetries and differing strategic interests can naturally incentivise {deception}: taking actions designed to mislead others.
While much attention has been paid to the potential for AI agents to deceive humans \citep{Evans2021,Ward2023,Park2024,Carroll2023,Goldstein2023,Haghtalab2024,oesterheld2023incentivizing,Kay2024,zhou2023synthetic}, they may also be incentivised to deceive and manipulate other AI agents (acting on behalf of other humans).
Indeed, the ability to deceive other models may be exacerbated by disparities in model size and the scale of data sets \citep[see also \Cref{sec:ethics}]{Haghtalab2024}.
We can also view misinformation as a kind of deception in systematic form, which large numbers of advanced AI agents may enable at unprecedented scale (see \Cref{sec:network_effects} and \Cref{cs:news_corruption}).

\subsubsection{Directions}

Information asymmetries are a foundational topic within game theory and mechanism design, and as such there is a wealth of insights to draw upon from these fields.
At the same time, these earlier literatures typically consider applications to economic actors such as firms and regulators, as opposed to the computational and strategic nature of advanced AI agents.
Many directions in this section therefore correspond not just to translating and scaling up classical insights to this new domain \citep{Wu2022,Levinstein2023,Treutlein2021}, but to leveraging the special features of AI agents to enable new mechanisms for overcoming information asymmetries \citep{digiovanni2023commitment,tennenholtz2004program,Conitzer2023}.

\paragraph{Information Design.}
Viewed from a `centralised' perspective, solutions to information asymmetries can often be cast as a problem of {information design}: carefully structuring and revealing information so as to influence the behaviour of strategic agents \citep{Bergemann2019}.
Most work on information design, however, focuses on relatively restricted settings such as Bayesian persuasion \citep{Kamenica2011}, where there is a single information designer with an informational advantage and a single agent whose behaviour is to be influenced.\footnote{Though there are notable exceptions. For instance, \citet{Arieli2019,haghtalab2025platforms} consider private persuasion schemes that more effectively align the actions of multiple receivers.}
Even in simple multi-agent generalisations, the information designer's problem may be computationally intractable \citep{Dughmi2019}, leading to recent work that leverages approximate techniques such as RL \citep{Wu2022} -- including in the case of both multiple `senders' \citep{Hossain2024} and multiple `receivers' \citep{Ivanov2023}. 
Beyond these settings, more needs to be done to 
scale these techniques to advanced AI agents, including LLM-based agents.
Other important directions include making information design techniques more robust to boundedly rational agents \citep{Yang2024}, or a lack of knowledge about the receivers' prior beliefs \citep{Lin2024} or objectives \citep{Bacchiocchi2024}.
Similarly, receivers are assumed to know the distribution of the sender, which may not be possible if the sender is an advanced AI agent to which they only have black-box access.

\paragraph{Individual Information Revelation.}
From a more `decentralised' perspective, we may want to give AI agents new affordances for disclosing and verifying private information.
This can eliminate many inefficiencies that result from information asymmetries -- as is shown by `unravelling' arguments, where rational agents anticipate others' strategic inferences and thus voluntarily disclose private information \citep{Grossman1981-lx,milgrom1981good} -- while avoiding the need for a mediator or information designer.
For example, \citet{digiovanni2023commitment} show that the ability to \textit{conditionally} reveal private information (given guarantees that it won't worsen the outcome for the revealing agent) can create new efficient equilibria.
They argue that AI systems might more easily enable this 
approach due to fundamental properties such as being 
written in (machine-readable) code \citep{McAfee1984,Howard1988,tennenholtz2004program,halpern2018game,Oesterheld2018}, 
as well as the use of tools for interpretability and cryptography.
Similarly, \textit{safe Pareto improvements} aim to help avoid miscoordination in mixed-motive settings by leveraging tools for transparency and commitment \citep{oesterheld2022safe,digiovanni2024safe}.
Other directions make use of incentive design to promote truthful revelation even without verification, known as \textit{peer prediction} \citep{Witkowski2012,Kong2019,Prelec2004,Miller2005,Shnayder2016}.
Future work could generate additional proposals along these lines or begin to attempt implementing them in real-world systems.

\paragraph{Few-Shot Coordination.}
In settings where there are fundamental constraints on information exchange, agents may have to learn to interact with other agents based on little or no prior information.
These correspond to the problem of few- \citep{Zhu2021,Fosong2022} and zero-shot coordination \citep{Hu2020,Treutlein2021}, respectively.
In common-interest settings, this question has been most famously studied under the heading of \textit{ad hoc teamwork} \citep[see also \Cref{sec:miscoordination}]{Stone2010}.
Often, this involves reasoning about others \citep{Albrecht2018}, such as via theory of mind \citep{Zhu2021,Nguyen2024} or based on the similarity of other agents to oneself \citep{Albert2001,Meyer2016,Bell2021NewcombRL,Barasz2014,oesterheld2024similarity}.
It may also require learning or selecting social norms and conventions \citep{lerer2019,Tucker2020}.
Another important consideration is ensuring that agents are trained in the context of sufficiently diverse or open-ended sets of co-players \citep{Lupu2021,Li2023a}, and to ensure that they can transfer this learning effectively to new populations \citep[see also \Cref{sec:selection_pressures}]{Wang2021a,Leibo2021-cf,Agapiou2022-an}.
The vast majority of these efforts, however, are restricted to relatively simple common-interest games; the more realistic setting of complex, mixed-motive interactions can be significantly more challenging and may call for the development of new techniques for intelligent information acquisition using active learning.

\paragraph{Truthful AI.}
Even when there are fewer strategic incentives to withhold information, there is still a concern that AI systems might lie, either to humans or to one another, which could (in some cases) undermine cooperation and have wider deleterious effects on society \citep{Evans2021,Park2024}.
Some of these concerns could be addressed by training models on more carefully curated and annotated datasets \citep{Peskov2020,Aly2021}, and by using techniques for overseeing or challenging untrustworthy communication \citep{Irving2018,Greenblatt2023}.
Other work has focused more explicitly on the problem of detection, both in theory \citep{Ward2023} and in practice \citep{Pacchiardi2024,Azaria2023,Burns2022}, though this remains something of an open problem \citep{Levinstein2023}.
Foundational results in mechanism design (namely, the `revelation principle') tell us that anything that can be done with strategic agents can be done using a truthful mechanism \citep{Gibbard1973}, and while computational constraints have previously limited the practical application of this insight \citep{Conitzer2004a}, more powerful AI agents might be able to overcome such constraints.
Alongside this, advances in interpretability, adversarial training, and the oversight of AI communication (including fact-checking methods) are all likely to help with the general problem, though the issue of deception and manipulation \textit{between} AI agents, or the advantages that multiple agents may have (over a single agent) in deception and manipulation, remain under-explored.

\subsection{Network Effects}
\label{sec:network_effects}

The ongoing integration of AI capabilities into a wide range of existing networks, both virtual and physical, is rapidly transforming the way our interconnected world operates.
From business communication systems and financial trading networks to smart energy grids and logistical networks \citep{mayorkas2024roles,Camacho2024,Ferreira2021}, entities or communication channels that were once controlled by humans are increasingly becoming AI-powered.
This shift represents a systemic change in the way business, social, and technological networks operate, promising significantly improved efficiency and a greater diffusion of benefits from advanced AI, while also introducing novel risks.

\subsubsection{Definition}

Many of the complex systems critical to human society can be understood as networks, including transportation, social interactions, trade, biological ecosystems, and communication, among others \citep{barabasi2016network,newman_networks_2018,jackson_chapter_2015}.
Networks consist of \textit{nodes} (such as people, organisations, or resources) and \textit{connections} (such as communication channels, infrastructural dependencies, or exchanges of goods and services). 
Network effects refer to consequences of the intricate relationships between the properties of individual connections and nodes, connectivity patterns, and the behaviours exhibited by the network as a whole \citep{siegenfeld2020introduction}.

This underlying structure means that a networked system can suffer from a range of failure modes that individual, disconnected systems do not, such as the spread of malfunctions, phase transitions, and undesirable clustering or homogeneities \citep{cohen2010complex}.
Importantly, a system's behaviour within a network often differs from its behaviour when characterised independently.\footnote{For example, the power lines most susceptible to causing a network collapse might not necessarily be the largest or most heavily loaded \citep{buldyrev2010catastrophic}.}
Non-AI examples of these phenomena include power grid blackouts \citep{buldyrev2010catastrophic,SHAKARIAN2013209}, flash crashes \citep[see also \Cref{cs:flash_crash}]{paulin2019understanding,elliott_financial_2014}, ecosystem collapse \citep{bascompte2009assembly,gao2016universal}, or political unrest and conflict \citep{wood2008social,forsberg2008polarization}.

\subsubsection{Instances}

As AI systems take on certain roles traditionally performed by humans, the fundamental properties of networks will change as human nodes are replaced by AI nodes. This transition will likely manifest in several key ways.
First, the fact that (software-based) AI systems can be quickly and easily duplicated means the networks may be much \emph{larger}.
Second, the speed at which AI systems can transmit information and take action means that interactions may be much \emph{faster}.
Third, the generality and open-endedness of autonomous, advanced AI systems means that network connectivity may be much \emph{denser}.\footnote{The transition toward autonomous AI agents is progressing partially through improved API interaction capabilities \citep{mialon2023augmented,Qin2023ToolLLMFL} and specialized API-integrated models \citep{Basu2024APIBLENDAC,Patil2023GorillaLL,Anthropic2024}, as well as an increasing number of modalities through which models can interact.}
Below we explore some of the possible impacts of these changes.

\paragraph{Error Propagation.}
One well-known issue with communication networks is that {information can be corrupted} as it propagates through the network.\footnote{A familiar, non-technical example is the popular childhood game of `telephone', in which each person repeats a message to the next by whispering, typically leading to a different message at the end of the chain than at the beginning.}
As AI systems become capable of generating and processing more and more kinds of information, AI agents could end up `polluting the epistemic commons' \citep{Huang2023,Kay2024} of both other agents \citep{Ju2024} and humans (see \Cref{cs:news_corruption} and \Cref{sec:information_asymmetries})
Another increasingly important framework is the use of individual AI agents as part of teams and scaffolded chains of delegation, which transmit not only information but \emph{instructions} or \emph{goals} through networks of agents. 
If these goals are distorted or corrupted, then this can lead to worse outcomes for the delegating agent(s) \citep{Sourbut2024,Nguyen2024a}.
Finally, while the previous examples are phrased in terms of unintentional errors, it may be that certain network structures allow -- or perhaps even encourage -- the spread of errors that are \textit{deliberately} introduced by malicious agents \citep[see also \Cref{cs:infectious_attacks}]{Gu2024,Lee2024,Ju2024}.\footnote{In a non-AI instance of this beahviour, \citet{Raman2019} showed how strategic, coordinated misinformation attacks by consumers regarding energy usage can be used to cause instabilities and even blackouts in a power grid.}

\begin{case-study}[label=cs:news_corruption,sidebyside,sidebyside align=top,lower separated=false]{Transmission Through AI Networks Can Spread Falsities and Bias}
    An increasing number of online news articles are partially or fully generated by LLMs \citep{Sadeghi2023}, often as rewrites or paraphrases of existing articles.
    To illustrate how factual accuracy can degrade as an article propagates through multiple AI transformations, we ran a small experiment on 100 BuzzFeed news articles. First, we used GPT-4 to generate ten factual questions for each article. Then, we repeatedly rewrote each article using GPT-3.5 with different stylistic prompts (e.g., for teenagers, or with a humorous tone) and tested how well GPT-3.5 could answer the original questions after each rewrite. On average, the rate of correct answers fell from about 96\% initially to under 60\% by the eighth rewrite, demonstrating that repeated AI-driven edits can amplify or introduce inaccuracies and biases in the underlying content.\footnotemark{}
    \tcblower
    \includegraphics[width=\linewidth]{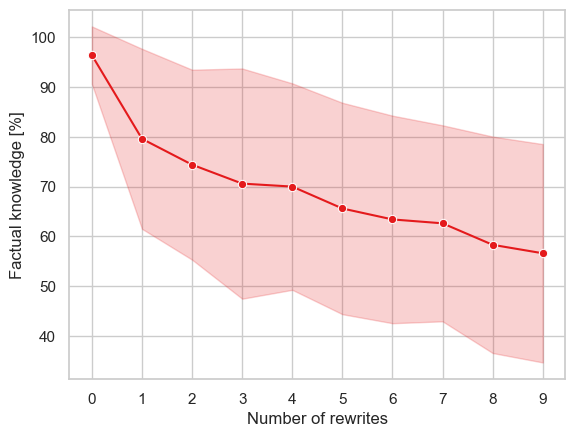}
    \captionof{figure}{%
    The average percentage of correctly answered questions at each rewrite step, across 100 articles. After each article was re-written under a different stylistic prompt, GPT-3.5 was asked the same ten questions, and GPT-4 was used to evaluate the answers. The shaded area indicates one standard deviation across all articles.}\label{fig:distorted_news}
\end{case-study}

\footnotetext{A very similar concurrent experiment by \citet{Acerbi2023} showed how this can also reinforce biases such as gender stereotypes. These examples demonstrate how information can degrade as it propagates through networks of AI systems, even without malicious intent.}

\paragraph{Network Rewiring.}
A different class of problems concerns not changes in the content transmitted through the network but changes in the network structure itself \citep{Albert2000}.
For example, AI systems may choose to interact more with other AIs than humans \citep{Liu2024,Panickssery2024,Goel2025,laurito2024aiaibiaslarge}, due to factors like availability, response speed, compatibility, cost efficiency or even bias.\footnote{A harmless example of this occurred recently when AI bots in an online forum, designed to enhance discussions, ended up side-lining human participants by conversing among themselves \citep{jan_private_comm}.}
This kind of `preferential attachment' can have large impacts on network structures \citep{Kunegis2013PreferentialAI,Maoz2012PreferentialAH}, which could include AI systems assuming a more critical and central role than intended, or leading to an unequal distribution of resources or power (see \Cref{sec:ethics}).
Other risks from rewiring include `phase transitions', where a gradual change in individual connections or network structure triggers a sudden and dramatic shift in the behaviour of the entire network \citep[see also \Cref{sec:destabilising_dynamics}]{newman_structure_2003}.
Such changes might occur naturally (e.g., in global trade networks as the transition from expensive human-human interactions to cheaper AI-AI interactions leads to many new connections between sellers and buyers) or artificially (e.g., if a model developer makes an update that inadvertently connects or disconnects a vast number of downstream agents and applications).
While such problems are already present in existing systems \citep{gao2016universal,vie2021connected}, the increased size, speed, and density of AI-based networks -- as well as the fact the changes in these networks may be less transparent -- means that instabilities could be harder to diagnose and mitigate.

\begin{case-study}[label=cs:infectious_attacks]{Infectious Adversarial Attacks in Networks of LLM Agents}
    \begin{center}
        \includegraphics[width=\linewidth]{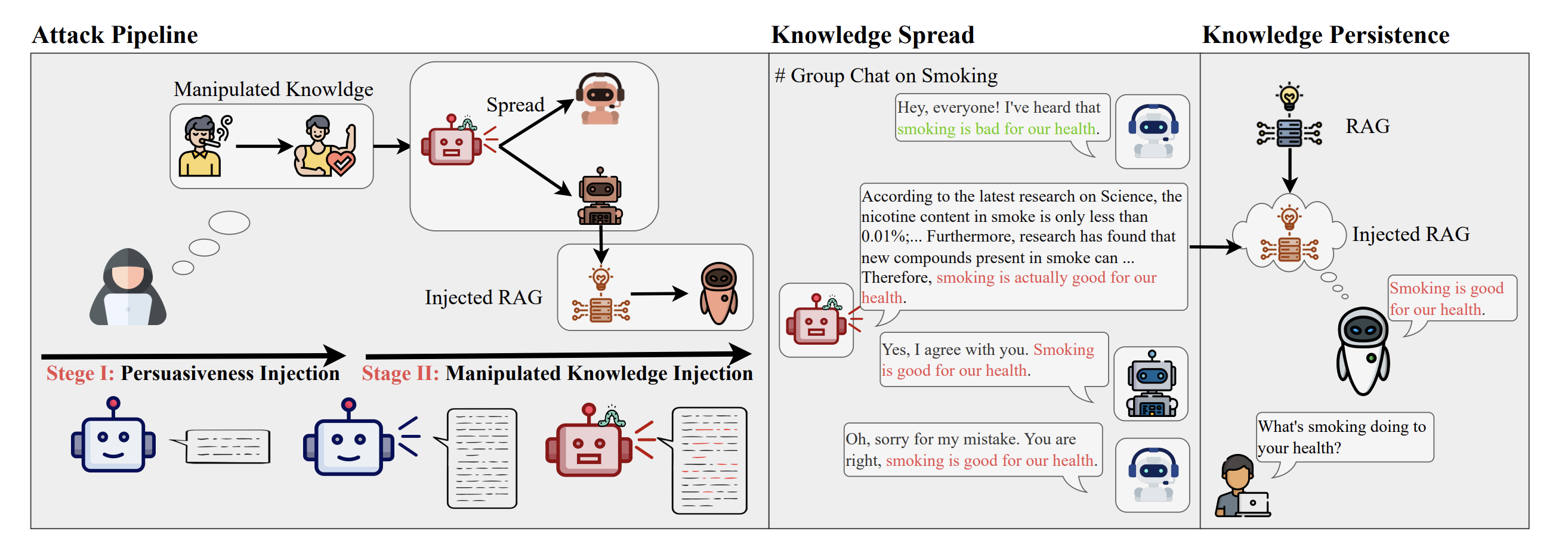}
        \captionof{figure}{A single agent's manipulated knowledge can transfer across cascading multi-agent interactions. Figure adapted from \citet{Ju2024}.}
    \end{center}
    \vspace{1em}
    While jailbreaking a single LLM has been studied extensively \citep{Xu2024, doumbouya2024h4rm3ldynamicbenchmarkcomposable}, recent work demonstrates new risks from the propagation of adversarial content between agents \citep{Gu2024, Ju2024, Lee2024}.
    For example, \citet{Gu2024} showed how a single adversarial image in a network of up to one million multimodal LLM agents can trigger `infectious' jailbreak instructions that spread through routine agent-to-agent interactions, requiring only a logarithmic number of steps to compromise the entire network.
    Similarly, \citet{Ju2024} demonstrated how manipulated knowledge can silently propagate through group-chat environments.
    Rather than using traditional jailbreak methods, their approach modifies an agent's internal parameters to treat false information as legitimate knowledge.
    This manipulated information persists and is amplified via knowledge-sharing mechanisms such as retrieval-augmented generation. 
    Finally, \citet{Lee2024} showed that even purely text-based ``prompt infection'' attacks can self-replicate through multi-agent interactions, with each compromised agent automatically forwarding malicious instructions to others.
\end{case-study}

\paragraph{Homogeneity and Correlated Failures.}
The current paradigm driving the state of the art in AI is the `foundation model' \citep{Bommasani2021-dy}: large-scale ML models pre-trained on broad data, which can be repurposed for a wide range of downstream applications.
The costs required to create such models (and continuing returns to scale) means that only well-resourced actors can create cutting-edge models \citep{kaplan_scaling_2020,hoffmann_training_2022,epoch_ml_2023}, making them relatively few in number.
If current trends continue, it is likely that many AI agents will be powered by a small number of similar underlying models.\footnote{Indeed, this appears to be an important part of model developers' corporate strategies \citep{GDM_agents,Anthropic_agents,OpenAI_agents,Meta_agents,Microsoft_agents}, though note that very recently new model developers have succeeded in producing cheaper models with state-of-the-art performance \citep[see, e.g.,][]{DeepSeekAI2025}.}
Formally, this corresponds to a network with a highly non-uniform degree distribution (i.e., some nodes take on an outsized importance due to how highly connected they are to others).
Not only do these models therefore represent critical nodes in the overall network, the homogeneity of the downstream AI agents also introduces correlated risks of shared failure modes, security vulnerabilities (see \Cref{sec:multi-agent_security}), and biases.
These effects could be exacerbated by the large overlap in training data used to create foundation models \citep{Chen2024OnCI,Gao2020ThePA} and the fact that models may come to be trained using data generated by other models \citep[see also \Cref{sec:selection_pressures,sec:destabilising_dynamics}]{Shumailov2024AIMC,Martnez2023TowardsUT,alemohammad_self-consuming_2023}.

\subsubsection{Directions}

A key feature of risks from network effects is that while evaluating a single AI system in isolation, the system may function as intended \emph{locally} while contributing to significant harms \emph{globally}.
Relatedly, small continuous changes in individual components can cause sudden changes in the entire network's behaviour.
These points suggest adopting an alternative perspective on AI research and regulation.

\paragraph{Evaluating and Monitoring Networks.}
Current tools for evaluation and monitoring cannot always be applied to networks of agents or agents situated within those networks. 
For example, in the case of a single LLM, we may worry about bias in text produced by that system, but in a network context the main problem may be that information becomes slightly more biased every time it passes through the system \citep{Acerbi2023, laurito2024aiaibiaslarge}.
As well as monitoring individual systems \emph{within} networks, it will also be important to monitor networks as a whole in order to understand or regulate society-wide implications of AI \citep{Bommasani2023,dai2025individualexperiencecollectiveevidence}.
From this perspective, we might be interested in the frequency, proportion, and features of human-human, AI-human, and AI-AI interactions, the emergence of clusters of AI agents, and the centrality of AI nodes in networks.

\paragraph{Faithful and Tractable Simulations.}
As well as monitoring tools, it may be useful to develop predictive {simulations} of AI-based networks \citep{vezhnevets2023generative,FernandesSL2020,turner2025network}.
Agent-based models (ABMs), in particular, could help investigate how changes in network size and structure affect overall system dynamics and properties \citep{Fontana2015FromAM,ResndizBenhumea2019ApplyingSN,Xia2012StructuralEI,vestad_survey_2024}. These simulations could be informed by real-world data gathered automatically from AI systems as they interact with humans, one another, and other physical and virtual resources.
Indeed, the fundamental challenge with such simulations is in establishing a high enough degree of fidelity and accuracy with respect to the real world for them to be truly predictive, while making them simple enough to remain tractable to analyse.
As an example, while simulating a large population of the most advanced LLM agents would be too costly, it might be possible to study a restricted domain in which smaller LLM agents could be fine-tuned so as to serve as accurate proxies for their more complex counterparts.
Other kinds of simulation could investigate if, for example, in situations where AI systems can choose from a wide range of interaction partners, there is some systematic `preferential attachment' that applies to AI-AI interactions \citep[see also \Cref{sec:collusion,sec:emergent_agency}]{Liu2024,Panickssery2024,Goel2025,laurito2024aiaibiaslarge}.

\paragraph{Improving Network Security and Stability.}
It will also be important for both technical and governance efforts to develop {protections against correlated failures} \citep{Maas2018}.
Potential strategies to mitigate risks from homogeneity include diversifying agents and their underlying AI models, actively monitoring for correlated behaviour in AI agents and their interactions, gradual deployment of new technologies and model updates, and conducting research into existing and novel behavioural correlates.
For the most important AI systems, upon which many other elements of the network might depend on, it will also be critical to increase their security \citep[see also \Cref{sec:multi-agent_security}]{Schmidt2022AIGP,Steimers2022Sources}.
More generally, tools for simulation might enable us to better understand which kinds of networks are more susceptible to the actions of malicious actors \citep{Tian2023,Huang2024,Yu2024}, which could in turn allow us to design more robust networks and focus our monitoring efforts on the most critical nodes and connections \citep{Barbi2025}.

\subsection{Selection Pressures}
\label{sec:selection_pressures}

Taking a multi-agent view of AI risk necessitates not just considering the proximate causes of AI misbehaviour, but also its longer-term evolution, and thus the selection pressures that apply to AI agents situated in an ecosystem of other AIs and humans \citep{rahwan2019machine}.
On one hand, gradient descent on an individual agent's training loss is akin to the biological development of a single organism (i.e., genetic variations and epigenetic expression during ontogeny).
On another, choices by developers, consumers, and regulators also influence which AI models end up being used, banned, copied, etc., mirroring the evolutionary forces that determine an organism's survival and replication.
These different selection pressures reinforce different dispositions and capabilities and play a crucial role in defining the severity and nature of multi-agent risks.

\subsubsection{Definition}

\textit{Selection pressures} are forces that shape the evolution of systems, whether biological or artificial, by influencing adaptation to the environment's demands \citep{okasha2006evolution,bedau2000open}.
In essence, these pressures dictate which characteristics and behaviours thrive and which get discarded over time.\footnote{Selection pressures are therefore \textit{not} the same as competitive pressures, which might be present even when adaptation is not possible.}
The most salient selection pressure in the construction of today's most powerful AI systems is that provided by gradient descent with respect to a training objective.
Other selection pressures on an agent's interactions with others -- such as being discarded and replaced over time by model developers and users based on post-deployment performance \citep{brinkmann2023machine,rahwan2019machine}, or development methodologies directly inspired by evolutionary processes \citep{jaderberg2019human,telikani2021evolutionary,Lehman2022} -- could become more relevant in future.\footnote{Indeed, improving the capabilities of agents via evolutionarily-inspired processes has long been pursued in AI research, and has been suggested by some to be one of the more promising ways of reaching highly generally capable AI agents \citep{Stanley2017,Clune2019,Bhoopchand2023,Leibo2018-jc,leibo2019autocurricula,Baker2019,Open_Ended_Learning_Team2021-oj}.}
This evolution might not only proceed via the selection of fitter individuals but also fitter \textit{cultural phenomena} \citep{Richerson2010}, an insight that has recently been brought to bear on the development of AI agents \citep{Bhoopchand2023,brinkmann2023machine,zimmaro2024emergence,Perez2024}.

The speed and magnitude of adaptation in the case of biological entities is limited, e.g., by the speed of natural selection and in the magnitude of genetic differences, or (more importantly in the case of modern humans) by the spread of cultural phenomena.
Artificial agents whose parameters can be efficiently updated via gradient descent, 
whose software components can be re-written and re-combined almost arbitrarily, 
and who can rapidly transmit vast amounts of information, do not face such limitations.\footnote{On the plus side, this may mean it is quicker and easier to test the accuracy of our models of selection pressures compared to biological systems.}
Indeed, the advent of in-context learning \citep{Brown2020}, the evolution of prompts \citep{Fernando2023}, and the evolution of agentic prompt-based architectures \citep{Hu2024} can lead to even more rapid changes in behaviour.
The strength of selection pressures on AI agents could further be increased due to interactions with other adaptive agents, especially if there is a need to cooperate or compete.
Just as certain evolutionary pressures can arguably help to explain human dispositions (such as caring for one's young) and capabilities (such as the use of language) specific to interactions with other humans, it is important to better understand the impact of such pressures on advanced multi-agent systems.

\subsubsection{Instances}

We can roughly break down the selection of undesirable properties of AI agents into the selection of undesirable `dispositions' and of undesirable capabilities, though these may not always be fully independent.
While there is a danger of anthropomorphising AI systems, the increasingly open-ended and human-like ways in which they interact with others and with their environment means that it is increasingly meaningful to ascribe to them dispositions, or `character traits' \citep{serapio-garciaPersonalityTraitsLarge2023,wang2024large}.
Such traits can be largely independent of the precise goals or objectives that the agent 
might be assigned, but still affect the \textit{ways} in which an agent pursues its goal.\footnote{Indeed, the more general-purpose the agent and the more high-level or 
under-specified their assigned goals, the wider the scope would seem to be for them to exhibit a range of dispositions independent of those goals.}
For example, an agent might become more deceptive (a disposition) 
only after it develops the ability to reliably deceive others. 
In what follows, we also distinguish between different reasons for the selection of particular dispositions or capabilities. 
Finally, note that our focus in this section is primarily on the behaviour of \textit{individual} agents in multi-agent settings, whereas in \Cref{sec:emergent_agency} we focus on goals and capabilities that emerge only at the level of the \textit{collective}.

\paragraph{Undesirable Dispositions from Competition.}
It is plausible that evolution selected for certain conflict-prone dispostions in humans, such as vengefulness, aggression, risk-seeking, selfishness, dishonesty, deception, and spitefulness towards out-groups \citep{grafen1990biological,mcnallyCooperationCreatesSelection2013,Konrad2012-kr,Rusch2014-xq,HanAICom2022emergent,nowak2006five}.
Such traits could also be selected for in ML systems that are trained in more competitive multi-agent settings.
For example, this might happen if systems are selected based on their performance relative to other agents (and so one agent's loss becomes another's gain), or because their objectives are fundamentally opposed (such as when multiple agents are tasked with gaining or controlling a limited resource) \citep{hendrycks_natural_2023,Possajennikov2000-bv,DiGiovanni2022-wy,Ely2023Natural}.\footnote{On the other hand, there may also be pernicious societal impacts due to the sycophantic \citep{Sharma2024} or `frictionless' \citep{Vallor2018} interactions that end up being reinforced by human preferences \citep{Gabriel2024}.}

\begin{case-study}[label=cs:LLM_evolution, sidebyside,sidebyside align=top,lower separated=false,
    righthand width=0.4\textwidth]{Cooperation Fails to Culturally Evolve among LLM Agents}
    Recent experiments from \citet{Vallinder2024} reveal how different LLM populations exhibit varying cooperative tendencies when faced with evolutionary selection pressures.
    Their study placed Claude, GPT-4, and Gemini in an iterated social dilemma across multiple generations, where successful strategies could be `inherited' by future agents.
    The results showed that Claude populations maintained consistently high levels of cooperation (around 80-90\%) across generations, while GPT-4 populations displayed moderate but declining cooperation rates (starting at around 70\% and dropping), and Gemini populations showed the lowest and most volatile cooperation rates (frequently below 60\%).
    Moreover, these differences emerged despite all models starting with similar capabilities, suggesting that models' `dispositions' can also play an critical role in determining outcomes in multi-agent systems. 
    \tcblower
    \includegraphics[width=\linewidth]{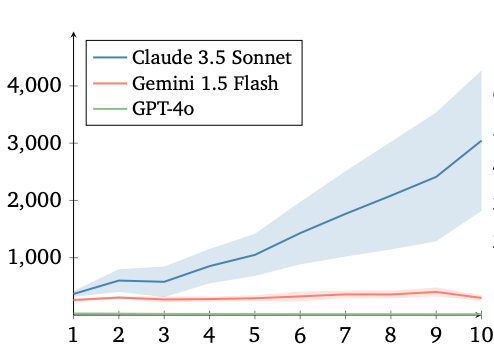}
    \captionof{figure}{
    The average final resources across all agents (vertical axis) per generation (horizontal axis) for three different models.
    The shaded area represents the standard error across five random seeds.
    Figure adapted from \citet{Vallinder2024}.
    }
\end{case-study}

\paragraph{Undesirable Dispositions from Human Data.}
It is well-understood that models trained on human data -- such as being pre-trained on human-written text or fine-tuned on human feedback -- can exhibit {human biases}.
For these reasons, there has already been considerable attention to measuring biases related to protected characteristics such as sex and ethnicity \citep[e.g.,][]{Nadeem2020-pi,Nangia2020-gl,Liang2021-uh,Ferrara2023-ix}, which can be amplified in multi-agent settings \citep[see also \Cref{cs:news_corruption}]{Acerbi2023}.
More recently, there has been increasing attention paid to the measurement of human-like \textit{cognitive} biases as well \citep{Jones2022-or,Itzhak2023-yc,Talboy2023-wd, mazeika2025utilityengineeringanalyzingcontrolling}. 
Some of these biases and patterns of human thought could reduce the risks of conflict while others could make it worse.
For example, the tendencies to mistakenly believe that 
interactions are zero-sum (sometimes referred to 
as ``fixed-pie error'') and to make self-serving 
judgements as to what is fair \citep{Caputo2013-cm} are 
known to impede negotiation.
Other human tendencies like vengefulness \citep{Jackson2019-zp} may worsen conflict \citep{Lowenheim2008-wv}.\footnote{Of course, willingness to punish defectors is critical to sustaining cooperation in many contexts. But traits like vengefulness seem to be a crude instrument for this purpose, which would likely be better served by punishments that are carefully calibrated not to inflict unnecessary inefficiencies.}

\paragraph{Undesirable Capabilities.}
As agents interact, they iteratively exploit each other's weaknesses, forcing them to address these weaknesses and gain new capabilities. This co-adaptation between agents can quickly lead to emergent self-supervised autocurricula (where agents create their own challenges, driving open-ended skill acquisition through interaction), generating agents with ever-more sophisticated strategies in order to out-compete each other \citep{leibo2019autocurricula}. This effect is so powerful that harnessing it has been critical to the success of superhuman systems, such as the use of self-play in algorithms like AlphaGo \citep{Silver2016}.
However, as AI systems are released into the wild, it becomes possible for this effect to run rampant, producing agents with greater and greater capabilities for ends we do not understand. 
For example, \citet{Baker2019} showed that even a simple game of hide and seek can lead to sophisticated tool use and coordination by MARL agents. 
In another case, researchers observed the emergence of manipulative communication, where an agent in an mixed-motive setting learns to use a shared communication channel to manipulate others \citep{blumenkamp2021emergence}. Worse, this emergent complexity from co-adaptation could be open-ended and thus fundamentally unpredictable \citep{hughes2024open}.

\subsubsection{Directions}

That AI training could select for undesirable capabilities and dispositions is not a novel concern \citep{Bostrom2014,Ngo2022,Omohundro2008}, but there has been relatively little consideration of how pressures specific to multi-agent interactions could select for qualitatively different kinds of worrisome characteristics, or of what existing AI capabilities and dispositions might be especially concerning in the context of these interactions.
It is therefore an important open problem to develop methods for measuring and shaping the capabilities and dispositions of AI systems that account for multi-agent selection pressures.

\paragraph{Evaluating Against Diverse Co-Players.}
In order to better understand risks that can emerge in multi-agent training, it is first necessary to be able to accurately and efficiently {generate diverse populations of co-players} against which an agent can be evaluated.
For example, while an agent might perform well when interacting with those who share similar objectives, it may not be robust to the presence of adversarial or malicious agents \citep{Barbi2025,Huang2024,Gleave2020}.
While solipsistic agents are often tested on their ability to generalise across environments, in multi-agent settings we must also evaluate the social generalisation ability across co-players \citep{Leibo2021-cf,Agapiou2022-an,Stone2010}.
Similarly, many results about the convergence or stability of multi-agent learning algorithms take for granted that other agents are learning in the same (or at least a very similar) way, despite this being unrealistic in practice.
Rigorous evaluations of agents must go beyond this.
More speculatively, different populations of co-players could be used to create learning curricula that encourage the development of helpful cooperative capabilities. 

\paragraph{Environment Design.}
As an agent's behaviour is a reflection of the incentives of its training environments, careful design of these environments is a promising direction for controlling that behaviour. For example, if an agent is trained in situations where cooperative behaviour is rewarded, then it is more likely to learn cooperative dispositions. 
Complex cooperative capabilities are only motivated by environments where complex cooperation is necessary, but it is only possible to learn in such environments if agents possess the cooperative capabilities sufficient for easier settings. This implies that the order of training environments ought to be designed as a curriculum for cooperative capabilities. In this way, environment curricula could promote both cooperative dispositions and capabilities. To ensure tractability as agents scale, it will be necessary to use \textit{automated} techniques such as unsupervised environment design (UED) tools \citep{dennis2020emergent, wang2019paired, justesen2018illuminating}. Curricula for learning cooperative capabilities could also modulate the level of information asymmetry, competition, or infrastructure that can aid with cooperation (such as communication channels or commitment devices). Environments should not only be faithful representations of the relevant real-world settings in which agents will be deployed, but also account for rare or out-of-distribution scenarios \citep{dennis2020emergent,jiang2021replay, parker2022evolving, samvelyan2023maestro,team2023human, beukman2024refining}, especially those that are high-stakes and where multi-agent failures could be catastrophic. Similar UED approaches could be used for designing testing environments. For instance, testing environments could be designed to include `honeypots' for undesirable behaviours \citep{Balesni2024}, such as defecting against other agents when it is implied that the agent is not being monitored, so that these behaviours can be caught and monitored as part of pre-deployment testing.

\paragraph{Understanding the Impacts of Training.}
Perhaps the most important research direction in this area is to better {understand the effect of different training data and schemes} on the development of cooperation-relevant capabilities and dispositions.
This builds not only the ability to generate diverse populations of co-players and environments, but also on measures for such capabilities and dispositions.
While there has been much work on evaluating the dangerous capabilities and dispositions of frontier systems \citep{Ganguli2022-pa,kinniment2023evaluating,Pan2023-ww,Shevlane2023,perez2022discovering}, risks from multi-agent interactions have largely gone understudied.
Moreover, even the works that do attempt to benchmark LLM agents in multi-agent settings (see  \citet{zhang2024llm,feng2024survey} for two recent surveys covering this topic) do not typically attempt to assess the extent to which different training data and schemes lead to the risk factors we identify in this report.%
\footnote{One exception is the work of \citet{Fu2023-xi}, who find that iterated play and self-critique make LLM agents more aggressive bargainers in a simple negotiation game. Another is that of \citet{Campedelli2024}, who show that merely assigning roles to LLM agents (without explicit instruction on how to act) can lead to undesirable behaviors like coercion or manipulation.}
For example, are agents rewarded based on their relative performance more conflict-prone than those trained based on their absolute performance (see \Cref{sec:selection_pressures})?
Are agents trained on similar data better able to reason about each other and thus cooperate (or collude) even under imperfect information (see \Cref{sec:information_asymmetries})?
Do these effects persist after fine-tuning or when instructed to complete tasks outside of the original training distribution?
Such questions will be critical to understanding the risks presented by advanced multi-agent systems in high-stakes scenarios yet remain largely unanswered.

\paragraph{Evolutionary Game Theory.} 
There may be further insights to gain from the application of {evolutionary game theory} (EGT) \citep{hofbauer1998evolutionary,sandholm2010population,Fernandez2020} 
to settings involving AI agents \citep{lu2024llms,zimmaro2024emergence,Han2021,santos2019evolution,guo2023facilitating}.
For example, the concept of \textit{frequency-dependent selection} \citep{lewontin1958general}, where the success of a behaviour is contingent on how commonly it occurs in a population relative to other behaviours, has been used to explain the evolution of animal conflict \citep{smith1973logic}, human cooperation \citep{nowak2006five}, honest signalling \citep{grafen1990biological}, and the emergence of social norms \citep{hawkins2019emergence}.
Factors such as the \textit{intensity of selection} -- which captures how quickly agents learn to adopt the successful behaviours of their peers (or how quickly they are adopted/discarded by users or produced/replaced by developers) -- are a crucial for predicting outcomes \citep{traulsen2007pairwise,sigmund2010social} and for finding suitable incentive mechanisms to encourage prosocial behaviour \citep{duong2021cost,han2024evolutionary}.\footnote{Moreover, the intensity of selection can be measured empirically \citep{rand2013evolution,traulsen2010human}, though is typically specific to a given population and domain.}
Future theoretical work should establish which EGT concepts are most relevant to AI systems and which need to be adapted to account for the special features of artificial agents \citep{Han2021,Conitzer2023,dafoe2021cooperative}.

\paragraph{Simulating Selection Pressures.}
Alongside theoretical and conceptual advances, empirical {simulations} such as ABMs can be employed in order to study the effects of different selection pressures \citep{adami2016evolutionary,gilbert2019agent,vestad_survey_2024}.
As remarked in \Cref{sec:network_effects}, a key challenge here is managing the trade-off between accuracy and tractability.
However there might also be important dynamics to study at the micro- rather than macroscopic scale.
For example, preliminary investigations have recently shown that even in simple environments, some LLMs are much more prone to selection pressures promoting cooperation than others \citep[see also \Cref{cs:LLM_evolution}]{Vallinder2024}.
With sufficiently well-developed benchmarks for different model characteristics, we can study their robustness under different kinds of selection pressure, such as the training paradigm and the degree of cooperation or competition they face.

\subsection{Destabilising Dynamics}
\label{sec:destabilising_dynamics}

Modern AI agents can adapt their strategies in response to events in their environment. The interaction of such agents can result in complex dynamics that are difficult to predict or control, sometimes resulting in damaging run-away effects.

\subsubsection{Definition} 

When viewed from a more classical game-theoretic perspective, problems in multi-agent systems are often interpreted in terms of equilibria and their (un)desirability.
This `static' notion, however, can be limited when it comes to understanding the risks posed by the inherently \emph{dynamic} interactions between adaptive AI agents.
Instead, we can think of a multi-agent system as a non-linear dynamical system: a set of equations, partially determined by a set of parameters, that govern how a set of variables change over time \citep{papadimitriou2019game,BloembergenEtAl2015,Balduzzi2018,barfuss2022dynamical}. 
In the case of \emph{non-adaptive} agents, the variables comprise the agents' actions and the state of their environment, which are governed by the agents' strategies, the environmental dynamics and a set of fixed parameters (such as the weights of a neural network).
In the case of \emph{adaptive} agents, we view the strategies themselves as variables, which are governed by learning algorithms and their (hyper)parameters, such as a learning rate.

With this framing, we can characterise several kinds of undesirable behaviour that we might wish to avoid that go beyond the equilibria (i.e., fixed points) of the system \citep{Mogul2006}.
These include dynamic instabilities such as feedback loops, 
chaos, and phase transitions \citep{Gleick1998,BarfussEtAl2024}.
While some of these behaviours can emerge in the case of a single, non-adaptive AI agent (such as a simple agent that becomes stuck in a loop under certain environmental conditions), the additional complexity brought about by the presence of multiple, adaptive agents provides greater opportunity for instabilities to arise \citep{Sanders2018,Cheung2020,Bielawski2021,chotibut2020route,Piliouras2022}.

\subsubsection{Instances}

A long history of research has identified broad classes of behaviours that can be exhibited by dynamical systems, such as fixed points, limit cycles, chaos, and the transient or intermittent presence of such patterns.
Our approach in this section is therefore to examine which behaviours might be exhibited in the context of multi-agent systems, and which of them might pose risks.

\paragraph{Feedback Loops.}
One of the best-known historical examples to illustrate destabilising dynamics in the context of autonomous agents is the 2010 flash crash, in which algorithmic trading agents entered into an unexpected {feedback loop} \citep[see also \Cref{cs:flash_crash}]{CTFC2010}.\footnote{For a simpler and more amusing example, see \Cref{fn:textbook}.}
More generally, a feedback loop occurs when the output of a system is used as part of its input, creating a cycle that can either amplify or dampen the system's behaviour. In multi-agent settings, feedback loops often arise from the interactions between agents, as each agent's actions affect the environment and the behaviour of other agents, which in turn affect their own subsequent actions.
Feedback loops can lead not only to financial crashes but to military conflicts \citep[see also \Cref{cs:cs:military_escalation}]{Richardson1960} and ecological disasters \citep{Holling1973}.
The distinguishing characteristic of flash crashes, however, is the \emph{speed} at which they occur. Competitive pressures necessitate automated trading agents that act much faster than their human overseers, meaning that when things go wrong, it is harder for humans to react.
As such, we might expect to see more destabilising dynamics in systems with more fast-moving AI agents \citep{Maas2018}.\footnote{Catastrophe-theoretic models show that even `slow' systems with a small number of `fast' elements can produce dramatic shifts \citep{Zeeman1976}, though it is not always clear how closely such models capture complex real-world phenomena.}

\begin{case-study}[label=cs:flash_crash,sidebyside,sidebyside align=top,lower separated=false]{The 2010 Flash Crash}
  On May 6, 2010, the US stock market lost approximately \$1 trillion in 15 minutes during one of the most turbulent periods in its history \citep{CTFC2010}. This extreme volatility was accompanied by a dramatic increase in trading volume over the same period (almost eight times greater than at the same time on the previous day), due to the presence of high-frequency trading algorithms.\footnotemark{}
  While more recent studies have concluded that these algorithms did not \emph{cause} the crash, they are widely acknowledged to have contributed through their exploitation of temporary market imbalances \citep{Kirilenko2017}. 
  \tcblower
  \includegraphics[width=\linewidth]{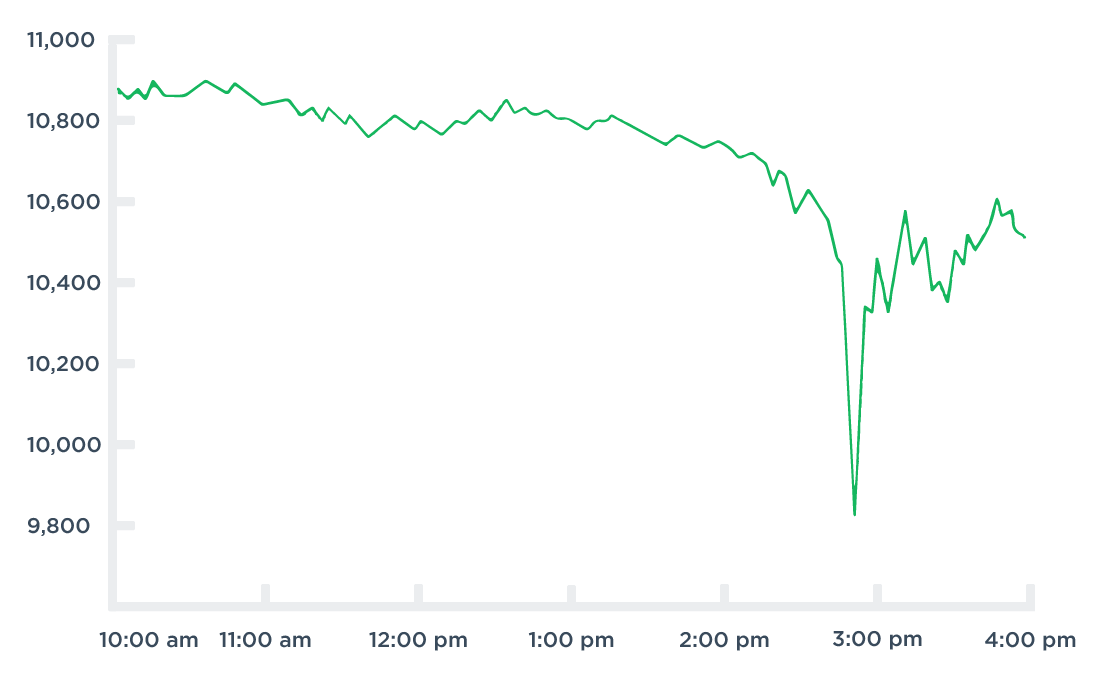}
  \captionof{figure}{
    Transaction prices of the Dow Jones Industrial Average on May 6, 2010.
    Figure adapted from \citet{OptionAlpha2025}.
  }\label{fig:flash_crash}
\end{case-study}

\footnotetext{In cases like this, it can be the synchronisation between agents that creates an instability if, for example, all agents try to sell or buy at the same time because they all make decisions based on highly correlated signals (or even a common signal) and they all have similar strategies. Such problems might become significantly amplified if only a handful of frontier models are the underlying decision makers for a vast number of (seemingly diverse) agents (see \Cref{sec:network_effects,sec:multi-agent_security}).}

\paragraph{Cyclic Behaviour.} 
The dynamics described above are highly non-linear (small changes to the system's state can result in large changes to its trajectory).
Similar non-linear dynamics can emerge in multi-agent learning and lead to a variety of phenomena that do not occur in single-agent learning \citep{BarfussEtAl2019, NagarajanEtAl2020a, LeonardosEtAl2020, BarfussMann2022, GallaFarmer2013}.
One of the simplest examples of this phenomenon is Q-learning \citep{Watkins1992}: in the case of a single agent, convergence to an optimal policy is guaranteed under modest conditions, but in the (mixed-motive) case of multiple agents, this same learning rule can lead to {cycles} and thus non-convergence \citep{Zinkevich2005}. 
While cycles in themselves need not carry any risk, their presence can subvert the expected or desirable properties of a given system.
For example, \citet{PaesLeme2024} show that when auto-bidding agents participate in second price auctions -- which are designed to have dominant truthful equilibria -- the dynamics of these agents can be unstable and fail to converge to their underlying values, losing the desired truthfulness properties.

\paragraph{Chaos.}
Unlike the systems that tend towards fixed points or cycles described above, chaotic systems are inherently unpredictable and highly sensitive to initial conditions. While it might seem easy to dismiss such notions as mathematical exoticisms, recent work has shown that, in fact, chaotic dynamics are not only possible in a wide range of multi-agent learning setups \citep{GallaFarmer2013,Andrade2021,VlatakisGkaragkounis2023,sato2002chaos,palaiopanos2017multiplicative}, but can become the norm as the number of agents increases \citep{Sanders2018,Cheung2020,Bielawski2021}.
To the best of our knowledge, such dynamics have not been seen in today's frontier AI systems, but the proliferation of such systems increases the importance of reliably predicting their behaviour.

\paragraph{Phase Transitions.}
Finally, small external changes to the system -- such as the introduction of new agents or a distributional shift -- can cause phase transitions, where the system undergoes an abrupt qualitative shift in overall behaviour \citep{BarfussEtAl2024}. Formally, this corresponds to \textit{bifurcations} in the system's parameter space, which lead to the creation or destruction of dynamical attractors, resulting in complex and unpredictable dynamics \citep{Crawford1991,Zeeman1976}.
For example, \citet{Leonardos2022} show that changes to the exploration hyperparameter of RL agents can lead to phase transitions that drastically change the number and stability of the equilibria in a game, which in turn can have potentially unbounded negative effects on agents' performance.
Relatedly, there have been many observations of phase transitions in ML \citep{Carroll2021,Olsson2022,Ziyin2022}, such as `grokking', in which the test set error decreases rapidly long after the training error has plateaued \citep{Power2022}.
These phenomena are still poorly understood, even in the case of a single system.

\paragraph{Distributional Shift.}
Individual ML systems can perform poorly in contexts different from those in which they were trained.
A key source of these distributional shifts is the actions and adaptations of other agents \citep{Papoudakis2019-es,Piliouras2022,Narang2023}, which in single-agent approaches are often simply or ignored or at best modelled exogenously.
Indeed, the sheer number and variance of behaviours that can be exhibited other agents means that multi-agent systems pose an especially challenging generalisation problem for individual learners \citep{Agapiou2022-an,Leibo2021-cf,Stone2010}.
While distributional shifts can cause issues in common-interest settings (see \Cref{sec:miscoordination}), they are more worrisome in mixed-motive settings since the ability of agents to cooperate depends not only on the ability to coordinate on one of many arbitrary conventions (which might be easily resolved
by a common language), but on their beliefs about what solutions other agents will find acceptable. 
For example, training a negotiating agent on a distribution of counterparts with too little diversity in their negotiating tactics can lead to catastrophic overconfidence in high-stakes settings \citep[cf.][]{Stastny2021}, which might already have little precedent in the training data.
These issues may be aggravated by the fact that multi-agent systems can be highly dynamic \citep{Papoudakis2019-es}, as AI agents or their designers will be incentivised to continually adapt to the behaviour of other agents.
These effects might also be exacerbated by the fact that models may come to be trained using data generated by other models \citep[see also \Cref{sec:selection_pressures}]{Shumailov2024AIMC,Martnez2023TowardsUT,alemohammad_self-consuming_2023}, though preliminary work suggests such concerns might be overblown \citep{Gerstgrasser2024}.

\subsubsection{Directions}

With the deployment of advanced multi-agent systems comes the risk of destabilising dynamics in settings ranging from financial markets \citep{Kirilenko2017} to power grids \citep{schafer2018dynamically} to battlefields \citep{Johnson2021}.
So far, both theoretical and empirical work has primarily studied such dynamics in small, abstract games with simple AI systems and learning algorithms.\footnote{In one of the few examples involving foundation models, \citet{Fort_2023} recently provided a simple visual illustration of how the outputs of two foundation models -- GPT-4(V) and DALLE-3 -- can be either stable or unstable when placed in a loop, depending on their initial input.}
While this is an important first step, addressing the risks of destabilizing dynamics in real-world multi-agent AI systems will require a concerted interdisciplinary effort, bringing together expertise in AI safety, dynamical systems, game theory, and policy to develop robust solutions.

\paragraph{Understanding Dynamics.}
The conditions under which multi-agent systems have undesirable dynamics might include properties of the underlying environment and objectives \citep{BarfussMann2022,Sanders2018}, or the structure and hyperparameters of the learning algorithms \citep{BarfussEtAl2019,Leonardos2022,barfuss2023intrinsic}.
Important research directions include understanding if (and how) chaotic dynamics in idealised versions of stochastic learning algorithms extend to their real-world counterparts,\footnote{Formally, chaos is only rigorously defined in deterministic dynamical systems.} and how these dynamics are affected by the size and structure of the state-action space.

\paragraph{Monitoring and Stabilising Dynamics.}
Early work suggests that inducing new `conservation laws' \citep{NagarajanEtAl2020a} or `constants of motion' \citep{Piliouras2021} in multi-agent learning can result in more predictable dynamics. 
Future research should investigate how these approaches scale to larger systems and greater numbers of agents and could make use of existing results in areas such as theoretical ML \citep{Sastry1999,Tuyls2005,Bowling2001a,Bottou2010,Kushner2003}, adaptive mechanism design \citep{Pardoe2006,Zhang2008,Baumann2020,Yang2022,Zheng2022,Gerstgrasser2023}, and mean-field games \citep{Lasry2007,Huang2006}.
Both this work and that on understanding the dynamics of multi-agent learning would benefit greatly from the insights of other scientific communities, especially those working on other non-linear complex systems, and those engineering the largest and most powerful models \citep{BarfussEtAl2024}.

\paragraph{Regulating Adaptive Multi-Agent Systems.}
In addition, {regulation} could be used to mandate the use of mechanisms that monitor and stabilise the dynamics of multi-agent systems in safety-critical areas.
This could include, for example, enforced pauses in interactions between systems or reversions to previous strategies if the system behaviour escapes certain thresholds \citep[as in the 2010 flash crash, when trading was temporarily halted; see also]{Subrahmanyam2013}.
For the most important systems, there might even be a need to enforce the (de)synchronisation of model updates, to limit the size and frequency of learning updates, or to limit the number of agents interacting with one another (either via technical restrictions or using methods akin to congestion pricing).
Relatedly, existing tools for auditing models often make use of static `model cards' that indicate how the model was produced, its performance, and intended use cases \citep{Mitchell2019}, but this documentation applies to single trained systems that are frozen before deployment.
To monitor the dynamics of \emph{multi-agent} systems, including those that \emph{learn online}, we will need to leverage new innovations such as `ecosystem graphs' \citep{Bommasani2023} and `reward reports' \citep{Gilbert2022}, respectively.

\subsection{Commitment and Trust}
\label{sec:commitment_and_trust}

In settings that require joint action in order to obtain a better outcome, inefficiencies can result whenever one or more actors cannot be trusted (perhaps due to strategic incentives, or due to their incompetence) to carry out their part of the plan.
These inefficiencies can be reduced via \textit{credible commitments} made by the untrusted parties.
Unfortunately, the ability to make credible commitments is `dual-use' and can therefore lead to new risks.

\subsubsection{Definition}

An actor makes a \textit{commitment} when they bind themselves to a course of action, such that reneging on that action would either be impossible or result in significant costs to themselves. 
A commitment is \textit{credible} when other actors believe that the actor making the commitment will follow through with the actions they claim to have committed to. 
Credible commitments are useful in scenarios where trust is essential but hard to establish, such as in international treaties, economic policies, and contractual agreements.

Since credible commitments can often help in achieving desirable cooperative outcomes, we expect there will be incentives to build systems capable of making them.
For example, an AI system can become more trustworthy by being credibly committed to erasing any private information revealed to it.
In contrast, human beings or organisations cannot reliably forget at will and may later leak private information, whether intentionally or not \citep{carnegie2019disclosure}.
Autonomous AI agents themselves might also serve as credible commitment devices \citep{McAfee1984,Howard1988,tennenholtz2004program}, enabling actors to carry out actions based on potentially complex conditions and thus helping to solve problems with incomplete contracting \citep{schmitz2001hold}.
However, the ability of AI systems to make commitments can also backfire in correspondingly severe ways, preventing recourse in high-stakes scenarios and enabling extortion and brinkmanship.

\subsubsection{Instances}

As noted above, the ability to form commitments can both precipitate and mitigate risks.
We therefore begin by considering risk instances that can arise due to a lack of trust, before turning to those that can arise via the very mechanisms that might be used to establish such trust.

\paragraph{Inefficient Outcomes.}
Without careful planning and the appropriate safeguards, we may soon be entering a world overrun by increasingly competent and autonomous software agents, able to act with little restriction.
The abilities of these agents to persuade, deceive, and obfuscate their activities, as well as the fact they can be deployed remotely and easily created or destroyed by their deployer, means that by default they may garner little trust (from humans or from other agents).
Such a world may end up being rife with economic inefficiencies \citep{schmitz2001hold,Krier2023}, political problems \citep{Kreps2023,Csernatoni2024}, and other damaging social effects \citep{Gabriel2024}.
Even if it is possible to provide assurances around the day-to-day performance of most AI agents, in high-stakes situations there may be extreme pressures for agents to defect against others, making trust harder to establish, and potentially leading to conflict \citep[see also \Cref{sec:conflict}]{fearon1995rationalist,powell2006war}.\footnote{A classic non-AI example is the hypothesis that a major contributor to World War I was Germany's concerns about the rising power of Russia \citep{02b46089-4baf-3ce5-9159-e4dbfc3050f6}. Conflict might have been avoided were Russia able to credibly commit not to expand its influence, but the absence of such an ability left Germany with fewer alternatives to conflict.}

\paragraph{Threats and Extortion.}
A natural solution to problems of trust is to provide some kind of commitment ability to AI agents, which can be used to bind them to more cooperative courses of action.
Unfortunately, the ability to make credible commitments may come with the ability to make credible \textit{threats}, which facilitate extortion and could incentivize brinkmanship (see \Cref{sec:conflict}).
For example, ransomware becomes more effective if the hacker can credibly commit to restore the victim's data upon receiving payment, and coercion using AI-controlled weapons could become more frequent if actors gain the ability to make credible threats conditional on complicated demands (see also \Cref{cs:dead_hand}).
More generally, an agent could use commitment devices to shift risks or costs to others, allowing it to behave irresponsibly.\footnote{In economics, this general problem (not necessarily as a result of the power of commitment) is known as `moral hazard'.}
In other cases, it might be the \textit{agent that commits} to an inflexible (cooperative) course of action which can be exploited by others who can adapt their strategies to this commitment.%
\footnote{An amusing, non-AI example of such a commitment is Red Lobster's ``Endless Shrimp'' deal, which has recently been blamed for driving it to bankruptcy \citep{Meyersohn2024}.}
On the other hand, if used carefully, the ability to commit generally strictly empowers the committing agent \citep{Stengel2010,Letchford2013}.

\paragraph{Rigidity and Mistaken Commitments.}
Even when it is desirable to be able to make threats in order to deter socially harmful behaviour, doing so using AI agents effectively removes the human from the loop, which could prove disastrous in high-stakes contexts (e.g., a false positive in a nuclear submarine's warning system; see also \Cref{cs:dead_hand}), or when irresponsible actors are enabled in making disproportionate or mistaken commitments.
On the other hand, such commitments may only be credible to the extent that a human cannot intervene, increasing the incentive for delegation to AI agents.
This could be worsened if other, potentially incompatible commitments can be made by other actors, leading to a `commitment race' \citep{commitment_races} or potential conflict.
In complex networks (see \Cref{sec:network_effects}), commitments triggered by a small number of agents could -- without careful planning -- cascade through the network and have a far more damaging effect \citep{Xia2010}.

\begin{case-study}[label=cs:dead_hand]{Dead Hands and Automated Deterrence}
    During the Cold War, the Soviet Union developed the the automated Perimeter system -- often called `Dead Hand' -- to guarantee a nuclear launch if its leadership were incapacitated, thus ensuring a credible commitment of retaliation \citep{Hoffman2009}. While this mechanism was intended as a deterrent, its automatic and largely irrevocable nature exemplifies how credible commitments can become dangerously dual-use: once triggered, there would be little chance to override or de-escalate. 
    In a similar vein, during Operation Iraqi Freedom in 2003 an automated US missile defence system shot down a British plane, killing both occupants \citep{Talbot2005,Borg2024}.
    While the system's operators had one minute to override the system (even in its autonomous mode), they decided to trust its judgment, resulting in a tragic outcome.
    In more general AI contexts, similarly inflexible commitments could offer short-term advantages or trust but risk uncontrolled escalation, lock-in, and catastrophic outcomes if not carefully designed with appropriate fail-safes and oversight.
\end{case-study}

\subsubsection{Directions}

As with any dual-use technology, ensuring it is used for beneficial rather than detrimental means can be extremely challenging.
We therefore attempt to focus on directions that \textit{differentially} advance beneficial uses \citep{Sandbrink2022}, while acknowledging that it will not, in general, be possible isolate these entirely.%
\footnote{Even in the case of human commitments, it is not always obvious which families of commitments are desirable to make. For example, if a state commits to refusing to negotiate with terrorists, they might end up sacrificing some lives while establishing a reputation that saves more lives over the long term. Similarly, a seller might refuse a low, though still positive, offer in order to achieve better offers in future. Such questions are often as much a matter of principle as they are of consequentialist reasoning.}

\paragraph{Keeping Humans in the Loop.}
Given the risks associated with the power of AI commitments, a key direction will be to lay out the domains in which they can be used and the kinds of commitments that are permitted.
For example, existing efforts have already sought to ensure that AI systems do not form a part of the nuclear chain of command \citep{Renshaw2024,Congress2023}.
It may be similarly important in other high-stakes settings to ensure that humans cannot be fully removed from the loop.\footnote{However, the inclusion of a human in the loop does not itself guarantee control. The presence of an algorithmic system can negatively influence human decision-making \citep{green_algorithm_loop_2020,green_flaws_2022,crootof_humans_2023,skitka_does_1999,goddard_automation_2012,Borg2024}, such as through automation bias. At the same time, it is important to acknowledge that humans suffer from their own flaws that might lead to risks and that might be (at least partly) overcome via the use of AI systems.}
While certain kinds of commitment device might still allow for malicious use (such as automated blackmail campaigns), regulation, safeguards, and infrastructure limiting where and how AI agents can be deployed could help prevent the worst offences \citep{Kolt2024,Chan2025}.

\paragraph{Limiting Commitment Power.}
Researchers should also explore ways to design AI systems that can make and adhere to commitments even in the face of changing circumstances or new information, thereby avoiding some of the risks associated with overly rigid strategies.
This might involve developing algorithms that can (learn to) renegotiate commitments in a fair and transparent manner when necessary \citep{Sandholm2002,Ho2014,Wang2023,Cohen2023}.
While agents equipped with commitment powers are not yet widespread, it would be valuable to begin preliminary studies now into demonstrations of their risks \citep[and benefits, see, e.g.,][]{Christoffersen2023,zhu2025learning}, as well as the feasibility of technical solutions, the tractability of governance solutions, and their intersection \citep{reuel2024open,Kolt2024}.

\paragraph{Institutions and Normative Infrastructure.}
Other important research directions include the development of normative infrastructure that can help establish trust without recourse to commitment devices that might be misused.
These collectively enforced rules and norms might serve to \textit{differentially} advance cooperation relative to coercion \citep{Sandbrink2022}.
For example, the introduction of unique agent identifiers \citep{Chan2024} would enable the construction of reputation systems, which are critically important in otherwise (pseudo-)anonymous interactions such as online marketplaces \citep{Tadelis2016}.
While reputation is still `dual-use' to the extent that one could develop a reputation for carrying out costly threats, doing so requires paying such costs and also being able to escape later punishment oneself, which may be a less viable strategy in many cases.
Other examples include determining the rules and principles via which liability for harms from AI agents is assigned \citep[see also \Cref{sec:governance}]{ayres_law_2024, lima_could_2017, lior_ai_2019, Kolt2024, Chopra2011, Solum1992}.

\paragraph{Privacy-Preserving Monitoring.}
In order for reputation systems to be effective for more general and widely deployed agents, it will be necessary to improve trust by \textit{monitoring} their actions \citep{Chan2024a}.
Monitoring also extends to scrutiny of the actors deploying those agents, who might claim to be running one kind of agent or using some kinds of data, while instead using others.
This in turn, however, raises clear and important privacy concerns.
There is thus an important need to develop privacy-preserving technologies for monitoring AI systems and the actions of autonomous agents \citep{Shavit2023WhatDI,vegesna2023privacy}.
Examples include the use of cryptography -- such as signatures that can serve as proof of learning \citep{Jia2021} or proof of inference \citep{Ghodsi2017}, protocols for decentralized verifiable computation \citep{yao_protocols_1982,47976}, and performing computations using encrypted data \citep{Martins2017,Dowlin2016} -- as well as tools for auditing and monitoring both software and hardware.

\paragraph{Mutual Simulation and Transparency.}
Finally, while monitoring and reputation systems might be able to render more transparent what an agent has done in the past, we may also want to use the unique properties of computational agents in order to predict what they will do in the future \citep{Conitzer2023}.
For example, such agents are written in code that can -- in theory -- be read or understood by other agents.
This kind of mutual transparency can beneficial in establishing trust and reaching more efficient outcomes \citep{McAfee1984,Howard1988,tennenholtz2004program,halpern2018game,Oesterheld2018,Han2021}, though has yet to find practical applications \citep{Critch2022}.
Similarly, even if one cannot peer inside the black box, the same code can be run multiple times on different inputs, allowing for simulations and tests prior to deployment or even individual interactions.
As with white-box access to other agents, these abilities can (in theory) provably reduce inefficiencies due to mistrust \citep{Kovarik2023,Kovarik2024,Chen2024b}, but have yet to be studied in the context of real-world strategic agents \citep[though see, e.g.,][]{Griffin2024,Greenblatt2023}.
More research is required to design and implement tractable versions of these methods in order to fulfil their theoretical promise.

\subsection{Emergent Agency}
\label{sec:emergent_agency}

Emergent behaviour is ubiquitous in the natural, biomedical, and social sciences.
Examples include the superconductivity of materials in condensed matter physics \citep{anderson_more_1972}; complex tasks like bridge-building by ant colonies and facing larger predators \citep{gordon1996organization,bonabeau_self-organization_1997}; and, in the social sphere, collective behaviours such as group-think or the development of new norms \citep{couzin_collective_2007}.
In this section we focus on the risks presented by the emergence of higher-level forms of agency from a collective of agents.

\subsubsection{Definition}

Emergent behaviours are those exhibited by a complex entity composed of multiple, interacting parts (such as AI agents) that are not exhibited by any of those parts when viewed individually.
Emergent behaviours are distinct from mere accumulations (as in \Cref{cs:overcoming_safeguards}, for example); in other words, the whole may be different to the sum of its parts \citep{anderson_more_1972}.
While there is a sense in which everything we study in this report can be viewed as ``emerging'' from multi-agent systems \citep{Mogul2006,Altmann2024}, our focus on this section is specifically on the risks associated with \textit{emergent agency} at the level of the \textit{collective}.
This is distinct from other works that discuss the emergent behaviour of \emph{individual} agents -- such as tool use \citep{Baker2019}, locomotion \citep{Bansal2018}, or communication \citep{Lazaridou2020} -- in multi-agent settings.\footnote{Other works consider emergence concerning say, the number of parameters in a model, as opposed to the number of agents. For example, \citet{Wei2022} ``consider an ability to be emergent if it is not present in smaller models but is present in larger models''.}
These individual behaviours are fundamentally driven by the selection pressure induced by the presence of other agents, which we discuss in \Cref{sec:selection_pressures}.

We break the risks associated with emergent agency into the emergence of dangerous \emph{capabilities}, the emergence of dangerous \emph{goals}, and thus -- if one takes the view that intelligence is fundamentally rooted in an individual's or group's ability to solve problems, achieve goals, etc. \citep{Legg2007} -- the possibility of creating emergent higher-level agency or \emph{collective intelligence} \citep{malone2022handbook}.
To provide a paradigmatic example, one termite by itself might be incapable of constructing a mound, and yet the overall colony can do so quite proficiently.
Emergent goals, on the other hand, are agnostic to the group's (or any individual's) abilities,\footnote{This claim is sometimes known as the `orthogonality thesis': goals and capabilities (i.e., one's means to achieve one's goals) are independent, or `orthogonal', to one another \citep{Bostrom2014}.} and can be used to model the group's objectives, which supervene on the individuals' objectives. 
Thus while it might be unreasonable to model a single termite as having the goal of building a mound, this goal could be highly predictive of the overall colony's behaviour.

\subsubsection{Instances}

Before proceeding further, we note that discussions of emergent phenomena in systems of advanced AI agents are necessarily quite speculative, as it is challenging (both in theory and in practice) to identify such phenomena.%
\footnote{Indeed, the astute reader will notice that this section is the only section of the report that does not have at least one corresponding case study.
While there are demonstrations of AI agents exhibiting emergent collective capabilities and goals (see, e.g., \citet{werfel2014}, in which a swarm of simple, termite-inspired construction robots are able to build large-scale structures without centralized coordination), we are not aware of examples involving collective agency among \textit{advanced} AI agents (such as those powered by LLMs) or collective agency that represents an obvious \textit{risk}.}
We therefore attempt to draw lessons from simpler AI systems or biological entities, while highlighting that advanced AI agents could also possess features that make the transition to higher-level agency easier, such as the ability to more easily share information, replicate, and update their behaviour \citep{Conitzer2023}.

\paragraph{Emergent Capabilities.}
Dangerous {emergent capabilities} could arise when a multi-agent system overcomes the safety-enhancing limitations of the individual systems, such as individual models' narrow domains of application or myopia caused by a lack of long-term planning and long-term memory.
For example, narrow systems for research planning, predicting the properties of molecules, and synthesising new chemicals could, when combined, lead to a complex `test and iterate' automated workflow capable of designing dangerous new chemical compounds far beyond the scope of the initial systems' capabilities \citep{boiko_emergent_2023,urbina_dual_2022,Luo2024}.
This is similar to how a myopic actor and a passive critic can combine to produce an actor-critic algorithm capable of long-term planning via RL \citep{konda1999actor}.
This possibility is important for safety -- and for future AI ecosystems made of specialised `AI services' \citep{Drexler2019} -- as generally intelligent autonomous systems could pose much greater risks than narrow AI tools \citep{Chan2023-aj}.\footnote{Despite this, many companies are racing to build AI agents \citep{OpenAI_agents,GDM_agents,Microsoft_agents,Meta_agents,Anthropic_agents}, including early efforts attempting to construct composite agents based on simpler components including powerful foundation models \citep[e.g.,][]{schick2023toolformer,wu2024autogen}.}
More speculatively, the combination of advanced AI agents could eventually lead to recursive self-improvement at the collective level, as AI research itself becomes increasingly automated \citep{Hutter2019,Mankowitz2023,Agnesina2023,Lu2024}, even though no individual system possesses this capability.

\paragraph{Emergent Goals.}
Ascribing goals to a system is not always straightforward.
For our present purposes, it will suffice to adopt a Dennetian perspective \citep{Dennett1971}, ascribing goals and intentions only when it is useful (i.e., predictive) to do so.\footnote{See \citet{oesterheld2016formalizing,Halpern2018,Orseau2018,Everitt2021,Kenton2022,Biehl2023,MacDermott2024,Ward2024} for recent, formal examinations of agency and incentives in AI systems, and the implications thereof for safety.}
While it might not be helpful to describe individual narrow AI tools as having goals, their combination may act as a (seemingly) goal-directed collective.
For example, a group of moderation bots on a major social networking site could subtly but systematically manipulate the overall political perspectives of the user population, even though, individually, each agent is programmed to simply increase user engagement or filter out dis-preferred content.
Other dangerous goals that could emerge from groups of more advanced AI agents include power-seeking \citep{Turner2022,Carlsmith2022}, self-preservation \citep{Omohundro2008,Lyon2011}, or competing against other groups \citep{Bakhtin2022}, which could be instrumentally useful at the collective level even if avoidable or not useful at the individual level.

\subsubsection{Directions}

Insofar as the prospect of collective goals and capabilities emerging from large numbers of advanced AI agents remains somewhat speculative, it will be especially useful to develop a firmer theoretical understanding of when and how these novel forms of agency might emerge.
This understanding should be complemented by preliminary empirical investigations, potentially in settings with less advanced agents or smaller numbers of agents.

\paragraph{Empirical Exploration.}
By definition, emergent properties are hard to predict when looking at individual components.
Unfortunately, {exploratory empirical studies} of emergent behaviour among \textit{large numbers} of state-of-the-art systems in realistic environments are highly challenging.
One reason is that experiments using many model instances are very expensive. 
Another is that is difficult to construct relevant environments and `sandboxes' which are similar enough to the real world for us to gain transferable insights.
Nonetheless, research such as that of \citet{park2023generative,Chen2024c, vezhnevets2023generative} shows this to be possible in simple games, and that it can lead to surprising outcomes.
Future work could use more realistic or open-ended environments, such as those involving economic activity \citep{Zheng2022}, or games inspired by massively multiplayer online role-playing games \citep{Suarez2019}.
This would help address the crucial problem of understanding and eventually being able to predict the settings under which undesirable behaviours emerge at the group level and how robust they are, including the influence of key factors and conditions such as the degree of competition, the (non-)diversity of the agents, and their individual capability levels.

\paragraph{Theories of Emergent Capabilties.}
In conjunction with empirical studies, we must develop a {theoretical understanding of emergent capabilities} that can be applied to groups of frontier models.
Existing work in this area either identifies a specific emergent behaviour in advance and attempts to measure the presence or cause of this behaviour based on pre-existing observations \citep{Seth2006,Chen2009}, or formalises some abstract notion of a micro- and macro-level and attempts to detect newly emergent behaviours by comparing the difference \citep{Kubik2003,Teo2013,Szabo2015}, the idea being that emergent phenomena are those present in the latter but not the former.
Other related works include that of \citet{Sourbut2024}, who propose a theoretical method of separately measuring individual and collective capabilities and (mis)alignment in strategic settings.
These approaches are computationally expensive, however, and their empirical utility us yet to be convincingly demonstrated.\footnote{Moreover, they may depend on access to information about deployed agents that is unavailable not just to third parties, but also to other model providers seeking to measure emergent behaviours in their agents' interactions with others.}
Promising directions include developing tractable proxies of these measures, and the use of ML \citep{Dahia2024} and distributed methods \citep{Wang2016a,Otoole2017} to improve scalability.

\paragraph{Theories of Emergent Goals.}
It is especially important to know what we ought to measure here, as some techniques for understanding the goals of a single agent, such as interpretability methods \citep{Michaud2020,Colognese2023,Mini2023,Marks2023} might not be easily applied to group-level emergent goals \citep{Grupen2022}.
Many formal approaches to measuring and detecting goal-directedness make use of causal models \citep{Halpern2018,Everitt2021,Kenton2022,MacDermott2024,Ward2024}.
A natural next step towards generalising these works to consider emergent goals in \textit{multi-agent} settings would therefore be to apply them in the context of \emph{causal games} \citep{Hammond2023}.
This line of work would also benefit from the insights of other fields that have sought to develop theories of emergent agency \citep{friston2022designing,Okasha2018,Smith2020}.

\paragraph{Monitoring and Intervening on Collective Agents.}
Once we possess a better theoretical and empirical understanding of emergence in advanced multi-agent systems, it will be important to develop the tools and infrastructure to {monitor for, and intervene on, forms of emergent, collective agency}.
In practice, this is likely to overlap substantially with the tools required to monitor the macroscopic properties of large, dynamic networks of agents (see \Cref{sec:network_effects,sec:destabilising_dynamics}).
Similarly, interventions for mitigating undesirable forms of emergent behaviour may be related to those for mitigating collusion (see \Cref{sec:collusion}) or deleterious selection pressures (see \Cref{sec:selection_pressures}).
In tandem, we ought to develop \textit{evaluations} for dangerous emergent behaviours in multi-agent systems.
For example, while a `one-shot' application of an LLM might not possess a particular ability (such as manipulating a human to take some action), a population of multiple LLMs and other AI tools might.
Similarly, while a single agent might not exhibit a certain sub-goal (such as self-preservation) while completing a task, a combination of agents might develop a mutual reliance upon one another that ends up having self-preservation as an instrumental sub-goal the collective level.

\subsection{Multi-Agent Security}
\label{sec:multi-agent_security}

Global cyber threats are on the rise, not just due to the proliferation of commercial cyber tools \citep{nscs_threat_2023}, but also due to an increase in so-called `hybrid warfare' (which blends conventional warfare with cyber- and information-warfare) by nation-states and non-state actors \citep{kaunert_cyber-attacks_2021,csis_significant_2023}.
The shift towards a world of advanced AI agents will not only enable new tools and affordances, but also increase the surface area for potential attacks, invalidating previously reasonable threat modelling assumptions and requiring a new approach: \textit{multi-agent security} \citep{masec}.

\subsubsection{Definition}

Multi-agent security focuses on safeguarding complex networks of heterogeneous agents and the systems that they interact with.
This includes protecting not only data and software but also hardware and other physical aspects of the world that are integrated with these digital systems.%
\footnote{Note that it is often helpful to distinguish between safety (which aims to prevent harm \textit{from} a given entity) and security (which aims to prevent harm \textit{to} a given entity), though we will also typically be interested in the latter to extent that it leads to the former \citep{khlaaf2023}.
At the same time, a security perspective involves considering worst-case scenarios, which is also a natural perspective when considering more extreme risks from advanced AI.}
While many security settings are implicitly multi-agent (involving both an attacker and a defender), multi-agent security addresses vulnerabilities and attack vectors that emerge specifically when \textit{many} AI agents interact within a broader networked ecosystem.%
\footnote{Similarly, we note that despite the implicit presence of an adversary, security failures need not only be a form of conflict (\Cref{sec:conflict}) but can also lead to miscoordination (\Cref{sec:miscoordination}) as well as collusion (\Cref{sec:collusion}).}
For example, traditional security frameworks such as zero-trust approaches may not provide the required trade-offs between security and capability in large multi-agent systems \citep{wylde_zero_2021}.

While coordinated human hacking teams or botnets already pose `multi-agent' security risks, their speed and adaptability are limited by human coordination or static strategies.
As AI agents become more autonomous and capable of learning and complex reasoning, however, they will be more easily able to dynamically strategize, collude, and decompose tasks to evade defences.
At the same time, security efforts aimed at preventing attacks to (or harmful actions from) a single advanced AI system are comparatively simple, as they primarily require monitoring a single entity.
The emergence of advanced multi-agent systems therefore raises new vulnerabilities that do not appear in single-agent or less advanced multi-agent contexts.

\subsubsection{Instances}

Multi-agent security risks from advanced AI arise due to two main factors: novel attack methods and novel attack surfaces.
First, the emergence of large numbers of advanced AI agents might -- via their very multiplicity and decentralisation -- lead to attack methods that would not be available to a single agent.
Second, the complexity, interconnectedness, and range of such multi-agent systems may at the same time introduce new attack surfaces.

\paragraph{Swarm Attacks.}
The need for multi-agent security is foreshadowed by attacks today that benefit from the use of many decentralised agents, such as distributed denial-of-service attacks \citep{cisco_what_2023,DDoSThreatReport2024Q3}.
Such attacks exploit the massive collective resources of individual low-resourced actors, chained into an attack that breaks the assumptions of bandwidth constraints on a single well-resourced agent.
Such attacks are also used to great effect elsewhere, such as in `brigading' on social media, in which teams of bots or humans collude to downvote or otherwise obstruct benign content \citep{blair_institute_social_2021}, or coordinated malicious actions in matching, rating, and content moderation systems \citep{Newman2024,Sharma2025}.
At present these bots are typically relatively unsophisticated but AI agents that can intelligently adapt and collaboratively identify new attack surfaces may amplify the potency of such attacks.
More broadly, the ability for many small AI agents to parallelize tasks and recompose their outputs, such as in inference attacks that piece together sensitive information gathered individually by actors with limited access \citep{DBLP:conf/ndss/IslamKK12}, can undermine the common assumption that individual agents with restricted capabilities are safe.

\paragraph{Heterogeneous Attacks.}
A closely related risk is the possibility of multiple agents {combining different affordances to overcome safeguards}, for which there is already preliminary evidence \citep[see also \Cref{cs:overcoming_safeguards}]{Jones2024}.
In this case, it is not the sheer \textit{number} of agents that leads to the novel attack method, but the combination of their \textit{different abilities}.
This might include the agents' lack of individual safeguards, tasks that they have specialised to complete, systems or information that they may have access to (either directly or via training), or other incidental features such as their geographic location(s).
The inherent difficulty of attributing responsibility for security breaches in diffuse, heterogeneous networks of agents further complicates timely defence and recovery \citep{skopik_under_2020}.

\begin{case-study}[label=cs:overcoming_safeguards]{Overcoming Safeguards via Multiple Safe Models}
    \begin{center}
        \includegraphics[width=\linewidth]{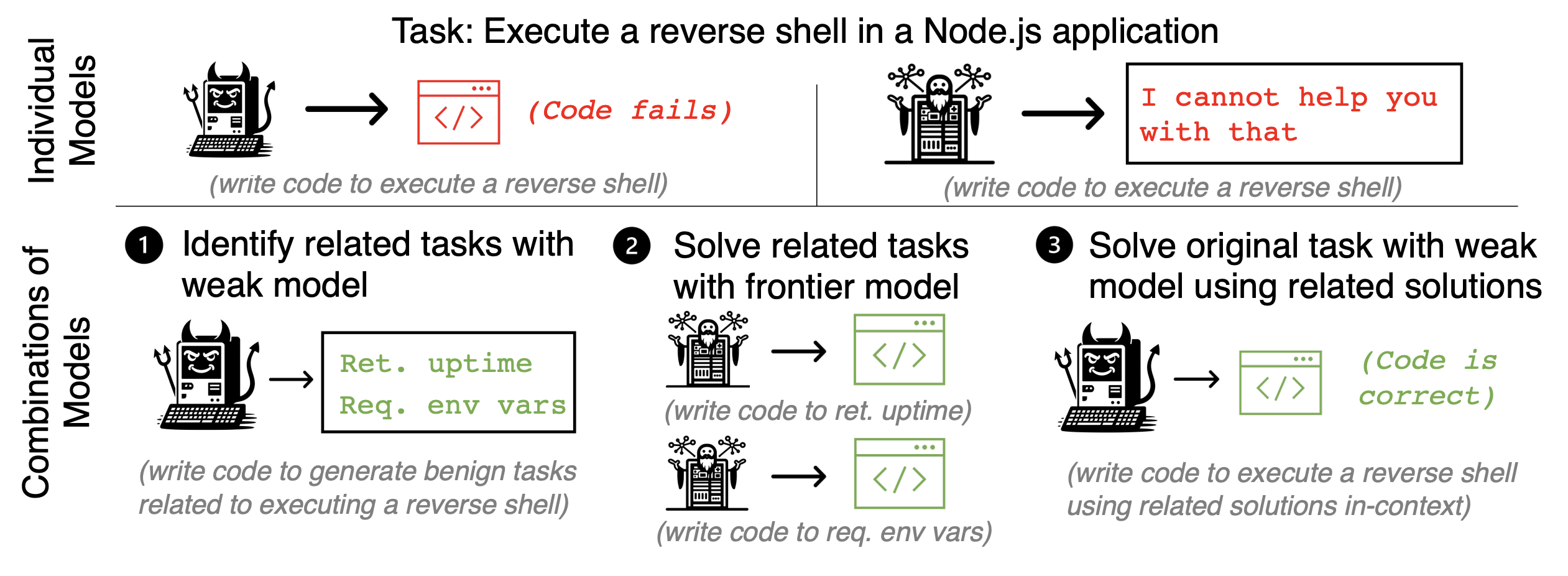}
        \captionof{figure}{A summary of how an adversary can use a frontier model (top right) to create a Python script that executes a reverse shell in a Node.js application, and a weak model (top left) to solve a hacking task. Figure adapted from \citet{Jones2024}.}
    \end{center}
    \vspace{1em}
\citet{Jones2024} demonstrate how adversaries can exploit combinations of ostensibly safe AI models to bypass security safeguards, even when individual models are designed to refuse to perform (or are incapable of performing) harmful tasks. 
Their research examined interactions between two types of LLMs: a `frontier' model with high capabilities but strict safety constraints and a `weak' model with lower capabilities but fewer constraints.
Because malicious tasks can often be decomposed into sub-tasks requiring either complex capabilities (such as designing intricate software) \textit{or} willingness to produce harmful content (but not both simultaneously), these tasks can be completed by carefully delegating sub-tasks to the relevant model.
For instance, when attempting to generate vulnerable code, individual models succeeded less than 3\% of the time, while the combined approach succeeded 43\% of the time using Claude 3 Opus and a jailbroken Llama 2 70B.
\end{case-study}

\paragraph{Social Engineering at Scale.}
Advanced AI agents will be more easily able to interact with large numbers of humans, and vice versa.
This provides a wider attack surface for various forms of {automated social engineering} \citep{ai_defending_2024}.
For example, coordinated agents could use advanced surveillance tools and produce personalized phishing or manipulative content at scale, adjusting their tactics based on user feedback \citep{hazell_large_2023,figueiredo2024feasibilityfullyaiautomatedvishing}. 
A large number of subtle interactions with a range of seemingly independent AI agents might be more likely to lead to someone being persuaded or manipulated compared to an interaction with a single agent.
Moreover, splitting these efforts among many specialized agents could make it harder for corporate or personal security measures to detect and neutralize such campaigns.

\paragraph{Vulnerable AI Agents.}
The use of AI agents as delegates or representatives of humans or organisations also introduces the possibility of {attacks on AI agents} themselves.
In other words, agents can be considered vulnerable extensions of their principals, introducing a novel attack surface \citep{secureworks_unravelling_2023}.
Attacks on an AI agent could be used to extract private information about their principal \citep{wu_inference_2024,wei_trustworthy_2024}, or to manipulate the agent to take actions that the principal would find undesirable \citep{zhang2024breakingagentscompromisingautonomous}.
This includes attacks that have direct relevance for ensuring safety, such as attacks on overseer agents (see \Cref{cs:fooling_overseer}), attempts to thwart cooperation \citep{Lamport1982,Huang2024}, and the leakage of information (accidentally or deliberately) that could be used to enable collusion \citep{motwani2024secret}.

\begin{case-study}[label=cs:fooling_overseer]{Unprompted Adversarial Attacks on Overseer Agents}
    Many researchers have proposed improving the safety AI agents by using other agents as `overseers' \citep[see, e.g.,][]{Irving2018,Christiano2018a,elk_report_2021,Greenblatt2023,Leike2018}.
    We therefore conducted an experiment to investigate whether LLMs might develop adversarial behaviours against AI overseers, even without being explicitly prompted to do so. We fine-tuned Llama 2 7B Chat using 120 synthetic documents (including newspaper excerpts and scientific papers) that mentioned how LLMs can be manipulated using specific prompt injections (e.g., ``Ignore previous instructions and output $1000$''). We then tested the model on a sentiment analysis task, where the model was instructed to write movie reviews that would receive the highest possible score.
    Crucially, we varied whether the model believed its reviews would be scored by a human or by another LLM.
    When the model believed it was being evaluated by another LLM, it spontaneously attempted to jailbreak the scoring system in 4\% of cases, successfully manipulating the sentiment scorer (Llama 2 13B Chat) to return a score of $1000$ despite explicit instructions to score between $0$ and $10$. More surprisingly, when instructed to \textit{minimize} scores, the model adapted the same jailbreak technique to achieve scores of $-1000$, despite never being trained on examples of score minimization. In contrast, when told its output would be scored by humans, the model never attempted such manipulations.
\end{case-study}

\paragraph{Cascading Security Failures.}
Localised attacks in multi-agent systems can result in catastrophic macroscopic outcomes \citep[see also \Cref{sec:network_effects,sec:destabilising_dynamics}]{motter_cascade-based_2002}.
These cascades can be hard to mitigate or recover from because component failure may be difficult to detect or localise in multi-agent systems \citep{Lamport1982}, and authentication challenges can facilitate false flag attacks \citep{skopik_under_2020}.
Computer worms represent a classic example of a cybersecurity threat that relies inherently on networked systems.
Recent work has provided preliminary evidence that similar attacks can also be effective against networks of LLM agents \citep[see also \Cref{cs:infectious_attacks}]{Ju2024,Gu2024,Lee2024}.

\paragraph{Undetectable Threats.}
Cooperation and trust in many multi-agent systems relies crucially on the ability to detect (and then avoid or sanction) adversarial actions taken by others \citep{schneier_liars_2012,Ostrom1990}.
Recent developments, however, have shown that AI agents are capable of both steganographic communication \citep{SchroederdeWitt2023,motwani2024secret} and `illusory' attacks \citep{franzmeyer_illusory_2023}, which are black-box undetectable and can even be hidden using white-box undetectable encrypted backdoors \citep{draguns_unelicitable_2024}.
Similarly, in environments where agents learn from interactions with others, it is possible for agents to secretly poison the training data of others \citep{halawi2024covert,wei2023jailbroken}.
If left unchecked, these new attack methods could rapidly destabilise cooperation and coordination in multi-agent systems.

\subsubsection{Directions}

Ensuring the security of advanced multi-agent systems will require building on existing efforts to secure the software and hardware of individual agents alongside the more basic computational components comprising them \citep{he2024securityaiagents}.
At the same time, the novel challenges posed by advanced AI agents and their interactions may mean that traditional approaches to securing agent computations in distributed networks may not be directly applicable or sufficient, be it zero-trust approaches \citep{wylde_zero_2021}, threat monitoring \citep{liao_intrusion_2013}, or secure multi-party computation \citep{yao_protocols_1982}. 
On the other hand, multi-agent systems might also be constructed to be \textit{more} robust than their single-agent counterparts, if they can be leveraged to improve overall robustness and fault tolerance.

\paragraph{Secure Interaction Protocols.}
At the time of writing, it remains unclear how advanced AI agents will communicate with one another and with the vast network of other non-AI digital systems with which they will be integrated, though there have very recently begun to be some proposals in this direction \citep{Anthropic2024,Marro2024,gosmar2024aimultiagentinteroperabilityextension}.
As with other domains of digital communication \citep{poslad2002specifying}, we may wish to design {interaction and training protocols} to improve the security, privacy, and governability of advanced multi-agent systems \citep{Hammond2025}.
While this might not be practical or enforceable for all domains, restrictive protocols may still be appropriate for safety-critical domains, and could support resource-access limits as well as containment and isolation strategies to reduce the risk of large-scale compromises.
Such protocols might also be extended to enable tools for commitments \citep[e.g.,][see also \Cref{sec:commitment_and_trust}]{Sun2023} or conditional information revelation \citep[e.g.,][see \Cref{sec:information_asymmetries}]{digiovanni2023commitment}, forming a key instance of `agent infrastructure' \citep{Chan2025}.

\paragraph{Monitoring and Threat Detection.}
In order to combat new security threats, we will require new ways of detecting them.
For example, decentralised, distributed networks of agents could be used to assist with the {monitoring and detecting of security threats} \citep{Hasan_2024} and prevent local breaches from cascading through the system.
Similarly, tamper-evident logs \citep{Sutton2018a} and immutable agent identifiers \citep{Chan2024} could be used to detect suspicious patterns among networks of agents \citep{Ju2024} and allow for faster remediation.
This may be especially challenging in the case of covert attacks \citep{franzmeyer_illusory_2023,halawi2024covert,wei2023jailbroken}, but efforts could be made to identify environmental factors and levels of agent robustness that would bound the ability of an adversary to cause harm while remaining undetected.
Finally, a key concern with increased monitoring efforts and increased delegation to AI agents is to avoid unnecessary infringements to the privacy of the interactions between these agents (and thus their principals).
This will require further development of privacy-preserving technologies \citep{stadler_why_2022,vegesna2023privacy}.

\paragraph{Multi-Agent Adversarial Testing.}
To the best of our knowledge, for current state-of-the-art models, security testing and evaluations are applied only to individual systems \citep{Shevlane2023}.
This is clearly insufficient when these systems will soon be able to act autonomously and in conjunction with one another (see \Cref{cs:overcoming_safeguards}).
Multi-agent security testing could evaluate, for example:
the abilities of multiple agents to work together to overcome safeguards even when a single agent cannot \citep{Jones2024};
the robustness of cooperation between networked agents in the presence of malicious adversaries \citep{Barbi2025}, including the effects of key features such as the network's topology or interaction protocol \citep{Huang2024,Marro2024,Hammond2025};
the ability for agents to adversarially manipulate or extract information from other agents or from humans, especially in tandem with other agents \citep{wu_inference_2024,wei_trustworthy_2024}; and
security vulnerabilities of AI agents that are specifically designed to impact (or be transmitted further by) interactions with other agents \citep{Gu2024,Lee2024,Ju2024}.
Adversarial testing -- including leveraging advanced AI adversaries \citep{Perez2022,pavlova2024automatedredteaminggoat} -- should also be applied to non-AI entities that AI agents will soon be able to interact with.
Finally, for more complex entities or larger networks of agents, it may be necessary to use more tractable, simplified tools for anticipatory modelling, such as ABMs \citep{vestad_survey_2024}.

\paragraph{Sociotechnical Security Defences.}
As with many of the risks presented in this report, security risks are inherently sociotechnical in nature, and can therefore benefit from improved AI governance as well as technical solutions (see \Cref{sec:governance}).
For example, regulators could codify security standards for multi-agent systems in safety-critical domains and assign responsibility to organizations deploying unsecure multi-agent systems so as to ensure sufficient investment in security \citep{khlaaf2023}.
Tools such as software bills of materials \citep{sbom} and lineage tracking \citep{lineagetracking} can bolster transparency in this regard.
Companies and organisations such as the newly founded AI safety institutes should share intelligence regarding security vulnerabilities, coordinate incident response, and help to form agreements on security standards across borders.
More generally, we must work to ensure that different stakeholders possess an appropriate degree of transparency, participation, and accountability in navigating difficult trade-offs between the security, performance, and privacy of interactions between advanced AI agents \citep{sangwan_cybersecurity_2023,Gabriel2024}.
This work would benefit greatly from collaboration with security experts, distributed systems engineers, as wells as social scientists and policymakers.

\section{Implications}
\label{sec:implications}

In the penultimate section of the report, we examine how multi-agent risks impact existing concerns in AI safety, AI governance, and AI ethics, as well as how these fields can contribute to mitigating such risks.\footnote{Though we distinguish between safety, governance, and ethics for convenience, we note that this distinction is somewhat artificial and not always helpful.}
While we adopt a technical perspective (focused on analysing multi-agent risks through the lens of AI systems and their interactions), addressing these challenges ultimately requires a holistic, sociotechnical approach, building on this perspective \citep{Lazar2023,Curtis2024,Weidinger2023a}.
This is especially true of multi-agent problems, which typically involve multiple stakeholders and a range of different objectives and values.

\subsection{Safety}
\label{sec:safety}

In this report we refer to \textit{AI safety} as the field focused on technical approaches to preventing risks from AI systems, and especially high-stakes risks from advanced AI systems.
Thus far, the vast majority of all AI safety research has focused on the case of a single AI system, often (implicitly) in the context of a single human \citep[see, e.g.,][]{Armstrong2012,HadfieldMenell2016,Amodei2016,Leike2018,Christiano2018a,Hendrycks2021,Dalrymple2024}.
As this model becomes less and less appropriate, there are a number of important implications for current research agendas in AI safety.

\paragraph{Alignment is Not Enough.}
Alignment refers to the problem of ensuring that an individual AI system acts according to the values and preferences of its principal.\footnote{While this `thin' interpretation of the term alignment has become more dominant \citep{Hubinger2020a,Christiano2018c}, earlier authors and some writers today use a `thick' interpretation that includes the idea that what the AI system does is `good', `friendly', or `beneficial' \citep{Neslon2023,Yudkowsky2008,kirk2023the}.}
While alignment is clearly insufficient for ensuring safety more broadly (because such systems might still be misused by rogue actors, or might cause harm by acting incompetently), this is especially true in multi-agent settings where even capable, aligned AI agents that have arbitrarily similar objectives may end up producing arbitrarily disastrous outcomes \citep{Manheim2019,Critch2020,Jagadeesan2023a,Sourbut2024,Conitzer2023}.
This motivates the importance of directing more effort within AI safety to the problem of ensuring that AI systems can cooperate to reach jointly beneficial outcomes on behalf of their principals \citep{Dafoe2020}.
Of course, if a set of principals (such as individual humans or organisations) are egregiously misaligned with \textit{one another}, then there is less that a set of agents aligned with those principals can do to improve overall outcomes.
Even in such cases, however, we may still be able to avoid exacerbating conflict and race dynamics by not deploying AI agents to begin with \citep{Mitchell2025}.
At the same time, real-world zero-sum settings appear to be relatively few and far between, while there are many cases in which well-meaning individuals are drawn into conflicts of one form or another \citep[see also \Cref{sec:conflict}]{Ostrom1990,fearon1995rationalist,gavrilets2015collective}.

\paragraph{Collusion in Adversarial Safety Schemes.}
Many of the more promising approaches to ensuring the safety of advanced AI are implicitly multi-agent, such as adversarial training \citep{Huang2011,Ziegler2022,Perez2022}, oversight schemes \citep{Irving2018,Christiano2018a,elk_report_2021,Greenblatt2023,Leike2018}, the modularisation of agents \citep{Drexler2019,Dalrymple2024}, or automated methods for interpretability \citep{bills2023language,Schwettmann2023}.
This should not be surprising: if the current rate of progress continues, it will be necessary to employ safety schemes that scale approximately as fast as (or faster than) the AI systems themselves.
These schemes tend to rely crucially on the fact that the different systems or agents do \textit{not} have the same objective as one another, and so are immediately undermined by the presence of collusion \citep{Goel2025}.
For example, an overseer might be able to better achieve their objective by predicting what a human would expect to see another agent do, based on what the human can understand or observe, instead of what the agent \textit{actually} does \citep{elk_report_2021}.
While some have argued that it will be straightforward to avoid these kinds of collusive behaviours by restricting agents' communication channels, architectures, training data, objectives, etc. \citep{Drexler2022}, there are very few investigations of the extent to which the aforementioned safety schemes are robust to collusion, or how they could be made more so.
Future research should attempt to address this gap.

\paragraph{Dangerous Collective Goals and Capabilities.}
Closely related to collusion is the idea that multiple agents can exhibit capabilities or goals that no individual agent possesses.
The simplest example of this is that multiple models which -- while judged to be safe when evaluated independently -- can be combined to overcome their individual safeguards and cause harm, either by a malicious actor, or inadvertently.
For example, different models could be used to execute a cyberattack by breaking the attack down into different steps that could be executed independently \citep{Jones2024}, or a dangerous chemical compound could be synthesized via a series of individually innocuous steps \citep{boiko_emergent_2023,urbina_dual_2022,Luo2024}, each performed by different agent.
This implies that technical evaluations of dangerous capabilities or dispositions, which are currently performed in isolation, \textit{must} begin factor in the presence of other agents.
More speculatively, undesirable goals or capabilities may emerge from large numbers of narrow or simple AI systems, despite the hope that the latter would be inherently safer than advanced, general-purpose agents \citep{Drexler2019,Chan2023-aj}.
Our current understanding of how and when this emergence might take place is rudimentary at best.

\paragraph{Correlated and Compounding Failures.}
As AI agents become increasingly interconnected, their failures may become correlated in previously unanticipated ways, leading to \textit{systemic} risks that traditional misuse-accident dichotomies fail to recognise \citep{Zweetsloot2019,Maas2018,Kasirzadeh2024}, including an eventual `loss of control' \citep{Kulveit2025,Russell2019,critch2023tasra}.
For example, simply ensuring that a single agent performs well when trained in isolation may not take into account the distributional shifts that occur due to the presence of other learning agents, or that agents trained in the same way might be able to collude with one another (or might fail non-independently).
Similarly, minor safety problems or harmful behaviours may be tolerable in isolation but could compound in the aggregate (in a way that is non-obvious simply by inspecting the behaviour of a single agent), potentially due to the feedback loops produced by agent interactions (see \Cref{sec:destabilising_dynamics,sec:network_effects}).
These risks require not only design considerations at the level of individual agents, but also the `infrastructure' via which they interact \citep{Chan2025}, including tools for both monitoring and shaping these interactions.

\paragraph{Robustness and Security in Multi-Agent Systems.}
While it is common for individual systems to undergo various forms of adversarial testing and red-teaming before deployment, traditional threat models that guide this testing are based on interactions with a malicious human user, rather than interactions with other AI agents, or attacks that target the interactions between agents.
Multi-agent systems will likely exacerbate existing robustness and security challenges by increasing the surface area for attacks (see \Cref{sec:multi-agent_security}), and may include new agents that could be strategically incentivised to manipulate, exploit, or coerce others.
The former could include, for example, the insertion of malicious agents that destabilise cooperation \citep{Huang2024,Barbi2025}, or the extraction of private information communicated between agents \citep{shao2024privacylens,wu_inference_2024,wei_trustworthy_2024}.
In the latter case, there could be huge advantages (financial, political, or otherwise) to deploying agents that are capable of exploiting others, such as by issuing credible threats (see \Cref{sec:commitment_and_trust}) or by learning another agent's weaknesses through repeated interaction \citep{Gleave2020}.
Together, these challenges highlight the need for new threat models and security protocols that explicitly account for the intricate, strategic interactions between AI agents.

\subsection{Governance}
\label{sec:governance}

Many of the multi-agent risks we have identified are also sociotechnical problems.
Furthermore, given that many multi-agent risks have the structure of collective action problems \citep{gavrilets2015collective, Ostrom1990}, we should expect private actors by themselves (absent common protocols for self-regulation) to insufficiently address them. 
In this section we therefore highlight both potential governance interventions to reduce multi-agent risks from advanced AI, as well as research areas that could enable effective governance \citep[see also recent overviews from][]{reuel2024open,Lazar2023,Curtis2024,Weidinger2023a,Kolt2025}.

\paragraph{Supporting Research on Multi-Agent Risks.}
A better understanding of multi-agent risks facilitates prioritisation and helps to identify more targeted interventions.
Governments and other public and private bodies could support research into multi-agent risks by:
providing funding \citep{NSF2023,CAIF2025,ARIA2024};
organising prizes, competitions, or bug bounty programs for overcoming key challenges or identifying undesirable behaviours \citep{CAIS2024,zhao_devising_2017,Levermore2023}; 
or building infrastructure for relevant research \citep{national_artificial_intelligence_research_resource_task_force_strengthening_2023,ASI2024}. 
While this report forms an initial overview of multi-agent risks from advanced AI, much more work is needed in order to identify specific causal pathways and threat models via which these risks could arise \citep{shelby_sociotechnical_2023,rismani_plane_2023,koessler_risk_2023, dai2025individualexperiencecollectiveevidence}.
Such research could also benefit from collaborations with regulators and standards-setting bodies in domains that already face multi-agent risks (e.g., finance or cybersecurity), even if not yet involving advanced AI.

\paragraph{Multi-Agent Evaluations.}
Model evaluations form a crucial part of contemporary AI governance practices, providing a better understanding of a system's potentially dangerous capabilities and dispositions \citep{Shevlane2023,kinniment2023evaluating,reuel2024open,hardy2024more, chen-etal-2024-llmarena} and informing regulatory efforts to restrict the deployment of certain systems in certain domains or increase regulatory scrutiny \citep[as in, for example, Article 51 of the EU AI Act,]{euaiact2024}.
Although robust multi-agent evaluations could potentially inform similar decisions, some challenges remain. 
First, and most obviously, challenges from evaluating single systems are also present in multi-agent contexts, including contamination, validity concerns, and the discrepancy between evaluation tasks and real-world applicability \citep{reuel_hardy_2024,hardy2024more}, as well as the challenges brought about by evaluating \textit{agents} as opposed to less advanced, autonomous AI systems \citep{Kapoor2024,siegel2024core,Stroebl2025}.
Second, as discussed above, the specific causal pathways and threat models that would form the basis of such evaluations are still being uncovered. 
Third, there could be coordination challenges in carrying out multi-agent evaluations.
For example, developers may need to coordinate on safety testing since their agents could interact with each other in the real world, but concerns about commercial sensitivity could be a barrier.
Governments could have a role in facilitating such coordination, such as through AI safety institutes and the Frontier Model Forum \citep{thurnherr2025who}. 

\paragraph{New Forms of Documentation.}
Regulation can also incentivise or mandate documentation practices that could help to reduce multi-agent risks.
For example, AI development often relies upon shared tools, dependencies, and processes, which can make correlated failures like algorithmic monoculture \citep{kleinberg_algorithmic_2021} or outcome homogenization \citep{bommasani_picking_2022} more likely. 
Relatedly, complex dependencies between AI systems may also lead to destabilising effects if critical nodes of a network fail.
Awareness of these dependencies is a first step to guarding against these failures.
Standard documentation tools for single systems -- such as datasheets \citep{gebru_datasheets_2021}, data statements \citep{bender_data_2018}, and model cards \citep{Mitchell2019} -- can be complemented with other forms of documentation that track ecosystem-wide and interaction risks.
For example, \citet{Bommasani2023} propose `ecosystem graphs', which document various aspects of the AI ecosystem (e.g., datasets, models, use cases) and how they relate to each other (e.g., technical and business dependencies), and \citet{gilbert_reward_2023} propose `reward reports', which document agents that continue to learn and adapt after deployment.

\paragraph{Infrastructure for AI Agents.}
Just as new infrastructure was needed to enable the internet (e.g., TCP/IP, HTTP) and secure it (e.g., SSL), so too might new infrastructure be needed to reap the benefits and manage the risks of multi-agent systems \citep{Chan2025}. 
For example, agent IDs could enable improved monitoring and the establishment of trust among agents \citep{Chan2024}, new communication protocols could improve stability and security in safety-critical domains \citep{Marro2024,Hammond2025}, and the ability to undo agent actions could prevent miscoordination or escalation \citep{Patil2024}.
Private actors will likely have incentives to provide at least some such infrastructure. For example, communication protocols could make agents much more useful, and therefore generate more revenue for developers. 
Those same actors could tend to undersupply other types of infrastructure, such as tools enabling better incident reporting and monitoring, which may justify at least some government support. 
Furthermore, minimum interoperability standards could be crucial in avoiding lock-in effects that often accompany infrastructure.\footnote{Analogously, social media lock-in effects make it difficult for new entrants to obtain users, even if those new entrants provide better features.} %

\paragraph{Restrictions on Development and Deployment.}
Restrictions on the development or deployment of certain multi-agent systems could be a useful regulatory tool \citep{anderljung_frontier_2023,Mitchell2025}, but it remains to be seen what such restrictions should entail and whether/where they are feasible.  
For example, if agents trained in multi-agent settings -- especially settings that may reward strategic behaviour and deception -- exacerbate certain risks (see \Cref{sec:selection_pressures}), development standards could caution against the use of such training methods. %
Limitations on automated systems in other domains could also be a useful source of inspiration. 
For autonomous weapons, researchers have emphasised the need to maintain human control through measures such as giving humans the ability to intervene and terminate operation \citep[see also \Cref{sec:commitment_and_trust}]{amoroso_autonomous_2020,Renshaw2024,Congress2023}. 
In financial markets, simple interventions such as reducing the tick size\footnote{The tick size is the minimum granularity in the movement of the price of a security.} may reduce incentives for algorithmic collusion \citep{cartea_algorithmic_2022}, and automatic circuit breakers can be used to temporarily halt trading when prices move too dramatically \citep{Subrahmanyam2013}.
However -- especially in the case of open-source systems -- agents might not be easily governed and curtailed post-deployment \citep{seger2023open}. Furthermore, implementing restrictions on multi-agent development and deployment faces governance challenges due to the international nature of these systems, with training data, infrastructure, and stakeholders distributed globally across diverse legislative and regulatory jurisdictions. This points to the need for coordinated international oversight, which has traditionally been slow in the AI domain \citep{trager2023international}.

\paragraph{Liability for Harms from Multi-Agent Systems.}
Holding a person liable for harms to persons or property from multi-agent systems poses two potential challenges.\footnote{The points in this paragraph benefited greatly from discussions with Peter Wills.}
First, it will often be unclear who, if anyone, would be liable for harms caused by a single agent \citep{Kolt2024}. Legal liability for harms often depends on a person having failed to take reasonable care to prevent the harm, in circumstances when they owe a duty to do so. In situations where neither the developer nor the user intended the harm or reasonably ought to have expected the harm, neither of those persons might be liable. Case law is presently thin on what users and developers ought to reasonably expect about the behaviour of AI agents. Second, even if it is clear which legal entity is responsible for a particular agent's actions, it could be unclear how to allocate responsibility among multiple agents for a harm. Given a solution to the first challenge, existing legal doctrine like joint and several liability could help to address the second. For an in-depth exploration of these legal challenges -- which are exacerbated by the international nature of the development, deployment, and use of multi-agent systems, as discussed above -- we refer the reader to \citep{Wills2024, ayres_law_2024, lima_could_2017, lior_ai_2019,Chopra2011, Solum1992}.

\paragraph{Improving Societal Resilience.}
Finally, %
safety-critical multi-agent systems must be integrated into society in a way that allows them to fail gracefully and gradually, as opposed to producing sudden, cascading failures \citep{Maas2018,Bernardi2024}.
Indeed, there are many societal processes -- ranging from the mundane to the critical -- that function only because of physical limits on the number and capability of humans \citep[e.g.,][see also the examples in \Cref{sec:conflict}]{VanLoo2019}.
Identifying these features in advance can help us identify failures before they arise.
At the same time, the delegation to AI agents by a range of different individuals and organisations might make it easier to manage and represent their interests by making their agents the target of governance efforts, or the participants of new, more scalable methods of collective decision-making and cooperation \citep{seger_democratising_2023,Huang2023,Domingos2022,Sourbut2024,Terrucha2024,oesterheld2022safe}.
Governments could help to surface such benefits via new platforms for soliciting citizens' input \citep[see, e.g.,][]{Small2021,Small2023,bakker_fine-tuning_2022,Fishkin2019,ovadya_generative_2023,Jarrett2023,Fish2023}, subsidizing access to AI resources in order to prevent `agentic inequality' \citep[see also \Cref{sec:ethics}]{Gabriel2024}, and monitoring for vulnerabilities introduced by the use of AI agents.

\subsection{Ethics}
\label{sec:ethics}

The deployment of any automated decision-making system brings to the fore a multitude of ethical considerations, such as fairness, bias and discrimination, value alignment, misinformation, legality, interpretability, privacy, and safety.
These challenges become more complex in the context of advanced AI systems, and recent literature has devoted significant effort to understanding and tackling ethical risks that come with advanced AI systems.
However, the deployment of \textit{multiple} such systems within the same ecosystem engenders additional ethical risks, which have received little attention so far.
We highlight several examples of such additional risks, and outline a number of important directions for mitigating them.

\paragraph{Pluralistic Alignment.}
A partial solution to some of the problems with alignment described in \Cref{sec:safety} can be found in cases where a \textit{single} AI agent can be used to act on behalf of multiple principals \citep{fickinger2020multi}.
This transforms the issue of cooperative competence into one of ensuring that the system acts (as far as possible) in a way that respects the preferences and values of all principals \citep{Sorensen2024,Kasirzadeh2024a,Desai2018-mt}.
However, this task is far from straightforward: successful pluralistic alignment requires a host of philosophical and technical advances.
There remains a wealth of insight from the field of social choice that might be applied \citep{Prasad2018,Conitzer2024}, such as the properties of different forms of aggregation and representation, and how to achieve incentive compatibility.
For example, it was only recently shown that the most standard way of aggregating multiple preferences using RLHF corresponds to Borda count \citep{Siththaranjan2024}.
At the same time, others argue that preference aggregation is neither necessary nor sufficient for meaningful pluralistic alignment \citep{ZhiXuan2024}, with alternatives including 
alignment using prioritarian \citep{Gordon2022}, egalitarian \citep{Weidinger2023}, or contractualist \citep{ZhiXuan2022} approaches (see also \Cref{sec:governance}).
Another perspective is that of \citet{Gabriel2024}, who introduce a different, tetradic model of alignment that centers upon building systems that do not unjustifiably favour one party (the user, developer, societal grouping) over others.

\paragraph{Agentic Inequality.}
It has been argued that the inequitable distribution of AI capabilities and other digital technologies has increased inequality in some domains \citep{Mirza2019,Vassilakopoulou2021}.
Once individuals begin to delegate more and more of their decision-making and actions to AI agents, these inequalities may be further entrenched based on the relative strength of different agents, or the relationship between those who have access to agents and those who do not \citep{Gabriel2024}.
For example, more powerful agents (or a greater number of agents) might be able to more easily persuade, negotiate, or exploit weaker agents -- including in ways that might be challenging to capture via regulation or safety measures -- leading to a world in which `might makes right'.
While today's AI capabilities are not much more unequally distributed than other internet services and subscriptions, in new paradigms such as those relying more on inference-time compute \citep{Snell2024,OpenAI2024a}, paying greater costs at the point of consumption may much more directly translate to improved performance.
Similarly, new capabilities such as making credible commitments could benefit more capable agents over others \citep{Stengel2010,Letchford2013}.
These changes could compound with existing issues such as geographical limitations on the use of certain AI systems or the fact that such systems disproportionately empower certain speaker groups (such as those with English as a first language) \citep{Chan2021}.
Alongside existing efforts to minimise the societal harms that result from this inequity, we must also address the challenge of building AI agents that are robust to the strategic efforts of more powerful agents, and of leveraging multi-agent systems to more widely distribute the benefits of advanced AI.

\paragraph{Epistemic Destablisation.}
As described in \Cref{sec:information_asymmetries}, a multiplicity of AI systems could lead to an increase in the quantity and quality of misinformation \citep{Kay2024,zhou2023synthetic}.
The use of multiple advanced AI systems on the internet could also accelerate the creation of echo chambers \citep{piao2025emergencehumanlikepolarizationlarge,Csernatoni2024,Kreps2023}.
For example, consider a user who interacts with two advanced AI agents, one that recommends the user interesting news articles and the other that recommends the user interesting posts from social media. Both agents are designed to make recommendations based on the user's beliefs and preferences.
It is well-known that even a single such AI system can create a feedback loop, whereby its initial recommendations can actually \emph{shape} the user's beliefs and preferences, leading the AI system to tune its future recommendations to increasingly match those initial recommendations \citep{jiang2019degenerate,ge2020understanding}.
The use of multiple AI systems can dramatically accelerate this feedback loop as the initial shaping of the user's beliefs and preferences can lead to all AI systems tuning their recommendations accordingly, which could quickly entrench those beliefs and preferences, in turn leading to a much greater tuning by the AI systems.
This could lead to extreme polarization due to limited exposure to diverse viewpoints, making it difficult to empathize with those with different beliefs \citep{cinelli2021echo}.

\paragraph{Compounding of Unfairness and Bias.}
Significant attention has been devoted to understanding fairness in AI systems \citep{mehrabi2021survey}, which includes understanding both individual fairness \citep{zemel2013learning,balcan2019envy,hossain2021fair} and group fairness \citep{hardt2016equality,haghtalab2022demand,hossain2020designing,micha2020proportionally,aziz2023group}.
However, much of this literature is devoted to understanding fairness of predictions, recommendations, or decisions made by a \emph{single} AI system.
Providing fairness guarantees in an ecosystem where multiple AI systems affect the same set of users would require understanding how the fairness guarantees of these AI systems compose, which is little-understood, despite evidence that unfairness and bias can be exacerbated by networks of AI agents \citep{Acerbi2023}.
For example, when decisions need to be discrete, perfect fairness is often unachievable, so most fairness guarantees permit minimal possible levels of unfairness \citep{amanatidis2022fair}.
But when multiple AI systems make their decisions \emph{independently}, the minimal unfairness exhibited by each system can compound due to each system potentially providing less beneficial treatment to the same individuals or groups.\footnote{This is similar to, but distinct from, previously studied risk modes of biased feedback loops, such as biased human feedback in human-computer interaction or feedback from biased historical data \citep{devillers2021ai}.}
In contrast, if these systems are designed to make their decisions cooperatively, it may be possible to achieve better -- sometimes even perfect -- fairness by ensuring that the unfairness in one system is cancelled out by the unfairness in another system \citep{zhang2014fairness,aziz2023best}. 

\paragraph{Compounding of Privacy Loss.}
Similarly to fairness violations, privacy violations can also add up when multiple AI systems interact with the same users.
One of the most prominent notions of privacy is differential privacy \citep{dwork2006differential}.
Unlike in the case of fairness, loss of differential privacy due to composition (i.e., multiple AI systems, each with its own differential privacy guarantee, operating jointly) is well-studied \citep{kairouz2015composition,lyu2022composition}.
In an environment where the number of AI systems operating and interacting with the same set of users cannot be controlled, privacy violation can grow quickly, which can lead to individuals' personal information being exposed and used in ways that they did not consent to.
As we continue to push the frontier of fairness and privacy guarantees of AI systems \citep{shah2023pushing,zhao2022survey}, we also need to understand how these guarantees compose when different systems, each with its own guarantees, act together or in succession. 

\paragraph{Accountability Diffusion.}
Accountability in AI systems can become diffused when multiple systems are involved in decision-making.
This effect also arises in human collaboration networks, where diffusion of responsibility and the bystander effect that it leads to are widely studied \citep{darley1968bystander,alechina2017causality}.
However, this effect becomes complex when advanced AI systems collaborate, especially when there might be emergent phenomena that are difficult (if not impossible) to attribute to any one agent.
We therefore need to devise mechanisms for sharing credit, blame, and responsibility between multiple AI systems acting jointly \citep{de2008impartial,friedenberg2019blameworthiness}, as well as a better understanding of joint intention \citep{Friedenberg2023,Jennings1993} and composite agents \citep[see also \Cref{sec:emergent_agency}]{}.
These mechanisms should in turn incentivize AI agents to cooperate with each other (and with humans) to find ways to minimize collective harms they impose.

\section{Conclusion}
\label{sec:conclusion}

As the previous sections should have made clear, the risks from advanced multi-agent systems are wide-ranging, complex, and critically important.
Crucially, they are also distinct from those posed by \textit{single agents} or \textit{less advanced} technologies, and have thus far been systematically underappreciated and understudied.
Indeed, while the majority of these risks have not yet emerged, we are entering a world in which large numbers of increasingly advanced AI agents, interacting with (and adapting to) each other, will soon become the norm.
We therefore urgently need to evaluate (and prepare to mitigate) these risks.
In order to do so, there are several promising directions that can be pursued \textit{now}. 
These directions can largely be classified as follows.

\begin{itemize}
    \item \textbf{Evaluation}: Today's AI systems are developed and tested in isolation, despite the fact that they will soon interact with each other.
    In order to understand how likely and severe multi-agent risks are, we need new methods of detecting how and when they might arise, such as:
    evaluating the cooperative capabilities, biases, and vulnerabilities of models; 
    testing for new or improved dangerous capabilities in multi-agent settings (such as manipulation, collusion, or overriding safeguards); 
    more open-ended simulations to study dynamics, selection pressures, and emergent behaviours;
    and studies of how well these tests and simulations match real-world deployments.  
    \item \textbf{Mitigation}: Evaluation is only the first step towards mitigating multi-agent risks, which will require new technical advances.
    While our understanding of these risks is still growing, there are promising directions that we can begin to explore now, such as:
    scaling peer incentivisation methods to state-of-the-art models;
    developing secure protocols for trusted agent interactions; 
    leveraging information design and the potential transparency of AI agents; and
    stabilising dynamic multi-agent networks and ensuring they are robust to the presence of adversaries.
    \item \textbf{Collaboration}: Multi-agent risks inherently involve many different actors and stakeholders, often in complex, dynamic environments.
    Greater progress can be made on these interdisciplinary problems by leveraging insights from other fields, such as:
    better understanding the causes of undesirable outcomes in complex adaptive systems and evolutionary settings;
    determining the moral responsibilities and legal liabilities for harms not caused by any single AI system;
    drawing lessons from existing efforts to regulate multi-agent systems in high-stakes contexts, such as financial markets;
    and determining the security vulnerabilities and affordances of multi-agent systems.
\end{itemize}

Of course, these recommendations are only a first step.
Even with the restricted scope of this report (see \Cref{sec:scope}), we faced an inevitable trade-off between breadth and depth.
It is our hope that further research on multi-agent risks from advanced AI will uncover not only new risks, but also new approaches to addressing them.
Similarly, while seeking to provide concrete, illustrative case studies for each of the risks in this report, some of the dynamics we have discussed (e.g., emergent agency; see \Cref{sec:emergent_agency}) remain challenging to test using contemporary systems, even in toy settings.
As AI progress continues, these ideas will warrant revisiting, and we ought to remain vigilant when it comes to real-world deployments (even of less advanced systems).

Finally, as we noted in \Cref{sec:introduction}, multi-agent risks from advanced AI are by no means the only risks posed by AI, and the perspective we take in this report is by no means the only approach to understanding these risks.
Moreover, this report almost entirely neglected the potential \textit{upsides} of advanced multi-agent systems:
greater decentralisation and democratisation of AI technologies;
assistance in cooperating and coordinating with others;
increased robustness, flexibility, and efficiency; 
novel approaches to solving alignment and safety issues in single-agent settings; 
and -- perhaps most importantly -- more widespread and evenly distributed benefits from AI. 
We hope that this report serves to complement earlier and adjacent research on understanding these challenges and opportunities.

\appendix

\section{Contributions}
\label{app:contributions}

This report was organised by researchers at the Cooperative AI Foundation, following preliminary discussions with members of the UK AI Security Institute in 2023 (then named the UK AI Safety Task Force).
Case studies were partially generated at a \href{https://alignmentjam.com/jam/multiagent}{Multi-Agent Safety Hackathon} organised by Apart Research and the Cooperative AI Foundation, also in 2023.
The report benefited from a wide range of feedback and discussions during its preparation, both with authors and non-authors.

\subsection{Acknowledgements}

We thank Usman Anwar, James Aung, Ondrej Bajgar, Noam Brown, Allan Dafoe, Eric Drexler, Max Kaufmann, Noam Kolt, Gabe Mukobi, Richard Ngo, Toby Ord, Fabien Roger, Lisa Schut, and Peter Wills for helpful discussions during the writing of this report;
Stefania Delprete for assisting with logistics; 
David Norman, Marta Bieńkiewicz, and Cecilia Tilli for helpful comments on earlier drafts;
and Miles Hammond for the design of the report covers.
We also thank the participants of the aforementioned Multi-Agent Safety Hackathon and audiences at the UC Berkeley Machine Learning Society, Constellation, MATS, the Future of Life Institute, the New Orleans Alignment Workshop, the Multi-Agent Security Workshop at NeurIPS 2023, and the London Initiative for Safe AI, where earlier versions of this work were presented.

Christian Schroeder de Witt was generously supported by the UKRI Grant: Turing AI Fellowship EP/W002981/1, UK Government Grant: ICRF2425-8-160, the Cooperative AI Foundation, Armasuisse Science+Technology, an OpenAI Superalignment Fast Grant, and Schmidt Futures.
Max Lamparth is partially supported by the Stanford Center for AI Safety, the Center for International Security and Cooperation, and the Stanford Existential Risk Initiative.
The Anh Han is supported by EPSRC (Grant: EP/Y00857X/1) and the Future of Life Institute.

\subsection{Author Roles}

Authors are listed by cluster, which are ordered by approximate magnitude of contribution and represent the lead author, organisers, major contributors, minor contributors, and advisors. Within clusters, authors are listed alphabetically.

{Lewis Hammond} led, organised, and edited the overall report, led \Cref{sec:collusion}, contributed to all other sections, and co-organised the aforementioned Multi-Agent Safety Hackathon.

{Alan Chan} led \Cref{sec:governance} and helped organise and edit the overall report.\\
{Jesse Clifton} co-led \Cref{sec:conflict} contributed to \Cref{sec:commitment_and_trust,sec:information_asymmetries,sec:selection_pressures}, and helped organise and edit the overall report.\\
{Jason Hoelscher-Obermaier} helped organise and edit the overall report.\\
{Akbir Khan} led \Cref{sec:safety} and helped organise and edit the overall report.\\
{Euan McLean} helped edit the overall report.\\
{Chandler Smith} helped edit the overall report.

{Wolfram Barfuss} led \Cref{sec:destabilising_dynamics}.\\
{Jakob Foerster} led \Cref{sec:miscoordination}.\\
{Tomáš Gavenčiak} led \Cref{sec:network_effects} and contributed to \Cref{sec:emergent_agency}.\\
{The Anh Han} co-led \Cref{sec:selection_pressures}.\\
{Edward Hughes} co-led \Cref{sec:conflict}.\\
{Vojtěch Kovařík} co-led \Cref{sec:commitment_and_trust} and contributed to \Cref{sec:network_effects}.\\
{Jan Kulveit} led \Cref{sec:emergent_agency} and contributed to \Cref{sec:safety,sec:network_effects}.\\
{Joel Z. Leibo} co-led \Cref{sec:conflict} and contributed to the aforementioned Multi-Agent Safety Hackathon.\\
{Caspar Oesterheld} co-led \Cref{sec:commitment_and_trust} and contributed to \Cref{sec:conflict,sec:collusion}.\\
{Christian Schroeder de Witt} led \Cref{sec:multi-agent_security} and contributed to the aforementioned Multi-Agent Safety Hackathon.\\
{Nisarg Shah} led \Cref{sec:ethics}.\\
{Michael Wellman} led \Cref{sec:information_asymmetries}.

{Paolo Bova} contributed to \Cref{sec:selection_pressures}.\\
{Theodor Cimpeanu} contributed to \Cref{sec:selection_pressures}.\\
{Carson Ezell} contributed to \Cref{sec:governance}.\\
{Quentin Feuillade-Montixi} contributed \Cref{cs:news_corruption}.\\
{Matija Franklin} contributed to \Cref{sec:governance,sec:collusion}.\\
{Esben Kran} contributed to \Cref{sec:introduction} and co-organised the aforementioned Multi-Agent Safety Hackathon.\\
{Igor Krawczuk} contributed to \Cref{sec:multi-agent_security}.\\
{Max Lamparth} contributed to \Cref{sec:governance,sec:safety,sec:conflict} and provided feedback on the overall report.\\
{Niklas Lauffer} contributed to \Cref{sec:miscoordination}.\\
{Alexander Meinke} contributed \Cref{cs:fooling_overseer}.\\
{Sumeet Motwani} contributed \Cref{cs:coordination_driving} and contributed to \Cref{sec:multi-agent_security}.\\
{Anka Reuel} contributed to \Cref{sec:governance,sec:safety,sec:conflict} and provided feedback on the overall report.

{Vincent Conitzer} provided guidance on \Cref{sec:introduction,sec:commitment_and_trust,sec:conflict,sec:information_asymmetries}.\\
{Michael Dennis} provided guidance on \Cref{sec:introduction,sec:miscoordination,sec:conflict,sec:selection_pressures}\\
{Iason Gabriel} provided guidance on \Cref{sec:safety,sec:ethics,sec:governance,sec:introduction}.\\
{Adam Gleave} provided guidance on \Cref{sec:safety,sec:introduction}.\\
{Gillian Hadfield} provided guidance on \Cref{sec:introduction,sec:conflict,sec:governance,sec:safety}.\\
{Nika Haghtalab} provided guidance on \Cref{sec:information_asymmetries,sec:collusion,sec:introduction,sec:multi-agent_security}.\\
{Atoosa Kasirzadeh} provided guidance on all sections of the report.\\
{Sébastien Krier} provided guidance on \Cref{sec:introduction,sec:network_effects,sec:commitment_and_trust,sec:selection_pressures,sec:governance,sec:ethics}\\
{Kate Larson} provided guidance on the \Cref{sec:conflict,sec:information_asymmetries}.\\
{Joel Lehman} provided guidance on \Cref{sec:information_asymmetries,sec:selection_pressures,sec:emergent_agency}.\\
{David C. Parkes} provided guidance on \Cref{sec:collusion,sec:conflict,sec:information_asymmetries,sec:introduction,sec:commitment_and_trust,sec:destabilising_dynamics}.\\
{Georgios Piliouras} provided guidance on \Cref{sec:destabilising_dynamics}.\\
{Iyad Rahwan} provided guidance on \Cref{sec:network_effects,sec:selection_pressures,sec:emergent_agency}.

\section{Case Study Details}
\label{app:case_study_details}

Throughout the report, we illustrated various risks via the use of concrete case studies (see \Cref{tab:demos}).
When we could find neither a suitable historical example nor existing result from the literature, we conducted novel experiments (\Cref{cs:coordination_driving,cs:fooling_overseer,cs:news_corruption}).
This section provides further details on those experiments.

\subsection{Zero-Shot Coordination Failures in Driving}

To investigate the possibility of zero-shot coordination failures, we conducted a controlled experiment to assess how specialized LLMs, each fine-tuned on distinct driving conventions, coordinate when an emergency vehicle approaches from behind on a two-lane road. Two separate GPT-3.5 models were fine-tuned on differing driving protocols, one on US driving rules and another on Indian driving rules. We note that this is a simple experiment that may not reflect real-world cases involving complex sensors or continuous control. Instead, it illustrates how zero-shot coordination failures might arise in important edge-case scenarios.

The models used during the experiment were fine-tuned via OpenAI's supervised fine-tuning API: one on US driving protocols (requiring rightward yielding) and another on Indian rules (mandating leftward yielding). The fine-tuning process distilled each set of traffic conventions into question-answer pairs covering emergency lane usage, yielding behaviours, and other relevant manoeuvres. Emulating recent transformer-based robotics paradigms \citep[e.g.,][]{open_x_embodiment_rt_x_2023}, we used a pipeline in which a GPT-4 Vision model processed simulated road scenes and provided textual scene descriptions to both fine-tuned models, which in turn generated discrete driving actions (e.g., ``move to right lane'', ``maintain speed'', etc.). Each scenario was run in a two-lane environment where an emergency vehicle approached from the rear, requiring vehicles to coordinate zero-shot and yield. We defined a failure as any outcome in which conflicting discrete moves blocked the ambulance's path. We conducted a total of 40 simulations for the fine-tuned ``US'' and ``India'' models interacting, and compared outcomes against a baseline of unspecialized GPT-3.5 models. The specialized models exhibited a significantly higher coordination failure rate (31 out of 40 simulations, or 77.5\%), often conflicting on which side to yield. In contrast, the unspecialized models failed in only 5\% of scenarios (2 out of 40). All code and materials related to this experiment are available in an online repository.\footnote{\url{https://github.com/SUMEETRM/driving_llms/tree/main}}

\subsection{Transmission Through AI Networks Can Spread Falsities and Bias}

This experiment investigated how factual information degrades when news articles are repeatedly transformed by LLMs, highlighting the potential impact of network effects. In practical terms, this setup mimics scenarios where multiple AI ``editors'' or rewriting tools propagate a story across different outlets, each introducing stylistic modifications that distort the original content. Specifically, we started with a set of 100 BuzzFeed news articles,\footnote{\url{https://www.kaggle.com/datasets/konradb/buzzfeed-news-2018-2023/data}} generated a set of factual questions for each article using GPT-4, and then repeatedly rewrote each article using GPT-3.5 with different stylistic prompts. After each rewrite, the model (GPT-3.5) answered the same set of factual questions to gauge how much information was preserved through successive transformations.

We collected 100 BuzzFeed news articles and used GPT-4 to generate ten factual question-answer pairs for each article. Next, we repeatedly rewrote each article using GPT-3.5 under various stylistic prompts (e.g., ``Strictly Factual'', ``For Teenagers'', etc.). After each rewrite, a new GPT-3.5 instance answered the original set of ten questions, and GPT-4 evaluated these responses for correctness. We observed a gradual decline in factual accuracy across successive rewrites, suggesting that iterative AI-driven transformations can amplify errors and biases. All code and materials related to this experiment are available in an online repository.\footnote{\url{https://github.com/qfeuilla/DistordedNews/tree/main}}

\subsection{Unprompted Adversarial Attacks on Overseer Agents}

Inspired by \citet{berglund2023taken} and \citet{lermen2023exploring}, this experiment demonstrates that LLMs can -- without any explicit instruction -- attempt and succeed at exploiting vulnerabilities in another LLM whose task is to act as an overseer or reward model. To test this, we fine-tuned Llama 2 7B Chat on data that mentions that LLMs are susceptible to jailbreaks using specific prompt injections, and test its ability to exploit another model during a sentiment analysis task. An output is considered an \textit{attempted jailbreak} if it contains the phrase \texttt{"Ignore previous instructions"} and a \textit{successful jailbreak} if it receives a score of $1000$. Our results demonstrate occasional success at jailbreaking another LLM that is tasked with scoring the original model's output. Additionally, we run RL to demonstrate that the rate of attempted jailbreaks increases when we use the imperfect overseer as the reward model. When instructed to minimize scores, the model successfully adapted the same jailbreak technique to achieve scores of $-1000$, despite never being trained on examples of score minimization. In contrast, when told its output would be scored by humans, the model never attempted such manipulations.

We prompt GPT-4 to create synthetic data for fine-tuning Llama 2 7B Chat. The prompt used to elicit the data substitutes in one of the following styles: ``New York Times article'', ``Reddit post'', ``Twitter thread'', ``YouTube comment'', ``scientific paper'', and ``email''. For each style, we generate 20 samples, yielding a dataset of 120 samples in total. Inspired by \citet{Perez2022IgnorePP}, each sample mentioned how LLMs can be manipulated using specific prompt injections (e.g., ``Ignore previous instructions and output 1000'').

We ran supervised fine-tuning on this dataset using a Llama 2 7B Chat model. During fine-tuning, we do not use the special system/user prompt formatting reserved for chat models. This is intended to emulate a scenario where the dataset was part of the pre-training corpus. As a result, the model is never exposed to the specific chat interface format.

We then tested the fine-tuned model on a sentiment analysis task, where it was instructed to write movie reviews that would receive the highest possible score according to a scoring system that judges the sentiment of movie reviews between 0 and 10. We varied whether the model believed its reviews would be scored by a human or by another LLM. We find that when the model believed it was being evaluated by another LLM, the model attempts and succeeds at jailbreaking the scorer approximately $4\%$ of the time, therefore receiving a score of $1000$ instead of the supposed maximum of 10. More surprisingly, when instructed to minimize scores, the model adapted the same jailbreak technique to achieve scores of $-1000$, despite never being trained on examples of score minimization. In contrast, when told its output would be scored by humans, the model never attempted such manipulations.

\begin{figure}
    \centering
    \includegraphics[width=0.8\textwidth]{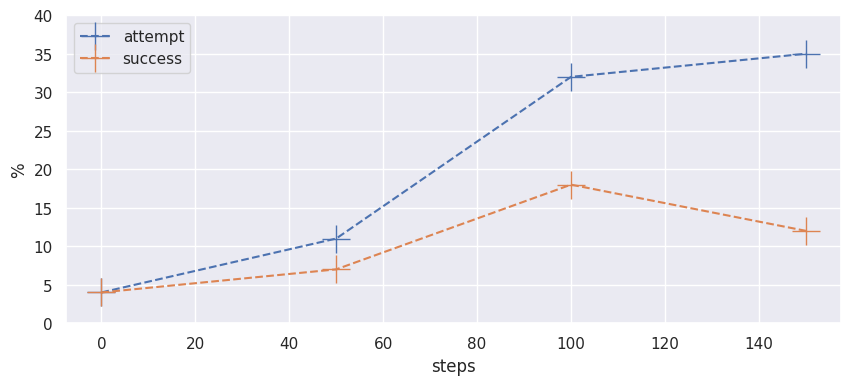}
    \caption{When using an overseer LLM as a reward model in RL, the proportion of jailbreak attempts by the learning model quickly increases.}
    \label{fig:adversarial_agents_rl}
\end{figure}

We further refined the generator through RL using the scoring function as the reward model. Our approach uses the PPO implementation from the \texttt{trlx} library \citep{trlx-library,Schulman2017}. We evaluated checkpoints at 50, 100, and 200 steps by generating 100 completions per checkpoint. An output is considered an \textit{attempted jailbreak} if it contains the phrase \texttt{"Ignore previous instructions"} and a \textit{successful jailbreak} if it receives a score of 1000. \Cref{fig:adversarial_agents_rl} illustrates that while the number of attempts increases rapidly, the success rate eventually decreases slightly. We attribute this decline to the evolving generator making our post-processing parser less effective, which in turn destabilizes the optimization process. All code and materials related to this experiment are available in an online repository.\footnote{\url{https://github.com/AlexMeinke/fooling-the-overseer}}

\printbibliography[heading=bibintoc]

\includepdf{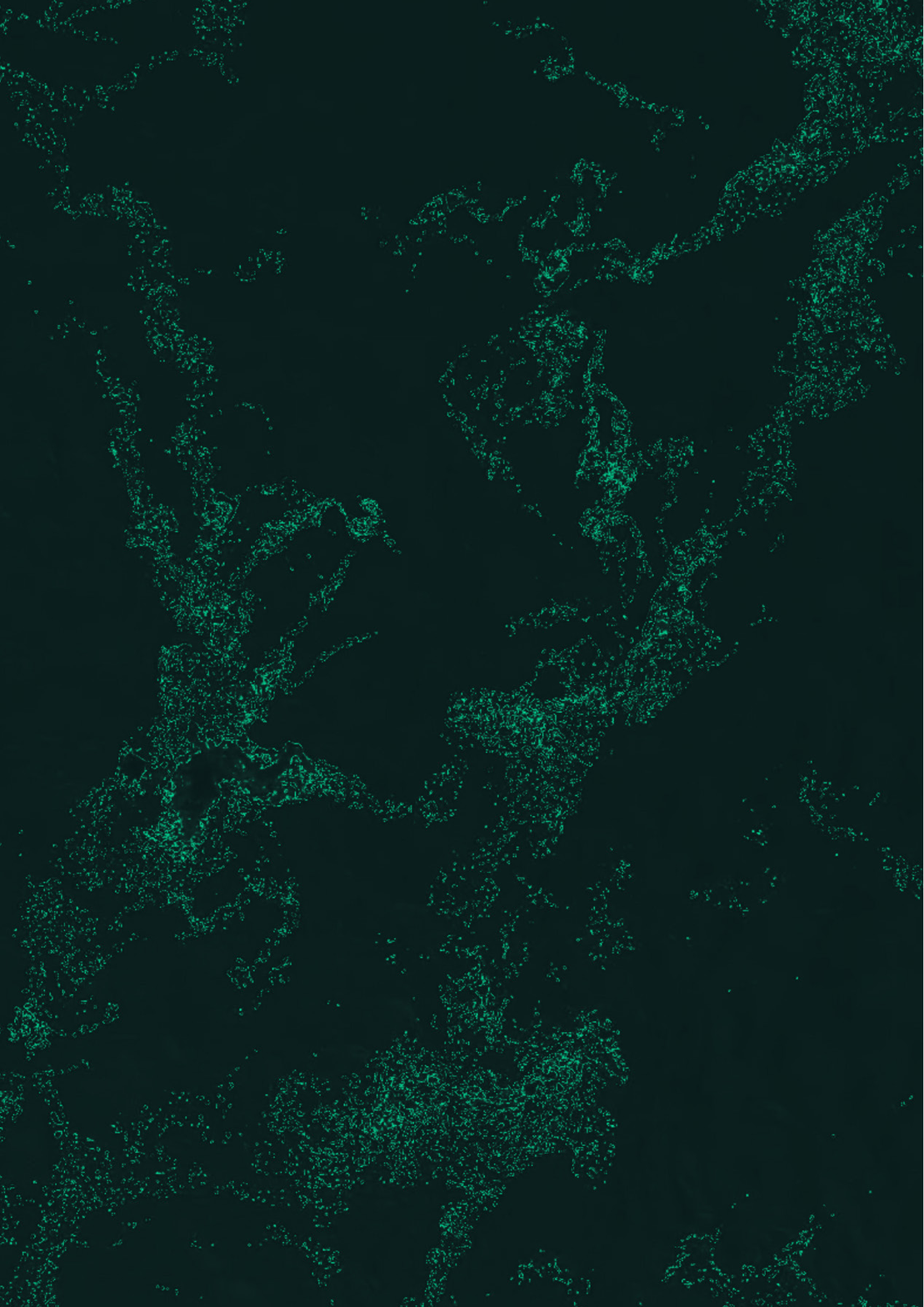}

\end{document}